\documentclass[twocolumn,showpacs,preprintnumbers,amsmath,amssymb,superscriptaddress,prc]{revtex4-2}

\usepackage{graphicx}
\usepackage{dcolumn}
\usepackage{bm}

\begin{document}

\title{Structure of even-even Zr isotopes with 52$\leq$N$\leq$58 neutrons.}

\author{W. Urban}
\author{A. Abramuk}
\author{T. Rz\c{a}ca-Urban}
\affiliation{Faculty of Physics, University of Warsaw, ulica Pasteura 5, PL-02-093 Warsaw, Poland}
\author{M.~Jentschel}
\author{P. Mutti} 
\author{U. K\"oster} 
\affiliation{Institut Laue-Langevin, Grenoble, France}
\author{G. de France}
\affiliation{Grand Acc\'el\'erateur National d'Ions Lourds (GANIL), CEA/DSM - CNRS/IN2P3, 
             Bd Henri Becquerel, BP 55027, F-14076 Caen Cedex 5, France}   
\author{C.A. Ur} 
\affiliation{INFN, Legnaro, Italy}
\date{\today}

\begin{abstract}

Excited levels in $^{92}$Zr have been studied in cold-neutron capture reaction using the EXILL
Ge array at ILL Grenoble. Excited levels in $^{94}$Zr and $^{98}$Zr nuclei have been studied
using high-statistics $\gamma$-coincidence data from neutron-induced fission of $^{235}$U
measured with EXILL. These data were complemented by measurements of $\beta$ decay of yttrium
isotopes produced in fission of $^{235}$U.

The goal of these measurements was to search for new levels in $^{92,94,98}$Zr and to
improve spin-parity assignments to excited levels. Total of 54 new levels, 180 new $\gamma$
transitions and 70 new or improved spin-parity assignments have been determined in the three
isotopes in the present work.

A precise neutron binding energy of 8634.81(2) keV has been determined in $^{92}$Zr. In
$^{94}$Zr a Gamow-Teller transition is proposed. The (9$^-$), 3894.1-keV level in $^{98}$Zr
is likely a few ns isomer.

Structures with characteristic 3$^+$ excitations, which may relate to $\gamma$ collectivity,
are identified in $^{92,94,98}$Zr. They are strongly mixed with collective structures based
on 0$^+_2$ levels.

A new technique of estimating half-lives in a picosecond range, developed in this work,
provided 61 half-lives and 60 B($\pi ,\lambda$) rates for transitions in $^{96,98}$Zr
and $^{94,96}$Sr.

New-type, phenomenological systematics, backed by Large Scale Shell Model calculations
done before have been used to classify 2$^+$ excitations in Zn-Zr even-even nuclei of the
50$\le$N$\le$60 range. This information and new phenomenological systematics of 0$^+$
excitations in Sr and Zr isotopes are used to explain the evolution of collectivity in
zirconium isotopes, showing the role of various single particle excitations in the phase
transition and coexistence in this region. In particular, the role of the $\nu$9/2$^+$[404]
extruder orbital as a catalyst in creating 0$^+$ excitations and the deformation change in
the region has been discussed.

\end{abstract}

\maketitle
\section{Introduction}

\begin{figure}
\centering
\scalebox{0.21}{\includegraphics{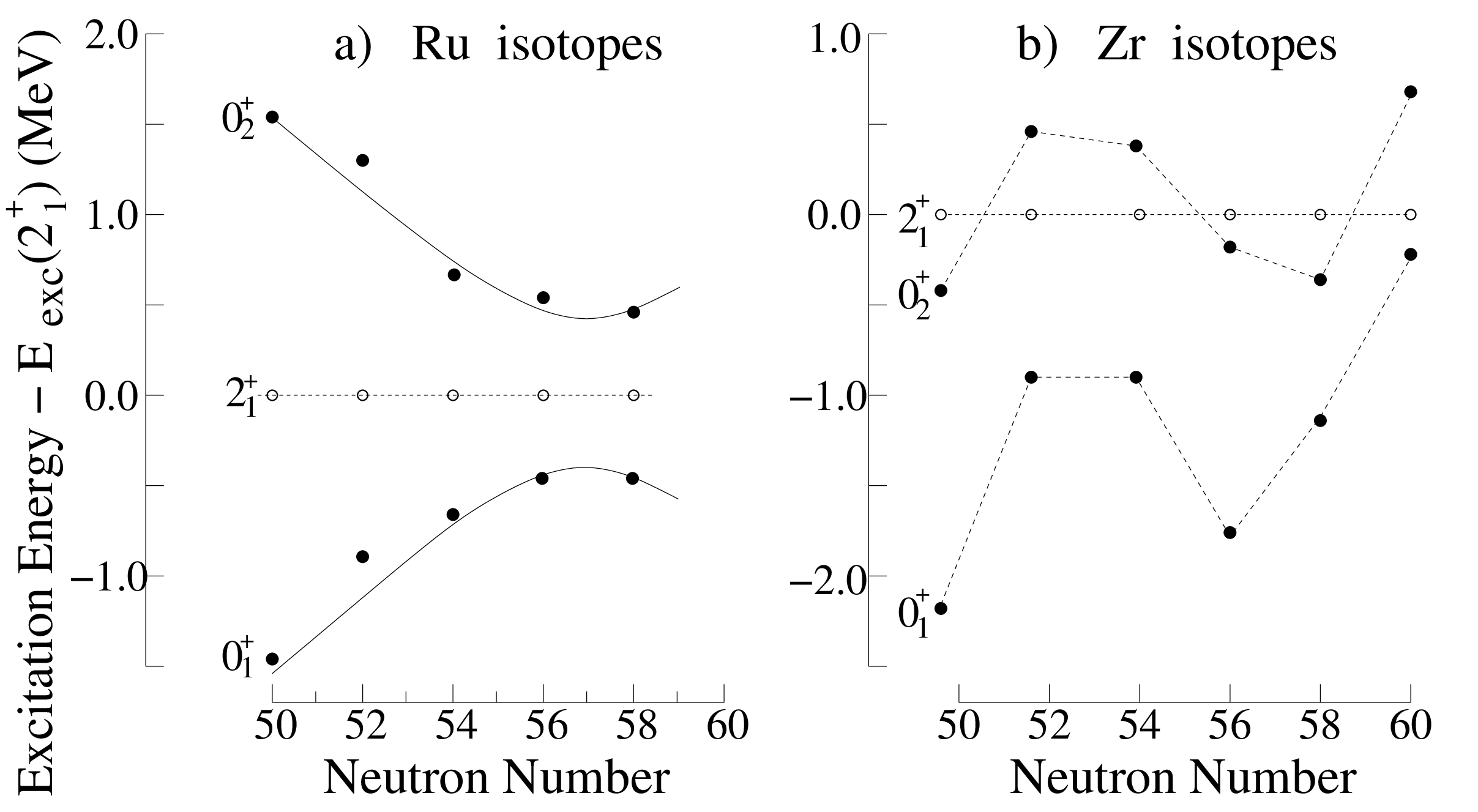}}
\caption{ Excitation energies of $0^+_1$ and $0^+_2$ levels in (a) Ru and (b) Zr isotopes, 
drawn relative to the respective 2$^+_1$ excitations.  Solid lines in (a) represent 
the two-level mixing calculation \cite{Urb13}. Dashed lines are drawn to guide the eye. The 
experimental data are taken from Refs. \cite{Urb13,ENSDF}. See text for further comments.}
\label{Zr_even_light_fig1}
\end{figure}

The mechanism of creation of 0$_2^+$ excited states, the internal structure of these states 
and their relation to the deformation change in the A$\approx$100 region is a subject of
intensive studies. Recently, we reported on the systematics of 0$_2^+$ excited states in Ru 
\cite{Urb13} and Sr \cite{Urb21} isotopes, which differ dramatically in the two isotopic
chains. The picture of the interacting 0$^+_1$ and 0$^+_2$ levels in Ru isotopes, drawn
relative to the 2$^+_1$ excitations, as shown in Fig. \ref{Zr_even_light_fig1}(a), may be
seen as an avoided crossing \cite{Lan81} of the 0$^+_1$ and 0$^+_2$ structures, with the
interaction strength, V$\approx$400 keV \cite{Urb13}. This picture suggests that 2$^+_1$
excitations in Ru isotopes correspond to collective, phonon excitations and the 0$^+_2$
levels in Ru may result from coupling of two of such phonons. An analogous picture drawn for
transitional Sr isotopes looks completely different, as seen in Fig. 1 of Ref. \cite{Urb19}.
Although still regular, it does not show any interaction between 0$^+_1$ and 0$^+_2$ levels
and no relation to 2$^+_1$ excitations. As discussed in our work on transitional Sr isotopes
\cite{Urb21}, the 2$^+_1$ excitations in Sr nuclei are dominated by a few single-particle
(s.p.) configurations and are not of a phonon nature.

\begin{figure}
\centering
\scalebox{.33}{\includegraphics{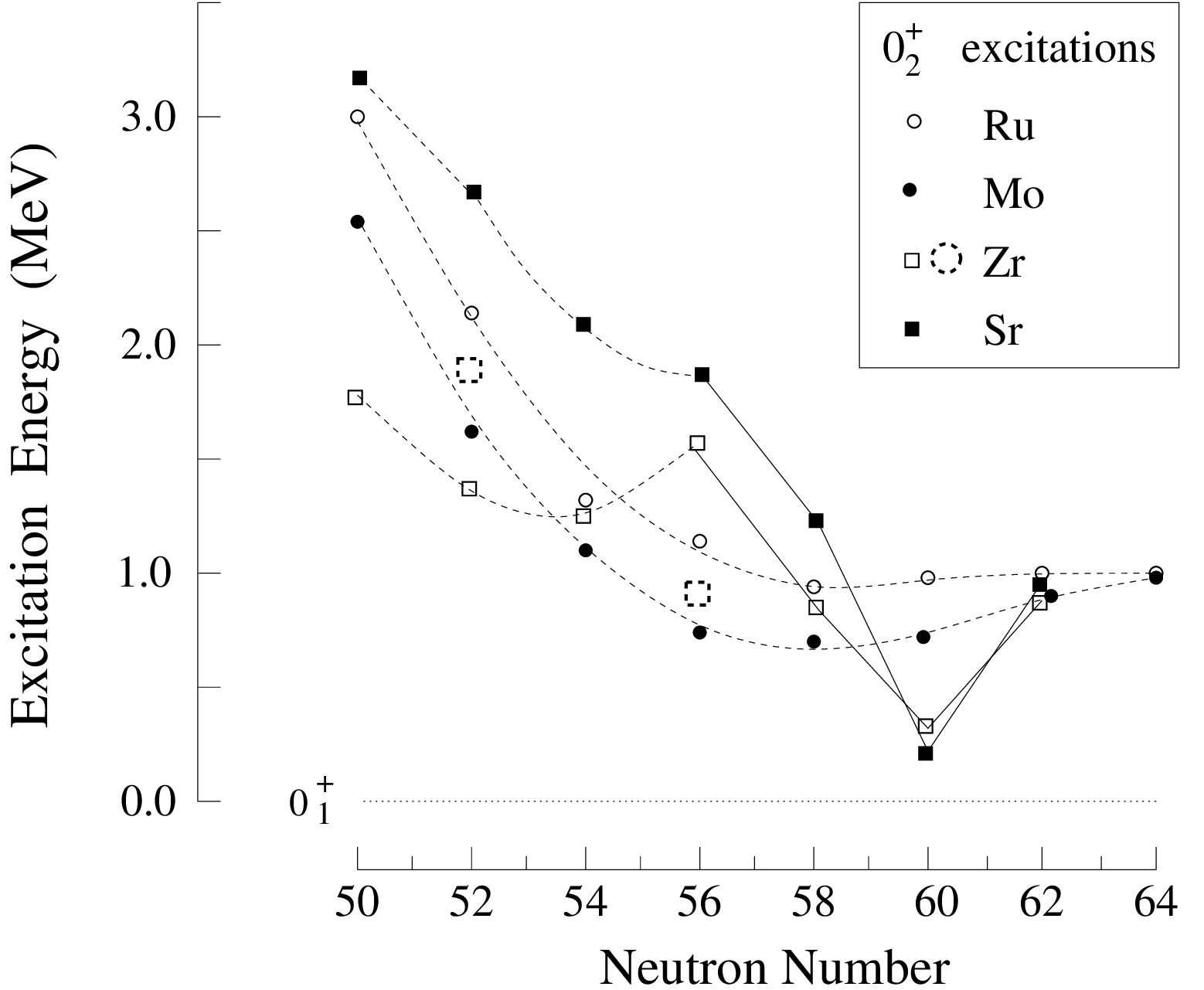}}
\caption{Excitation energies of $0^+_2$ levels relative to the 0$^+_1$ ground state in the 
50$\leq$N$\leq$62 isotopes of Sr, Zr, Mo and Ru. For Mo and Ru points at N=64 are also shown.
The data are taken from Refs. \cite{ENSDF,Cru20}. Lines are drawn to guide the eye.}
\label{Zr_even_light_fig2}
\end{figure}

Compared to Ru and Sr pictures the trend for 0$^+_1$ and 0$^+_2$ levels in Zr transitional 
isotopes shown in Fig. \ref{Zr_even_light_fig1}(b) seems chaotic and without any correlation
between 0$^+$ and 2$^+$ levels, although there is some correlation between the 0$^+_1$ and
0$^+_2$ levels. This suggests displaying 0$^+_2$ excitation energies relative to the 0$^+_1$
levels. Indeed, trends shown in Fig. \ref{Zr_even_light_fig2} are more regular than in Fig.
\ref{Zr_even_light_fig1}. There is some similarity between systematics of 0$^+_2$ levels in
Sr and Zr isotopes, which are clearly different from trends of 0$^+_2$ levels in Mo
and Ru isotopes. In Ref. \cite{Urb19} we asked whether this difference could be due an
incomplete experimental data and suggested searching for 0$^+$ levels around positions shown
in Fig. \ref{Zr_even_light_fig2} by dashed circles.  Our recent study has not revealed any
candidate for a 0$^+$ level around 1 MeV in $^{96}$Zr \cite{Wis23} .

The present investigation of Zr isotopes, analogous to our work on even-Sr isotopes in the
50$\leq$N$\leq$58 range \cite{Urb21}, reports on studies of low-spin, single-particle (s.p.)
as well as collective excitations, including $\gamma$ and octupole vibrations
and rotations of deformed structures. Among others we present
the ``complete spectroscopy'' measurement of $^{92}$Zr using the $^{91}Zr(n,\gamma )$
reaction, aimed at searching for a 0$^+$ excitation around 2 MeV in $^{92}$Zr, the other
suggestion marked by a dashed circle in Fig. \ref{Zr_even_light_fig2}.

Section II presents the measurements performed in this work. New experimental results
obtained for $^{92}$Zr, $^{94}$Zr and $^{98}$Zr are described in Section III and in
Section IV the new data in Zr nuclei are discussed together with other excited levels in the 
region and new interpretations are proposed. Section V summarizes the work.

\section{Experimental details}

\subsection{Measurements and analysis techniques}

New experimental results on even-even Zr isotopes with 50$\leq$N$\leq$60 neutrons have been
obtained from a campaign of measurements of $\gamma$ rays following neutron-induced fission
of $^{235}$U and cold-neutron capture reactions on zirconium targets. The measurements have
been performed at the PFB1 cold-neutron beam facility \cite{Urb13b} of the Institut
Laue-Langevey (ILL) in Grenoble using the EXILL array \cite{Jen17} consisting of 16 large
germanium detectors, including eight Clovers.

For $^{98}$Zr and $^{96}$Zr prompt-$\gamma$ decays following neutron-induced fission of
$^{235}$U have been observed using EXILL. The measurement lasting 21 days provided
information on medium-spin excitations in the two nuclei. One of the $^{235}$U targets used
was sandwiched between two $^{Nat}$Zr foils to stop fission fragments. Neutron capture on
this backing material provided high-statistics data from the $^{91}$Zr($n,\gamma )^{92}$Zr
reaction. The fission run provided also high-statistics data from $\beta^-$ decay of
yttrium isotopes produced in fission of $^{235}$U however, immersed in the complex data
from neutron-induced fission of $^{235}$U. Favourably, the 934.51-keV main transition in
$^{92}$Zr forms a well separated peak in this complex $\gamma$ spectrum.

After the fission run a 4 hour measurement of target activity was taken. Strong production
of mass A=92 fission fragments resulted in high cumulative yield of $^{92}$Y and the intense
$\beta^-$ decay with half-life of T$_{1/2}$=3.5 h to levels in $^{92}$Zr. The measurement
provided a clean set of data, competitive to the previous study \cite{Tal70}.

\begin{table}[]
\caption{The Zr target used in the present neutron-capture measurement. Neutron-capture cross 
sections are in barn for thermal neutrons \cite{ENSDF,NGATLAS}.}
\begin{center}
\begin{tabular}{l c c c c c c c c}
\hline
           & & & & & & & & \\
Isotope    & $^{90}$Zr&$^{91}$Zr&$^{92}$Zr&$^{93}$Zr&$^{94}$Zr&$^{95}$Zr&$^{96}$Zr&$^{97}$Zr\\
           & & & & & & & & \\
Fraction   & & & & & & & & \\
in target  &~~~19.3\%  & ~5.1\% & ~7.8\%  &  -      & 8.2\%   &   -     &59.6\%   &   -     \\
           & & & & & & & & \\
Cross      & & & & & & & & \\
section    & ~0.078  & 0.83     &  0.23   &  -      & 0.050   &   -     & 0.022   &   -     \\
           & & & & & & & & \\
Capture    & & & & & & & & \\
yield      &    -    &  16\%    & 45\%    &  19\%   &    -    &  5\%    &   -     &   15\%  \\
\hline
\end{tabular}
\end{center}
\label{Zr92_pub_tab1}
\end{table}

Using EXILL we also performed a neutron-capture measurement on Zr target enriched in $^{96}$Zr.
It was dedicated to study excited states in $^{97}$Zr \cite{Rza18} but provided additional
information on other Zr isotopes due to admixtures in the target. The target contained 700 mg
of zirconium dioxide (ZrO$_2$) powder contained in two bags of negligible-mass, made of
fluorinated-ethylene-propylene copolymer. The isotopic composition of the target is shown in
Table \ref{Zr92_pub_tab1}. The measurement lasting 21 hours delivered about 3.3$\cdot$10$^{10}$
trigger-less events.

\subsection{Calibrations of EXILL}

Precise energy and efficiency calibrations of the EXILL array have been performed in the
energy range from 30 keV to 10 MeV, using standard calibration sources of $^{133}$Br,
$^{60}$Co and $^{152}$Eu as well as data from the $^{27}$Al({\it n},$\gamma$)$^{28}$Al and
$^{35}$Cl({\it n},$\gamma$)$^{36}$Cl neutron-capture reactions. The accuracy of the obtained
EXILL energy calibration is 0.03 keV and the uncertainty of the efficiency calibration is
about 3$\%$ as described in Ref. \cite{Jen17}.

To include in the analyzed histograms the energy range up to 9 MeV we applied
``constant-peak-width'', nonlinear energy calibration, which maintains approximately constant
peak width over the whole energy range. The calibration enhances the visibility of high-energy
$\gamma$ lines without loosing the resolving power there while not compromising the energy
resolution at low energies \cite{Urb13b}. 

\subsection{Angular correlations and directional-linear polarization with EXILL.}

To determine spins and parities of excited states in the studied Zr isotopes we performed 
measurements of angular correlations and directional-linear polarization using the eight 
Clover detectors of the EXILL array. These detectors, mounted in one plane in octagonal 
geometry, provided the possibility of measuring $\gamma$-$\gamma$ angular correlations at 
three different angles between detectors, 0$^o$, 45$^o$ and 90$^o$ using angular-correlation
techniques developed for EXILL-type arrays \cite{Urb13b,Cze15,Jen17}.

For strong transitions in $^{92}$Zr we could also determine linear polarization by
measuring directional-polarization correlations in $\gamma\gamma$ cascades, using the 
EXOGAM Clover detectors as Compton polarimeters. In the analysis we used techniques
for directional-polarization correlations with EXILL-type arrays \cite{Urb16,Urb23},
based on formulas and conventions of Refs. \cite{Kra73,Ham75}.

\subsection{Doppler-Broadening half-life Estimates (T$_{dbe}$)}

Electromagnetic decays of short-lived, excited levels populated in the fission fragments may
happen before fragments stop inside the fission target/source. The initial velocity of fission
fragments is as high as 4$\%$ of the speed of light and the stopping time may be of the
order of picoseconds, depending on the target composition. Fission fragments can fly in any
direction. Consequently, $\gamma$ lines corresponding to radiation emitted from flying fission
fragments will show significant broadening, due to the  Doppler effect. The shape of the
resulting broad $\gamma$ lines may provide information on half-lives of levels decaying during
flight, provided the make-up of the target and the stopping times of materials used are known.

For measurements with the EUROGAM Ge array \cite{Nol94,Urb97}, where the structure of the
fission source containing the $^{248}$Cm  isotope was known, the peak-shape analysis called
Doppler Profile Method (DPM) was developed \cite{Smi94} and used to determine half-lives of
levels in a range 0.5$< T_{1/2} <$5 ps for several nuclei \cite{Smi96,Urb01,Smi12}.

\begin{figure}[]
\centering
\scalebox{.73}{\includegraphics{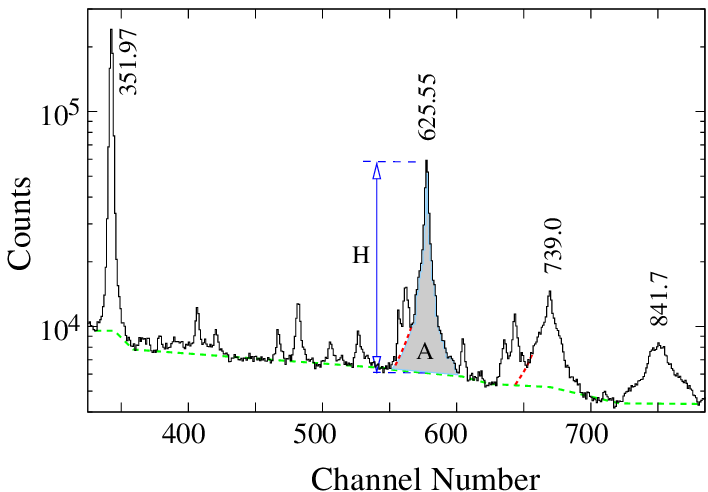}}
\caption{Fragment of a $\gamma$ spectrum, double-gated on the 212.61- and 497.36-keV lines of
$^{100}$Zr in the 3D histogram containing data from the EXILL fission measurement, shown to
illustrate the T$_{dbe}$ derivation technique. More details in the text.}
\label{Zr_even_light_TDBE}
\end{figure}

In the fission data from EXILL one clearly observes Doppler broadening of $\gamma$ lines.
However, the complex structure of the EXILL fission target \cite{Jen17} disabled the DPM
analysis. In the present work a simplified analysis, using Doppler broadening of $\gamma$ lines
to estimate half-lives, has been developed. The Doppler Broadening Estimate (DBE) analysis is
based on the observation that half-lives correlate well with a simple peak parameter, which
is the ratio of the peak height, H to the total peak area, A above the background line.

The DBE technique is illustrated in Fig. \ref{Zr_even_light_TDBE} showing an excerpt of a
$\gamma$ spectrum with lines in the ground-state cascade of $^{100}$Zr. The spectrum,
doubly-gated on the 212.61- and 497.36-keV lines in a triple-$\gamma$ histogram containing
data from the EXILL measurement of neutron-induced fission of $^{235}$U, shows the 351.97-,
625.55-, 739.0- and 841.7-keV lines of $^{100}$Zr \cite{ENSDF} (note the logarithmic scale on
the ordinate axis). The 351.97-keV line corresponding to $\gamma$ transition from the 4$^+_1$
level at 564.57 keV with the 7.0(4) ps half-life \cite{ENSDF} is not broadened, because this
transition happens mostly after stopping of fission fragments. However lines at 625.55, 739.0
and 841.7 keV, corresponding to decays from the 8$^+_1$, 10$^+_1$ and 12$^+_1$ levels, with
their respective half-lives of 1.75(17) ps, 0.75(9) ps and 0.37(4) ps are visibly wider, the
broadening increasing as the half-life decreases. The constant peak-width energy calibration
(Sect. II.B) facilitates the analysis, compensating for the increasing peak width with
increasing $\gamma$ energy due to Fano factor in Ge detectors.

\begin{table}[]
\caption{Values of the R=H/A ratio, obtained for $\gamma$ lines corresponding to decays of 
short-lived states in Sr, Zr and Mo isotopes. Energies of levels and $\gamma$ transitions,
spins of levels and their half-lives are taken from the Adopted and XUNDL datasets of Ref.
\cite{ENSDF} and from Refs. \cite{Urb01,Urb21,Smi12,Reg17,Ans17,Esm21,Ral17,Kru01}.}
\begin{center}
\begin{tabular}{c c c c c l }
\hline
  Isotope  & Level   &    Spin   & T$_{1/2}$  &E$_{\gamma}$ &R=H/A     \\
           & (keV)   & ($\hbar$) &   (ps)     & ~~~(keV)~~~ &          \\
\hline
           &         &           &            &             &          \\

 $^{96}$Sr & 1792.8  &   4$^+$   &  7(4)      &  977.8      & 0.27(1)  \\
           & 3126.0  &   8$^+$   &  3.15(44)  &  659.3      & 0.20(1)  \\
           & 3887.0  & (10$^+$)  &  1.50(21)  &  761.0      & 0.11(1)  \\
           & 4725.5  & (12$^+$)  &  0.90(12)  &  838.5      & 0.07(1)  \\
           &         &           &            &             &          \\
 $^{98}$Sr &  867.37 &   6$^+$   &  7.86(24)  &  433.3      & 0.27(1)  \\
           & 1433.65 &   8$^+$   &  2.97(48)  &  566.3      & 0.19(1)  \\
           & 2123.15 &  10$^+$   &  1.07(17)  &  689.5      & 0.09(1)  \\
           & 2927.7  & (12$^+$)  &  0.46(7)   &  804.5      & 0.045(9) \\
           &         &           &            &             &          \\
$^{100}$Zr &  564.5  &   4$^+$   &  37(3)     &  352.0      & 0.31(1)  \\
           & 1061.87 &   6$^+$   &  5.2(11)   &  497.4      & 0.24(1)  \\
           & 1687.42 &   8$^+$   &  1.75.17)  &  625.6      & 0.14(1)  \\
           & 2426.41 &  10$^+$   &  0.75(9)   &  739.0      & 0.06(1)  \\
           & 3268.1  & (12$^+$)  &  0.37(4)   &  841.7      & 0.040(5) \\
           &         &           &            &             &          \\

$^{102}$Zr &  478.4  &   4$^+$   &  31.9(5)   &  326.4      & 0.31(1)  \\
           & 1594.9  &   8$^+$   &  1.39(21)  &  630.1      & 0.11(1)  \\
           & 2351.5  &  10$^+$   &  0.53(10)  &  756.6      & 0.048(5) \\
           &         &           &            &             &          \\

$^{102}$Mo &  743.7  &   4$^+$   & 11.0(8)    &  447.1      & 0.29(1)  \\
           & 1327.9  &   6$^+$   & 4.6(+5,-21)&  584.2      & 0.24(1)  \\
           & 2018.8  &   8$^+$   &  1.8(3)    &  690.9      & 0.14(1)  \\
           & 2790.3  & (10$^+$)  &  1.03(18)  &  771.5      & 0.08(1)  \\
           & 3532.2  & (12$_2^+$)&  0.66(12)  &  842.0      & 0.05(1)  \\
           &         &           &            &             &          \\

$^{104}$Mo &  560.7  &   4$^+$   &  26.1(8)   &  368.4      & 0.30(1)  \\
           & 1080.0  &   6$^+$   &  4.73(15)  &  519.2      & 0.23(1)  \\
           & 1721.8  &   8$^+$   &  2.21(11)  &  641.7      & 0.14(1)  \\
           & 2455.4  &  10$^+$   &  1.08(7)   &  733.6      & 0.09(1)  \\
           &         &           &            &             &          \\
\hline
\end{tabular}
\end{center}
\label{Zr_even_light_table_TDBE}
\end{table}

The main sources of the uncertainty of the peak area, A, shown for the 625.55-keV line as
the shaded region, is the background-line estimate, shown by the green, dashed line in
Fig. \ref{Zr_even_light_TDBE} and possible contaminating $\gamma$ lines, shown for the
625.55-keV line above the red, dashed line. The two contributions to the uncertainty of
the peak height, H are the level of the background line under the peak and, more
importantly, the position of the peak top. The latter is presently determined as the
average count in the two channels of a peak with the highest counts.

\begin{figure}[]
\centering
\scalebox{.31}{\includegraphics{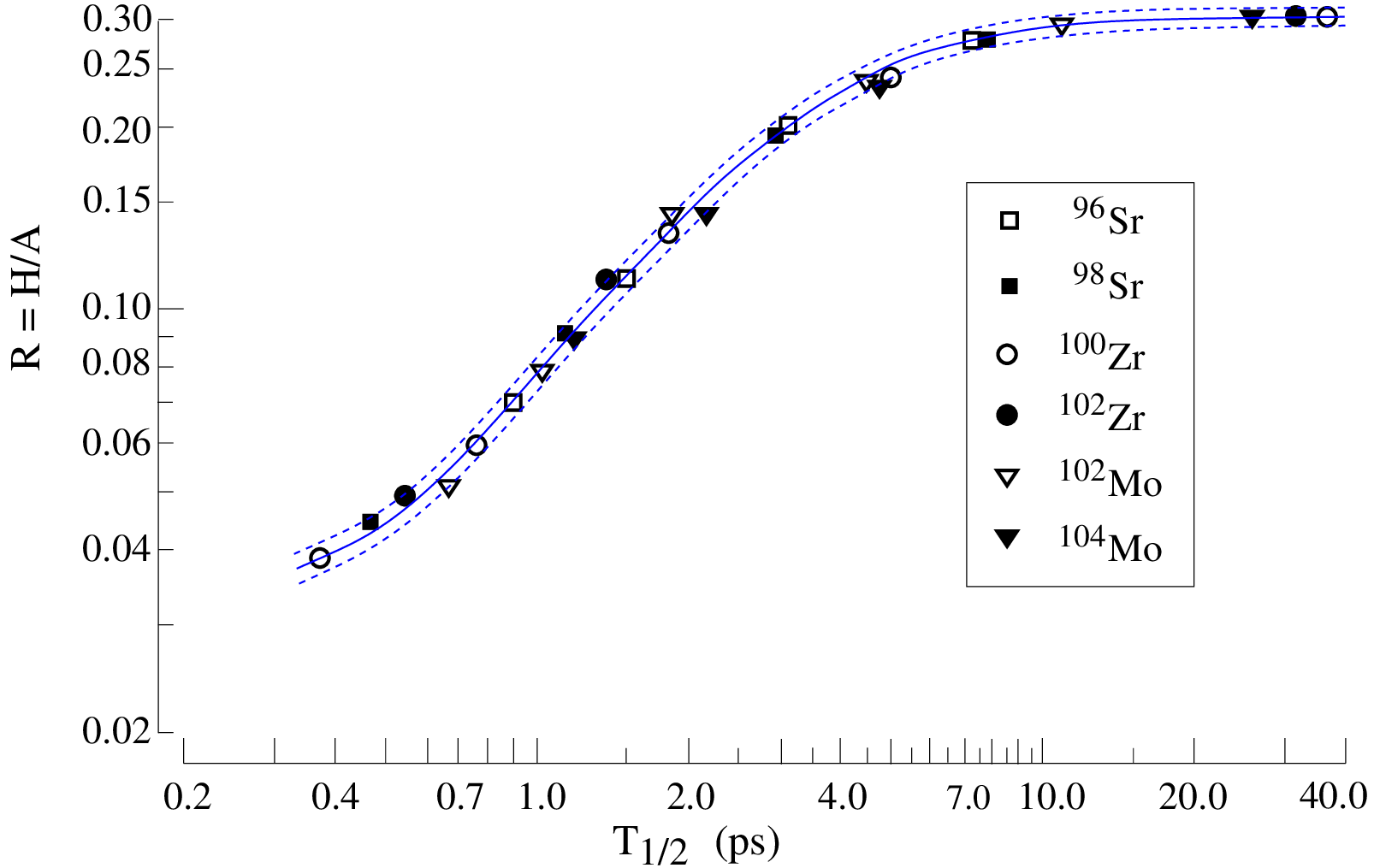}}
\caption{The DBE calibration curve for EXILL2012 fission measurement, as determined in
this work. Further explanations in the text.}
\label{Zr_even_light_TCAL}
\end{figure}

The resulting values of the R=H/A ratio, obtained for several peaks in Sr, Zr and Mo nuclei
strongly populated in fission of $^{235}$U, are listed in Table \ref{Zr_even_light_table_TDBE}
together with the corresponding literature half-lives \cite{ENSDF,Urb01,Urb21,Smi12}. The
obtained R values, drawn in Fig. \ref{Zr_even_light_TCAL} on a $log$-$log$ scale versus
respective half-lives, reveal a regular dependence on T$_{1/2}$. This may serve as
a calibration for half-life estimates, T$_{dbe}$, based on the R values determined
for $\gamma$ lines observed in the EXILL fission data. The analytical form of the calibration
curve, drawn as a solid line in Fig. \ref{Zr_even_light_TCAL}, is not known and has been
sketched as a spline. For ``fully stopped lines'', corresponding to T$_{1/2}>$ 15 ps, the
ratio R saturates at 0.30(1). At low T$_{1/2}<$0.4 ps the R ratio saturates at about 0.03
because H values become similar for very broad peaks.

In Fig. \ref{Zr_even_light_TCAL} uncertainties of the R and T$_{1/2}$ values listed in Table
\ref{Zr_even_light_table_TDBE} are not shown. Instead, we mark by dashed lines a corridor of
errors by moving the calibration line up and down by an averaged uncertainty of
the R value at the corresponding T$_{1/2}$ in Table \ref{Zr_even_light_table_TDBE}.

An important advantage of DBE analysis, which uses the $empirical$ calibration shown in Fig.
\ref{Zr_even_light_TCAL} is that it automatically includes corrections applied in the DPM
analysis \cite{Smi94}, like the amount of fast side feeding or feeding from the level above
in a cascade. Therefore, for levels in rotational cascades similar to those listed in Table
\ref{Zr_even_light_table_TDBE} the T$_{dbe}$ values will represent their half-lives. For
levels in non-rotational cascades or levels with extra feeding the T$_{dbe}$ values give
definite, $upper$ limits for their half-lives. For ``fully stopped lines'' the $lower$
limit of 20 ps for half-lives of levels in question can be set.

The analysed spectra should be gated on lines below the line of interest or
lines in the complementary fission fragments. Full account of the DBE analysis technique
will be reported in another work, after a sufficient number of analysed cases provides
a satifying conversion of T$_{dbe}$ values to T$_{1/2}$ half-lives.

\section{Data analysis and results}

\subsection{Excitations in $^{92}$Zr}

The $^{92}_{40}$Zr$_{52}$ nucleus is a subject of intensive research, aimed at explaining
the emerging collective effects in transitional nuclei of the region.
To properly describe possible excitation modes in $^{92}$Zr one should identify all low-spin
levels up to about 4 MeV and, importantly, determine their spins and parities. A suitable
process, seen as a ``complete spectroscopy'' tool, is the neutron-capture reaction. Neutron
capture data can be complemented by a $\beta$-decay measurement, which often populates the
same excited states but with different intensity. While neutron capture will populate
collective states via $\gamma$ decays of the complex, neutron-capture level, the $\beta$
decay favors the population of single-particle states.

\begin{figure}[]
\centering
\scalebox{.33}{\includegraphics{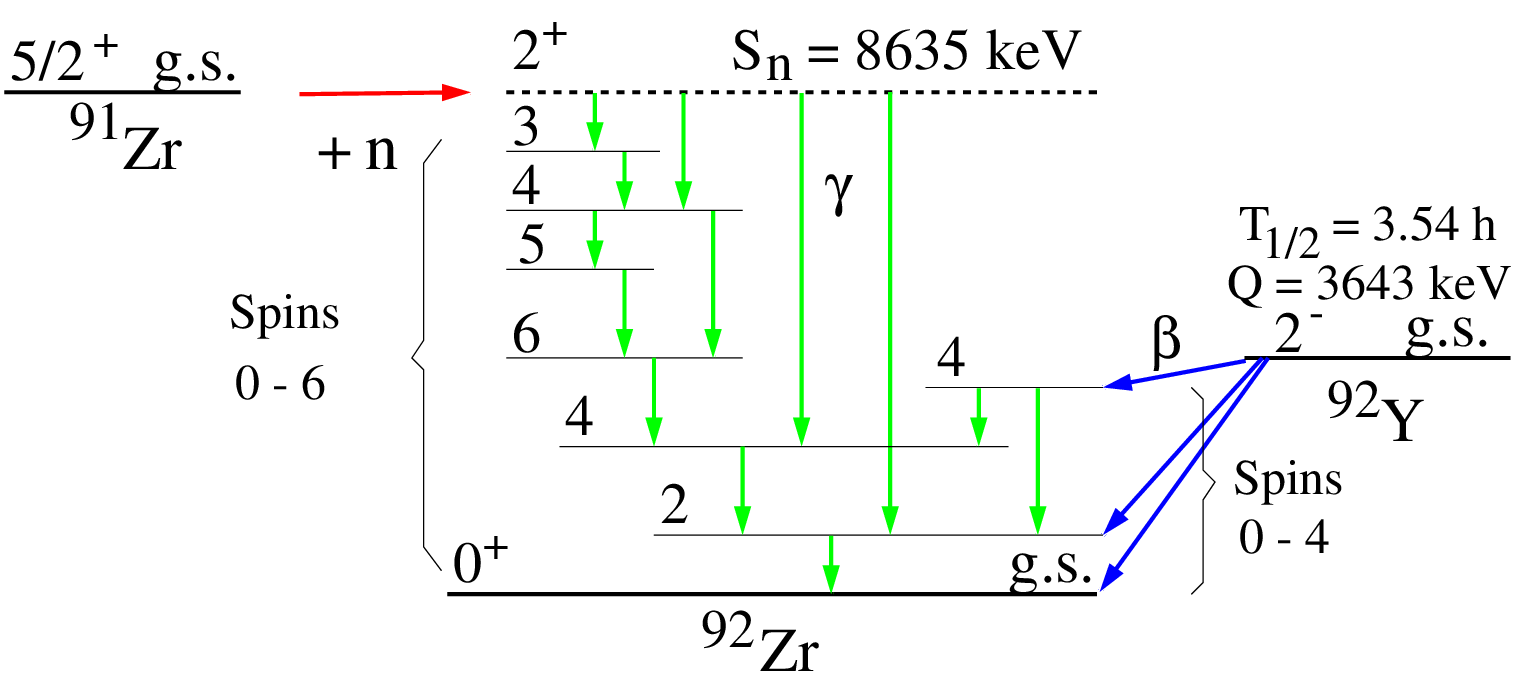}}
\caption{Schematic representation of neutron-capture and $\beta^-$ decay processes
populating excited states in $^{92}$Zr. The data are taken from Ref. \cite{NDS_Zr92}.}
\label{Zr_even_light_fig5}
\end{figure}

Figure \ref{Zr_even_light_fig5} shows schematically the two processes populating levels
in $^{92}$Zr. The $\gamma$ cascades decaying from the high-energy, neutron-capture level in
$^{92}$Zr with spin-parity I$^{\pi}$=2$^+$, may populate levels of both parities and spins
in a range from 0 up to 6 $\hbar$. The $\beta^-$-decay of the I$^{\pi}$=2$^-$ ground state
of $^{92}$Y will populate states with spins in a range from 0 up to 4 $\hbar$, preferably
with negative parity. It is expected that all 0$^+$ levels in $^{92}$Zr with excitation
energy below 3 MeV should be populated following either the $(n,\gamma)$ or $\beta^-$-decay
reactions.

The present work reports on the experimental study of low-spin excitations in $^{92}$Zr
populated in the cold-neutron capture on $^{91}$Zr and in $\beta^-$ decay of $^{92}$Y.

Prior to this work low- and medium-spin levels in $^{92}$Zr were reported in over forty
studies listed in the compilation \cite{NDS_Zr92}, including $\beta^-$ decay of $^{92}$Y 
\cite{Tal70,Mac90} and three $^{91}$Zr(n,$\gamma$)$^{92}$Zr neutron-capture works compiled 
in Ref. \cite{NDS_Zr92}. The yrast structure of $^{92}$Zr was observed up to spin
I$^{\pi}$=18$^+$ in heavy-ion-induced fission works \cite{Fot02,Pan05} with an intriguing
lifetime measurement of the 10$^+_1$ level \cite{Sug17}. Important contributions came from
inelastic photon scattering \cite{Wer02} and transfer reactions compiled in \cite{NDS_Zr92}.
One should also mention crucial lifetime measurements of Ref. \cite{Pet13} and references
therein.

\subsubsection{Cold-neutron capture on enriched Zr target}

\begin{figure*}[]
\centering
\scalebox{0.97}{\includegraphics{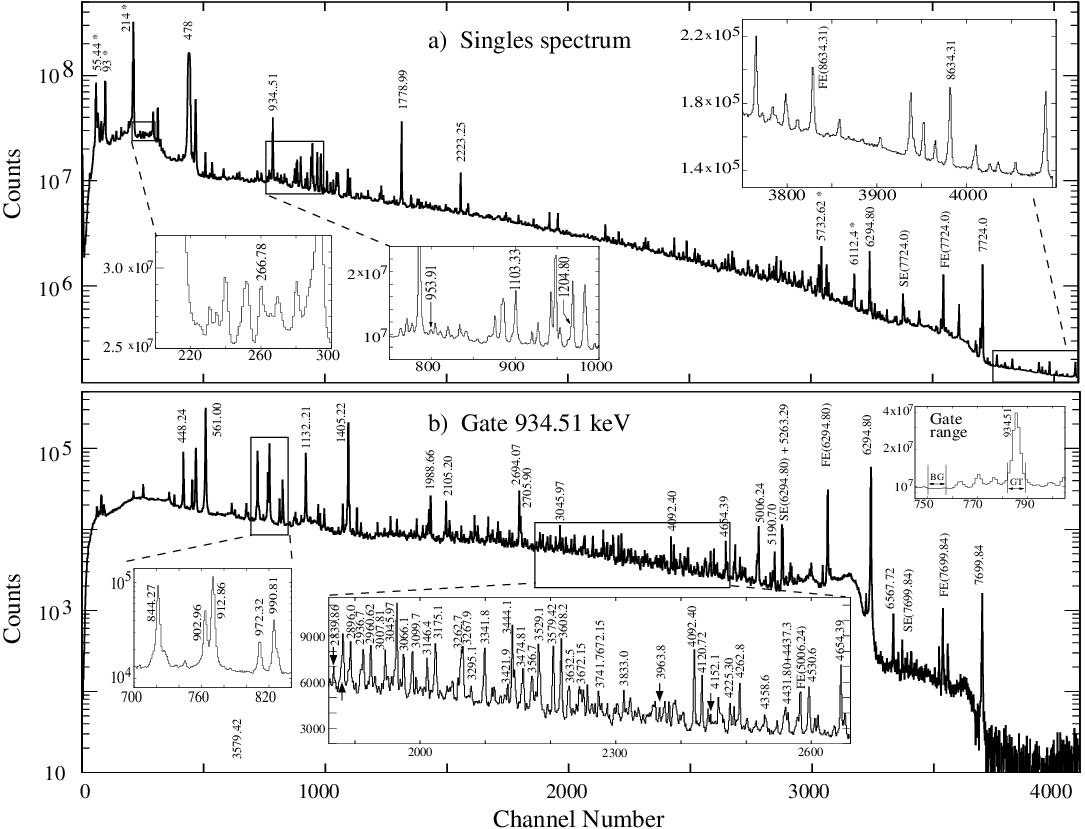}}
\caption{(a) Singles $\gamma$ spectrum from the cold-neutron capture on the $^{96}$Zr
enriched target. The insets show parts of the spectrum with lines of Zr isotopes. First
escape (FE) and second escape (SE) lines are marked. Asterisks indicate lines of Hf
isotopes (see text for further description of lines). (b) $\gamma$ spectrum gated on the
934.51-keV line with gate (GT) and background (BG) regions shown in the ``Gate range''
inset. The other inset shows fragment around the 912.86-keV line. In spectra (a) and (b)
constant-peak-width energy calibration has been applied:
E$_{\gamma}$=C$_0$+C$_1*(channel)$+C$_2*(channel)^2$, with C$_0$=-0.41 keV,
C$_1$=0.9507076 keV/channel and C$_2$=0.0003059994 keV/channel$^2$. Lines are labeled with
their $\gamma$ energies in keV. Energies of $\gamma$ lines shown in the spectra are taken
from Refs. \cite{NDS_Zr91,NDS_Zr92,NDS_Zr93,NDS_Zr95,Rza18} and the present work.}
\label{Zr_even_light_fig6}
\end{figure*}

Figure \ref{Zr_even_light_fig6}(a) shows the singles $\gamma$ spectrum from the 21-hour
measurement of neutron capture on the enriched Zr target. The $\gamma$ lines corresponding
to ground-state transitions in various Zr isotopes produced in the reaction, are marked by
their energies in keV, including the 1204.80-keV line of $^{91}$Zr \cite{NDS_Zr91}, the
934.51-keV line of $^{92}$Zr, the 266.78-keV line of $^{93}$Zr \cite{NDS_Zr93}, the
953.91-keV line of $^{95}$Zr \cite{Rza23} and the 1103.33-keV line of $^{97}$Zr
\cite{Rza18}. Also marked are the 6294.88- and 8634.4-keV lines corresponding to primary
transitions in $^{92}$Zr \cite{NDS_Zr92}.

The spectrum in Fig. \ref{Zr_even_light_fig6}(a) is dominated by Hf lines (the strongest
line at 214 keV), which is due to the 0.3$\%$ $^{Nat}$Hf admixture to the target mass and
their high neutron-capture cross section. A wide peak at 478 keV is due to (n,$\alpha\gamma$)
reaction on
traces of the $^{10}$B contamination of the ZrO$_2$ material. There are also lines of
$^2$H (2223.25-keV \cite{NDS2H}), $^{28}$Si (1778.99-keV) and $^{28}$Al (7724.03-keV)
\cite{NDS28Al}. Lines of $^{28}$Al with known, precise energies \cite{Sch82} have been
used for an internal energy calibration.

Various 2D and 3D histograms have been used to build excitation scheme of $^{92}$Zr.
Figure \ref{Zr_even_light_fig6}(b) shows the $\gamma$ spectrum doubly gated on the 934.51-keV
line of $^{92}$Zr (gating region shown in the inset) and on the ``prompt''-time peak in
the $gg$DT histogram, where DT is the time difference between the two $\gamma$ signals in a
coincidence event. Gating on the clean 934.51-keV peak eliminated strong contaminating
events seen in the singles spectrum in Fig. \ref{Zr_even_light_fig6}(a).

\begin{figure*}
\centering
\scalebox{.54}{\includegraphics{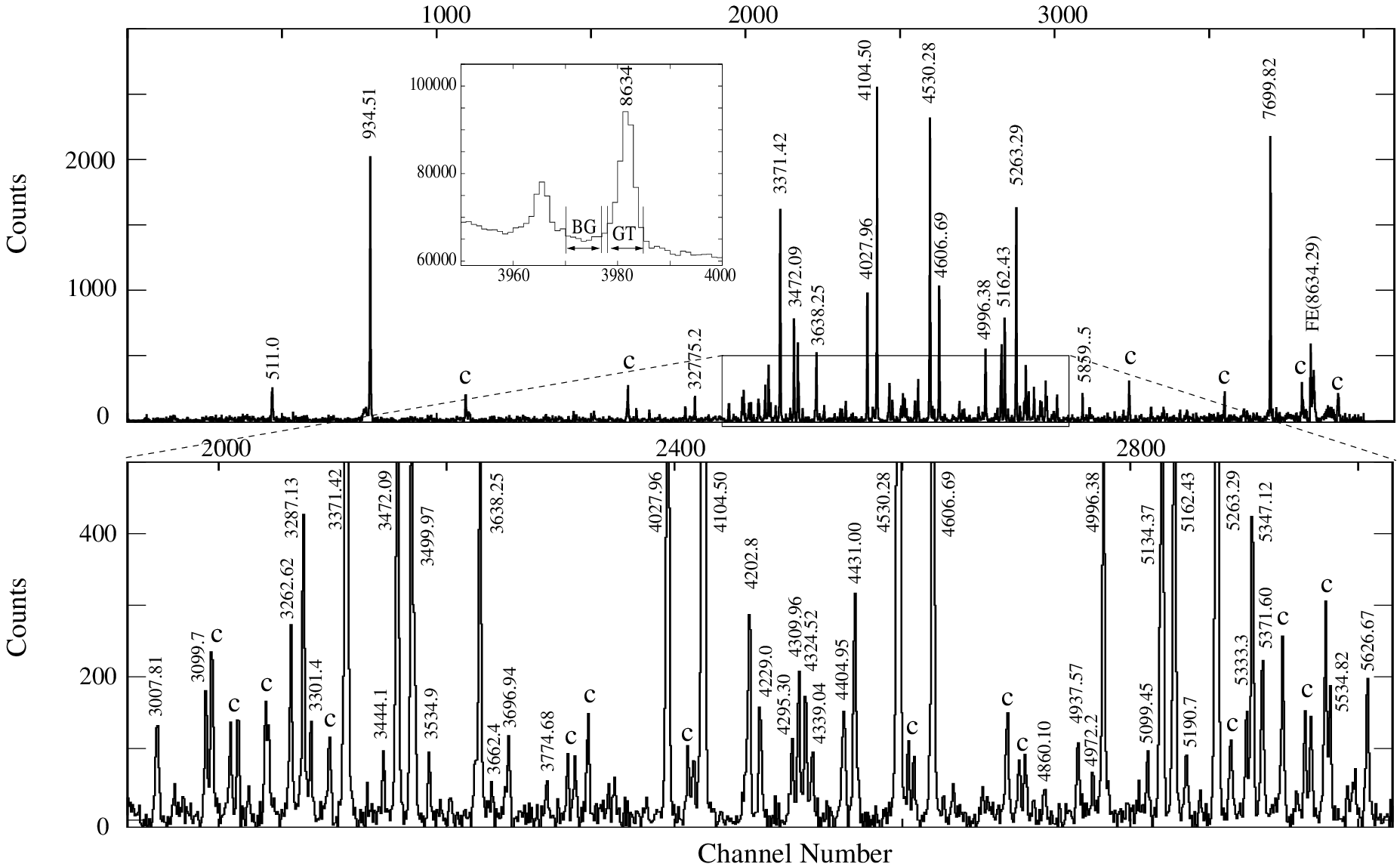}}
\caption{The upper panel shows the spectrum gated on the 8634-keV line on the $S2$ axis of
the $gS2$, histogram (gate region shown in the inset). The marked excerpt of the
spectrum is shown enlarged in the lower panel. Lines are labeled with their energies in
keV. The constant-peak-width calibration is applied. Label ``c'' denotes contaminating
lines. More comments in the text.}
\label{Zr_even_light_fig7}
\end{figure*}

Particularly useful was the $gS2$, histogram sorted out of $\gamma\gamma$ coincidences,
with the two $\gamma$ energies sorted along the $g$ axis and their sum along the $S2$.
Figure \ref{Zr_even_light_fig7} (a) displays the $\gamma$ spectrum gated on the $S2$ axis on
the 8634.31-keV line, corresponding to the neutron-capture energy in $^{92}$Zr (gate (GT)
and background (BG) regions shown in the inset). The spectrum contains pairs of $\gamma$
lines with summed energy within the GT range. Because the GT range at 8634 keV is about
20 keV, the spectrum may contain contaminating $\gamma\gamma$ cascades with their summed
energies within the (8634 +/- 10)-keV range.

In the $gS2$ histogram, the number of counts in the two $\gamma$ lines of a pair is the
same. This, combined  with the contstant-peak-width calibration applied, produces nearly
equal amplitudes of the two lines. The advantage of the $gS2$ analysis is a significant
reduction of Compton background in gated $\gamma$ spectra as seen by comparing Figs.
\ref{Zr_even_light_fig7} and \ref{Zr_even_light_fig6} (b). The spectrum in Fig.
\ref{Zr_even_light_fig6}(b) is background subtracted but this does not remove
Compton-scattered events forming ``Compton ridges'' in a conventional $\gamma\gamma$
histogram. Such ridges are not present in the $gS2$ analysis (see Ref. \cite{Hon96} for
more information (n,$\gamma$) analysis).

The $gS2$ matrix allowed identification of new, primary $\gamma$ transitions deexciting
the capture state in $^{92}$Zr. In Fig. \ref{Zr_even_light_fig7} we marked $\gamma$ lines
corresponding to pairs of transitions in $\gamma\gamma$ cascades deexciting the capture
level to the ground state in $^{92}$Zr. Gating on other histograms helped to find out,
which of the two transitions is the primary one.

The primary decays in cascades involving three or more $\gamma$ transitions are not
present in Fig. \ref{Zr_even_light_fig7}. They can be found by gating other lines on
the $S2$ axis. For example, $\gamma\gamma$ cascades feeding the 934.52-keV level in
$^{92}$Zr are found in a spectrum gated on the 7699.8-keV (= 8634.31-934.51) line on
the $S2$ axis.

\begin{figure*}
\centering
\scalebox{.33}{\includegraphics{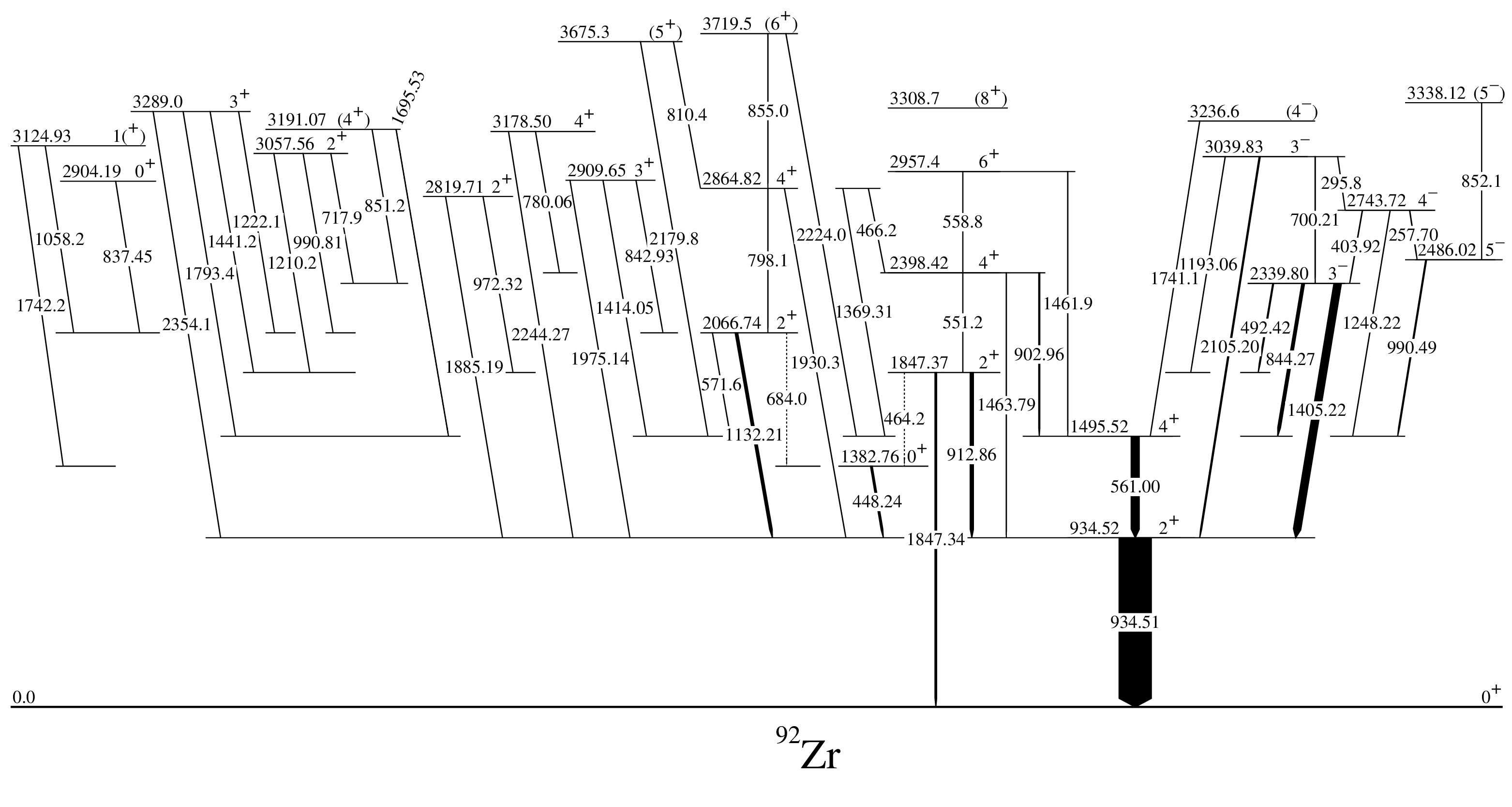}}
\caption{Partial level scheme of $^{92}$Zr populated in $^{91}$Zr(n,$\gamma$)$^{9}$Zr
reaction, as obtained in the present work. The arrow width is proportional to the $\gamma$
intensity. More comments in the text. The (8$^+$), 3308.7-keV level is shown according to
\cite{NDS_Zr92} to help the discussion.}
\label{Zr92_scheme.eps}
\end{figure*}

Figure \ref{Zr92_scheme.eps} shows a partial scheme of levels in $^{92}$Zr populated in
the $^{91}$Zr(n,$\gamma$)$^{92}$Zr, cold-neutron capture reaction in the present work.
Information on all excited states and their decays in $^{92}$Zr observed in the present
work in the $^{91}$Zr(n,$\gamma$) reaction is given in Tables \ref{Zr92_table_ng_1} -
\ref{Zr92_table_ng_5}.

Relative $\gamma$ intensities for transitions in $^{92}$Zr have been estimated by
combining decay branchings of excited levels with relative $\gamma$ intensities observed
in the singles spectrum and $\gamma\gamma$ coincidence spectra. They mostly agree with
previous thermal-neutron capture data \cite{NDS_Zr92} except I$_{\gamma}$(rel.) values
of the 990.49- and 2132.90-keV transitions.

The present work confirms excited levels reported in thermal-neutron capture works
\cite{NDS_Zr92} except the 2473.64-keV level. Our coincidences show the 2473.53-keV
(2473.6-keV in \cite{NDS_Zr92}) decay from the 3407.94-keV level. We add 40 new
levels, 148 new transitions including 53 new primary transitions and 59 new or revised
spin-parity assignments to the excitation scheme of $^{92}$Zr.

Important new results are the 551.2-558.8-keV and 798.1-855.0-keV cascades on top of
the 2$^+_2$, 1847.37-keV and 2$^+_3$, 2066.74-keV levels, respectively. We also propose
464.4- and 684.0-keV decays of these 2$^+$ levels to the 0$^+_2$ level at 1382.76 keV.
Both decays are on the limit of observation and are introduced tentatively.

No candidate for a 0$^+$ excitation around 1.9 MeV, suggested in Fig.
\ref{Zr_even_light_fig2}, has been found.

\begin{table}[]
\caption{Properties of excited states and their decays in $^{92}$Zr populated in the
$^{91}$Zr(n,$\gamma$) cold-neutron capture reaction, as observed in the present work.
Levels and transitions newly observed in this work, compared to previous (n,$\gamma$)
studies \cite{NDS_Zr92}, and spin-parity assignments new compared to the compilation
\cite{NDS_Zr92} are marked with asterisks.}
\begin{center}
\begin{tabular}{l c l r r r}
\hline
& & & & & \\
~E$_i$(keV)~~ &J$^{\pi}_i$ &E$_{\gamma}$(keV&I$_{\gamma}$(rel.)&~E$_f$(keV)&~~J$^{\pi}_f$\\
\hline
 ~934.52(3)   &   2$^+$    &  934.51(3)       & 1000(30)   &    0.00  &   0$^+$    \\
 1382.76(4)   &   0$^+$    &  448.24(3)       &    47(3)   &  934.52  &   2$^+$    \\
 1495.52(4)   &   4$^+$    &  561.00(3)       &   226(6)   &  934.52  &   2$^+$    \\
 1847.37(4)   &   2$^+$    &  464.4(3)*       &$\le$0.2    &  934.52  &   2$^+$    \\
              &            &  912.86(3)       &    90(3)   &  934.52  &   2$^+$    \\
              &            & 1847.34(4)       &    48(2)   &    0.00  &   0$^+$    \\
 2066.74(4)   &   2$^+$    &  219.5(3)        &     0.6(2) & 1847.37  &   2$^+$    \\
              &            &  571.6(3)        &     0.5(2) & 1495.52  &   4$^+$    \\
              &            &  684.0(4)*       &$\le$0.2    & 1382.76  &   0$^+$    \\
              &            & 1132.21(3)       &    78(3)   &  934.52  &   2$^+$    \\
 2339.80(4)   &   3$^-$    &  492.42(3)       &    25(1)   & 1847.37  &   2$^+$    \\
              &            &  844.27(4)       &    63(3)   & 1495.52  &   4$^+$    \\
              &            & 1405.22(3)       &   235(8)   &  934.52  &   2$^+$    \\
              &            & 2340.0(2)        &     0.5(2) &    0.00  &   0$^+$    \\
 2398.42(5)   &   4$^+$    &  551.2(4)*       &     0.3(1) & 1847.37  &   2$^+$    \\
              &            &  902.96(3)       &    31(2)   & 1495.52  &   4$^+$    \\
              &            & 1463.79(5)       &    10(1)   &  934.52  &   2$^+$    \\
 2486.02(6)   &   5$^-$    &  990.49(4)       &    19(1)   & 1495.52  &   4$^+$    \\
 2743.72(5)   &   4$^-$    &  257.70(5)       &     6.5(5) & 2486.02  &   5$^-$    \\
              &            &  403.92(6)       &     4.7(4) & 2339.80  &   3$^-$    \\
              &            & 1248.22(7)       &     6.8(5) & 1495.52  &   4$^+$    \\
 2819.71(5)   &   2$^+$    &  972.32(4)       &    11.0(6) & 1847.37  &   2$^+$    \\
              &            & 1885.19(8)       &     7.1(4) &  934.52  &   2$^+$    \\
 2864.82(7)   &   4$^+$    &  466.2(2)        &     2.5(5) & 2398.42  &   4$^+$    \\
              &            &  798.1(3)*       &     0.5(2) & 2066.74  &   2$^+$    \\
              &            & 1369.31(5)       &    13.0(5) & 1495.52  &   4$^+$    \\
              &            & 1930.3(2)        &     5.8(6) &  934.52  &   2$^+$    \\
 2904.19(12)  &   0$^+$    &  837.45(11)      &     2.1(3) & 2066.74  &   2$^+$    \\
 2909.65(5)   &   3$^+$    &  569.8(4)        &     0.4(2) & 2339.80  &   3$^-$    \\
              &            &  842.93(6)       &     5(1)   & 2066.74  &   2$^+$    \\
              &            & 1414.05(7)       &     7.8(5) & 1495.52  &   4$^+$    \\
              &            & 1975.14(7)       &    13(1)   &  934.52  &   2$^+$    \\
 2957.4(1)    &   6$^+$    &  558.8(3)*       &     0.4(1) & 2398.42  &   4$^+$    \\
              &            & 1461.9(1)        &     1.5(3) & 1495.52  &   4$^+$    \\
 3039.83(6)   &   3$^-$*   &  295.8(3)        &     0.3(1) & 2743.72  &   4$^-$    \\
              &            &  700.21(8)       &     4.8(4) & 2339.80  &   3$^-$    \\
              &            & 1193.06(29)      &     0.7(2) & 1847.37  &   2$^+$    \\
              &            & 2105.20(5)       &    22(2)   &  934.52  &   2$^+$    \\
 3057.56(7)   &   2$^+$    &  717.9(1)        &     2.2(4) & 2339.80  &   3$^-$    \\
              &            &  990.81(7)*      &     5.6(3) & 2066.74  &   2$^+$    \\
              &            & 1210.2(2)*       &     1.0(2) & 1847.37  &   2$^+$    \\
 3124.93(8)   &  1$^{(+)}$ & 1058.2(1)        &     2.5(5) & 2066.74  &   2$^+$    \\
              &            & 1742.2(1)*       &     4.3(3) & 1382.76  &   0$^+$    \\
 3178.50(7)   &   4$^+$    &  780.06(6)       &     5.9(4) & 2398.42  &   4$^+$    \\
              &            & 2244.27(21)      &     2.2(3) &  934.52  &   2$^+$    \\
 3191.07(8)   &  (4$^+$)*  &  851.2(2)*       &     1.9(3) & 2339.80  &   3$^-$    \\
              &            & 1695.53(7)       &     7.7(5) & 1495.52  &   4$^+$    \\
 3236.8(2)*   &  (4$^-$)*  & 1741.3(2)*       &     2.4(3) & 1495.52  &   4$^+$    \\
 3262.82(8)   &   2$^+$    & 2328.30(9)       &    13.5(5) &  934.52  &   2$^+$    \\
              &            & 3262.62(15)      &     4.2(4) &    0.00  &   0$^+$    \\
 3276.00(7)   &  (3$^+$)*  &  877.76(11)      &     2.5(3) & 2398.42  &   4$^+$    \\
              &            & 1209.18(7)       &     8.4(7) & 2066.74  &   2$^+$    \\
              &            & 2341.24(17)      &     4.0(3) &  934.52  &   2$^+$    \\
\hline
\end{tabular}
\end{center}
\label{Zr92_table_ng_1}
\end{table}

\begin{table}[]
\caption{Continuation of Table \ref{Zr92_table_ng_1}.}
\begin{center}
\begin{tabular}{l c l r r c}
\hline
& & & & & \\
~E$_i$(keV)~~&J$^{\pi}_i$ &E$_{\gamma}$(keV&I$_{\gamma}$ (rel.)&~E$_f$(keV)&~~J$^{\pi}_f$\\
\hline
 3287.17(13)* &  (2$^+$)*  & 2353.2(3)*       &  $\ge$2.3  &  934.52  &   2$^+$    \\
              &            & 3287.13(13)*     &  $\ge$2.3  &    0.00  &   0$^+$    \\
 3289.0(2)    &   3$^+$    &  379.6(2)        &     1.1(2) & 2909.65  &   3$^+$    \\
              &            & 1222.1(2)        &     3.2(5) & 2066.74  &   2$^+$    \\
              &            & 1441.2(3)        &     0.9(2) & 1847.37  &   2$^+$    \\
              &            & 1793.4(2)        &     1.5(2) & 1495.52  &   4$^+$    \\
              &            & 2354.1(3)        &     7.8(5) &  934.52  &   2$^+$    \\
 3338.12(11)* &  (5$^-$)*  &  852.1(1)*       &     1.5(2) & 2486.02  &   5$^-$    \\
 3371.43(6)   &   1$^-$*   & 1032.4(3)        &     0.6(2) & 2339.80  &   3$^-$    \\
              &            & 1988.66(6)       &    23(1)   & 1382.76  &   0$^+$    \\
              &            & 2436.84(7)       &    11.2(6) &  934.52  &   2$^+$    \\
              &            & 3371.42(9)       &    11(1)   &    0.00  &   0$^+$    \\
 3407.94(9)   &   2$^-$*   & 1068.03(11)      &     3.6(3) & 2339.80  &   3$^-$    \\
              &            & 2473.53(12)      &     5.3(4) &  934.52  &   2$^+$    \\
 3452.47(7)   &  (3$^-$)*  &  709.1(2)*       &     0.5(1) & 2743.72  &   4$^-$    \\
              &            & 1112.9(2)        &     0.7(2) & 2339.80  &   3$^-$    \\
              &            & 1604.92(8)       &     4.9(4) & 1847.37  &   2$^+$    \\
              &            & 1957.15(13)      &     3.8(3) & 1495.52  &   4$^+$    \\
              &            & 2517.90(8)       &     7.6(4) &  934.52  &   2$^+$    \\
 3463.1(2)*   &  (4)$^+$   & 1967.6(2)        &     4.4(5) & 1495.52  &   4$^+$    \\
              &            & 2528.3(3)        &     0.9(3) &  934.52  &   2$^+$    \\
 3472.30(8)   &    1$^+$   & 1405.7(2)*       &     2.7(4) & 2066.74  &   2$^+$    \\
              &            & 2089.59(12)      &     1.8(2) & 1382.76  &   0$^+$    \\
              &            & 2537.88(12)      &     4.8(4) &  934.52  &   2$^+$    \\
              &            & 3472.09(9)       &    10.2(7) &    0.00  &   0$^+$    \\
 3481.55(16)  &  (5$^-$)*  &  737.8(2)        &     0.5(2) & 2743.72  &   4$^-$    \\
              &            &  995.56(19)      &     5.2(5) & 2486.02  &   5$^-$    \\
 3500.04(12)  &    2$^+$   & 681.0(3)         &     0.5(2) & 2819.71  &   2$^+$    \\
              &            & 1159.92(21)      &     0.7(2) & 2339.80  &   3$^-$    \\
              &            & 1433.3(3)        &     1.0(2) & 2066.74  &   2$^+$    \\
              &            & 1652.90(14)      &     2.2(3) & 1847.37  &   2$^+$    \\
              &            & 2565.3(3)*       &     1.1(2) &  934.52  &   2$^+$    \\
              &            & 3499.97(9)       &     3.9(4) &    0.00  &   0$^+$    \\
 3628.50(7)   &   3$^+$ *  & 1229.52(17)      &     1.4(2) & 2398.42  &   4$^+$    \\
              &            & 1780.6(2)*       &     0.7(2) & 1847.37  &   2$^+$    \\
              &            & 2132.90(10)      &     2.4(2) & 1495.52  &   4$^+$    \\
              &            & 2694.07(5)       &    37(2)   &  934.52  &   2$^+$    \\
 3638.33(9)*  &   1$^-$    & 1791.2(3)*       &     0.7(2) & 1847.37  &   2$^+$    \\
              &            & 2255.51(15)*     &     1.3(2) & 1382.76  &   0$^+$    \\
              &            & 3638.25(14)*     &     6.7(7) &    0.00  &   0$^+$    \\
 3640.48(10)  & (2,3)$^+$* & 1300.8(2)        &     0.8(2) & 2339.80  &   3$^-$    \\
              &            & 2705.90(9)       &    14(1)   &  934.52  &   2$^+$    \\
 3649.03(10)  &   3$^+$    &  784.7(2)*       &     2.7(3) & 2864.82  &   4$^+$    \\
              &            & 1250.53(17)      &     1.9(2) & 2398.42  &   4$^+$    \\
              &            & 1801.6(2)        &     2.8(3) & 1847.37  &   2$^+$    \\
              &            & 2153.56(13)      &     3.3(3) & 1495.52  &   4$^+$    \\
              &            & 2714.46(21)      &     4.2(5) &  934.52  &   2$^+$    \\
 3675.3(2)*   &  (5$^+$)*  &  810.4(3)*       &     0.4(1) & 2864.82  &   4$^+$    \\
              &            & 2179.8(2)*       &     0.9(2) & 1495.52  &   4$^+$    \\
 3697.06(15)* &  1$^{(+)}$ & 2762.41(19)*     &     2.7(3) &  934.52  &   2$^+$    \\
              &            & 3696.94(21)*     &     1.7(2) &    0.00  &   0$^+$    \\
 3719.5(2)*   &  (6$^+$)*  &  855.0(3)*       &     2.0(2) & 2864.82  &   4$^+$    \\
              &            & 2224.0(2)*       &     1.2(2) & 1495.52  &   4$^+$    \\
 3774.74(17)* & 1$^{(+)}$* & 1708.2(3)*       &     1.0(2) & 2066.74  &   2$^+$    \\
              &            & 1927.0(2)*       &     0.7(2) & 1847.37  &   2$^+$    \\
              &            & 2839.86(24)*     &     2.3(3) &  934.52  &   2$^+$    \\
              &            & 3774.68(27)*     &     0.9(2) &    0.00  &   0$^+$    \\
\hline
\end{tabular}
\end{center}
\label{Zr92_table_ng_2}
\end{table}

\begin{table}[]
\caption{Continuation of Table \ref{Zr92_table_ng_1}.}
\begin{center}
\begin{tabular}{l c l r r c}

\hline
& & & & & \\
~E$_i$(keV)~~&J$^{\pi}_i$ &E$_{\gamma}$(keV&I$_{\gamma}$ (rel.)&~E$_f$(keV)&~~J$^{\pi}_f$\\
\hline
 3830.9(1)    &(1$^-$,2$^+$)& 378.5(2)       &     1.5(3) & 3452.57  &  (3$^-$)   \\
              &            &  791.13(19)     &     3.7(3) & 3039.83  &   3$^-$    \\
              &            & 1491.09(15)     &     5.4(4) & 2339.80  &   3$^-$    \\
              &            & 1765.2(3)       &     0.9(3) & 2066.74  &   2$^+$    \\
              &            & 2449.5(4)       &     0.3(1) & 1382.76  &   0$^+$    \\
              &            & 2896.0(2)       &     5.2(3) &  934.52  &   2$^+$    \\
 3847.2(3)*   &  (3$^-$)*  & 1103.5(2)*      &     1.0(2) & 2743.72  &   4$^-$    \\
 3980.51(7)*  &  (2$^+$)*  & 2133.08(8)*     &     4.9(6) & 1847.37  &   2$^+$    \\
              &            & 3045.97(9)*     &     7.0(8) &  934.52  &   2$^+$    \\
 4005.6(2)*   &  (4$^+$)*  & 2510.1(2)*      &     2.8(3) & 1495.52  &   4$^+$    \\
 4070.73(12)* &  (3$^+$) * & 2223.30(12)*    &     2.8(3) & 1847.37  &   2$^+$    \\
              &            & 3136.3(2) *     &     0.9(2) &  934.52  &   2$^+$    \\
 4083.1(1)*   & (1,2$^+$)* & 2700.36(14)*    &     2.6(3) &  1382.76 &   0$^+$    \\
              &            & 3148.6(2)*      &     1.1(2) &   934.52 &   2$^+$    \\
              &            & 4082.9(3)*      &     1.2(2) &    0.00  &   0$^+$    \\
 4103.9(1)*   &   2$^+$*   &  867.2(2)*      &     0.8(2) & 3236.8   &  (4$^-$)   \\
              &            & 1764.0(2)*      &     1.8(3) & 2339.80  &   3$^-$    \\
              &            & 2608.1(2)*      &     1.3(3) & 1495.52  &   4$^+$    \\
              &            & 3169.3(2)*      &     1.5(3) &  934.52  &   2$^+$    \\
              &            & 4103.8(1)*      &     16(2)  &    0.00  &   0$^+$    \\
 4107.6(2)*   & (3$^+$)*   & 2612.1(2)*      &     4(1)   & 1495.52  &   4$^+$    \\
 4202.8(2)*   &(1,2$^+$)*  & 2355.5(2)*      &     1.7(2) & 1847.37  &   2$^+$    \\
              &            & 3267.9(2)*      &     7.1(5) &  934.52  &   2$^+$    \\
              &            & 4202.8(2)*      &     3.2(4) &    0.00  &   0$^+$    \\
 4229.3(2)*   & (2)$^+$*   & 2733.6(3)*      &     0.8(2) & 1495.52  &   4$^+$    \\
              &            & 3295.1(2)*      &     2.3(3) &  934.52  &   2$^+$    \\
              &            & 4229.0(3)*      &     1.2(3) &    0.00  &   0$^+$    \\
 4310.07(18)* &(1,2$^+$)*  & 4309.96(18)*    &     1.1(2) &    0.00  &   0$^+$    \\
 4409.40(9)*  &(2,3)$^+$*  & 2562.00(12)*    &     2.2(2) & 1847.37  &   2$^+$    \\
              &            & 3474.81(14)*    &     5.3(5) &  934.52  &   2$^+$    \\
 4514.00(9)*  &(2,3)$^+$*  & 3579.42(8)*     &     8(1)   &  934.52  &   2$^+$    \\
 4542.80(12)* &   3$^+$*   & 2203.2(2)*      &     0.7(2) & 2339.80  &   3$^-$    \\
              &            & 2476.1(2)*      &     1.0(2) & 2066.74  &   2$^+$    \\
              &            & 3047.3(2)*      &     0.8(2) & 1495.52  &   4$^+$    \\
              &            & 3608.2(1)*      &    11(1)   &  934.52  &   2$^+$    \\
 4606.75(8)*  &(1,2$^+$)*  & 3672.15(15)*    &     4.3(4) &  934.52  &   2$^+$    \\
              &            & 4606.69(9)*     &    11(1)   &    0.00  &   0$^+$    \\
 4702.1(2)*   &  (3$^+$)*  & 2635.3(3)*      &     1.2(2) & 2066.74  &   2$^+$    \\
 4774.33(12)* & 1,2$^+$*   & 4774.21(12)*    &     3.0(5) &    0.00  &   0$^+$    \\
 4776.3(2)*   & 1,2$^+$*   & 3841.7(2)*      &     2.3(3) &  934.52  &   2$^+$    \\
 4893.3(2)*   & (3,4$^+$)* & 2494.7(2)*      &     1.0(2) & 2398.42  &   4$^+$    \\
              &            & 3397.8(2)*      &     0.9(2) & 1495.52  &   4$^+$    \\
 5105.6(2)*   &  (3$^+$)*  & 3038.6(2)*      &     1.4(2) & 2066.74  &   2$^+$    \\
              &            & 4171.0(1)*      &     2.2()4 &  934.52  &   2$^+$    \\
 5128.1(2)*   &   2$^+$*   & 3632.5(3)*      &     3.0(4) & 1495.52  &   4$^+$    \\
              &            & 3744.7(2)*      &     4.0(7) & 1382.76  &   0$^+$    \\
              &            & 4194.2(3)*      &     2.5(5) &  934.52  &   2$^+$    \\
              &            & 5128.0(3)*      &     0.8(3) &    0.00  &   0$^+$    \\
 5190.70(12)* &  1,2$^+$*  & 3343.2(2)*      &     1.4(2) & 1847.37  &   2$^+$    \\
              &            & 3807.9(2)*      &     1.2(2) & 1382.76  &   0$^+$    \\
              &            & 5190.7(2)*      &     0.5(2) &    0.0   &   0$^+$    \\
 5197.4(2)*   &  (3$^+$)*  & 4262.8(2)*      &     4.1(4) &  934.52  &   2$^+$    \\
 5212.77(15)* &  (3$^+$)*  & 3145.97(13)*    &     1.7(2) & 2066.74  &   2$^+$    \\
 5267.7(2)*   &(1,2$^+$)*  & 3884.9(2)*      &     1.4(2) & 1382.76  &   0$^+$    \\
 5293.1(2)*   &  (3$^+$)*  & 3226.5(2)*      &     1.6(3) & 2066.74  &   2$^+$    \\
              &            & 3444.6(2)*      &     1.4(2) & 1844.37  &   2$^+$    \\
              &            & 4358.6(2)*      &     3.5(7) &  934.52  &   2$^+$    \\
\hline
\end{tabular}
\end{center}
\label{Zr92_table_ng_3}
\end{table}

\begin{table}[]
\caption{Continuation of Table \ref{Zr92_table_ng_1}.}
\begin{center}
\begin{tabular}{l c l r r c}
\hline
& & & & & \\
~E$_i$(keV)~~&J$^{\pi}_i$ &E$_{\gamma}$(keV&I$_{\gamma}$ (rel.)&~E$_f$(keV)&~~J$^{\pi}_f$\\
\hline
 5459.2(2)*   &  (3$^+$)*  & 3963.8(2)*      &     2.6(3) & 1495.52  &   4$^+$    \\
              &            & 4525.2(2)*      &     2.8(3) &  934.52  &   2$^+$    \\
 5488.8(14)*  &(1,2$^+$)*  & 3421.9(2)*      &     1.4(2) & 2066.74  &   2$^+$    \\
              &            & 4106.2(2)*      &     1.3(2) & 1382.76  &   0$^+$    \\
              &            & 5488.8(2)*      &     1.5(5) &    0.00  &   0$^+$    \\
 5535.0(2)*   & (1,2$^+$)* & 4152.1(2)*      &     0.8(2) & 1382.76  &   0$^+$    \\
              &            & 4600.9(2)*      &     0.6(2) &  934.52  &   2$^+$    \\
              &            & 5534.8(2)*      &     1.4(2) &    0.00  &   0$^+$    \\
 5563.8(2)*   &  (3$^+$)*  & 4629.2(2)*      &     3.8(3) &  934.52  &   2$^+$    \\
 5568.4(2)*   &(3,4$^+$)*  & 3501.6(2)*      &     1.8(3) & 2066.74  &   2$^+$    \\
 5626.80(13)* &(1,2$^+$)*  & 3560.0(2)*      &     1.1(2) & 2066.74  &   2$^+$    \\
              &            & 5626.67(17)*    &     1.4(2) &    0.00  &   0$^+$    \\
 5661.6(2)*   & (1,2$^+$)* & 4278.6(2)*      &     1.5(2) & 1382.76  &   0$^+$    \\
              &            & 4727.2(2)*      &     0.5(2) &  934.52  &   0$^+$    \\
 5674.12(15)* & (3$^+$)*   & 4739.47(13)*    &     4.3(4) &  934.52  &   2$^+$    \\
 5698.2(2)*   & (3$^+$)*   & 2788.5(2)*      &     0.8(2) & 2909.65  &   3$^+$    \\
              &            & 3850.8(2)*      &     1.2(2) & 1847.37  &   2$^+$    \\
 5765.0(2)*   & (3$^+$)*   & 4830.4(2)*      &     3.5(7) &  934.52  &   2$^+$    \\
 5859.8(2)*   & (1,2$^+$)* & 5859.6(2)*      &   $\ge$0.9 &    0.00  &   0$^+$    \\
 5897.8(2)*   &   (3)*     & 3831.0(2)*      &     1.6(2) & 2066.74  &   2$^+$    \\
 5923.8(2)*   &  (1,2)*    & 5923.6(2)*      &   $\ge$0.7 &    0.00  &   0$^+$    \\
 6147.2(3)*   & (3$^-$)*   & 3807.3(2)*      &     0.8(2) & 2339.80  &   3$^-$    \\
 6158.9(2)*   &(1,2$^+$)*  & 4092.4(3)*      &     1.5(2) & 2066.74  &   2$^+$    \\
              &            & 6158.3(3)*      &     0.5(2) &    0.00  &   0$^+$    \\
 6172.8(2)*   & (3$^-$)*   & 3833.0(2)*      &     2.5(4) & 2339.80  &   3$^-$    \\
 8634.81(3)   &   2$^+$    & 2462.0(1)*      &     2.0(3) & 6172.8   &  (3$^-$)   \\
              &            & 2476.0(2)*      &     0.9(2) & 6158.9   & (1,2$^+$)  \\
              &            & 2487.6(3)*      &     0.8(2) & 6147.2   &  (3$^-$)   \\
              &            & 2710.9(2)*      &     0.7(2) & 5923.8   &  (1,2)     \\
              &            & 2736.8(1)*      &     0.9(2) & 5897.8   &    (3)     \\
              &            & 2775.2(2)*      &     0.9(2) & 5859.8   & (1,2$^+$)  \\
              &            & 2869.6(2)*      &     3.5(6) & 5765.0   &  (3$^+$)   \\
              &            & 2936.7(2)*      &     1.7(3) & 5698.2   &  (3$^+$)   \\
              &            & 2960.62(16)*    &     4.0(5) & 5674.12  &  (3$^+$)   \\
              &            & 2974.2(2)*      &     1.4(3) & 5661.6   & (1,2$^+$)  \\
              &            & 3007.81(15)*    &     2.5(5) & 5626.80  & (1,2$^+$)  \\
              &            & 3066.1(3)*      &     1.5(3) & 5568.4   & (3,4$^+$)  \\
              &            & 3071.0(2)*      &     3.0(5) & 5563.8   &  (3$^+$)   \\
              &            & 3099.7(2)*      &     0.9(2) & 5535.0   & (1,2$^+$)  \\
              &            & 3146.3(2)*      &     1.6(3) & 5488.8   & (1,2$^+$)  \\
              &            & 3175.1(1)*      &     5.0(6) & 5459.6   &  (3$^+$)   \\
              &            & 3341.8(1)*      &     5.8(8) & 5293.1   &  (3$^+$)   \\
              &            & 3367.3(1)*      &     1.3(3) & 5267.7   & (1,2$^+$)  \\
              &            & 3421.91(11)*    &     1.6(3) & 5212.77  &  (3$^+$)   \\
              &            & 3437.3(1)*      &     3.8(5) & 5197.4   &  (3$^+$)   \\
              &            & 3444.1(2)*      &     3.5(5) & 5190.70  &  1,2$^+$   \\
              &            & 3506.7(2)*      &     3.2(3) & 5128.1   &   2$^+$    \\
              &            & 3529.2(2)*      &     3.3(5) & 5105.6   &  (3$^+$)   \\
              &            & 3741.7(2)*      &     1.7(3) & 4893.3   & (3,4$^+$)  \\
              &            & 3858.5(2)*      &     1.5(2) & 4776.3   &  1,2$^+$   \\
              &            & 3860.50(11)*    &     2.5(5) & 4774.33  &  1,2$^+$   \\
              &            & 3932.8(2)*      &     0.7(2) & 4702.1   &  (3$^+$ )  \\
              &            & 4027.91(12)*    &     9.1(8) & 4606.85  & (1,2$^+$)  \\
              &            & 4091.90(14) *    &     7.5(7) & 4542.80  &   3$^+$   \\
              &            & 4120.72(12) *    &     5(1)   & 4514.00  & (2,3)$^+$ \\
              &            & 4225.30(14) *    &     2.4(3) & 4409.40  & (2,3)$^+$ \\
 \hline
\end{tabular}
\end{center}
\label{Zr92_table_ng_4}
\end{table}

\begin{table}[]
\caption{Continuation of Table \ref{Zr92_table_ng_1}.}
\begin{center}
\begin{tabular}{l c l r r c}
\hline
& & & & & \\
~E$_i$(keV)~~&J$^{\pi}_i$ &E$_{\gamma}$(keV&I$_{\gamma}$ (rel.)&~E$_f$(keV)&~~J$^{\pi}_f$\\
\hline
              &            & 4324.52(18) *    &     1.1(2) & 4310.07  & (1,2$^+$)  \\
              &            & 4404.95(25) *    &     2.6(3) & 4229.3   &  (2)$^+$   \\
              &            & 4431.80(20) *    &     2.2(3) & 4202.8   & (1,2$^+$)  \\
              &            & 4526.90(20) *    &     2.0(5) & 4107.6   &  (3$^+$)   \\
              &            & 4530.6(15) *     &     14(2)  & 4103.9   &   2$^+$    \\
              &            & 4551.52(16) *    &     2.5(3) & 4083.1   & (1,2$^+$)  \\
              &            & 4564.09(11) *    &     3.5(3) & 4070.73  &   (3$^+$)  \\
              &            & 4629.1(2) *      &     1.5(3) & 4005.6   &   (4$^+$)  \\
              &            & 4654.39(9) *     &     11(1)  & 3980.51  &  (2$^+$)   \\
              &            & 4787.4(2) *      &     0.7(2) & 3847.2   &  (3$^-$)   \\
              &            & 4804.03(13)      &     6.6(4) & 3830.9   &(1$^-$,2$^+$)\\
              &            & 4860.10(20) *    &     0.9(2) & 3774.74  & 1$^{(+)}$  \\
              &            & 4937.57(21) *    &     1.5(2) & 3697.06  & 1$^{(+)}$  \\
              &            & 4985.46(13)      &     5.0(4) & 3649.03  &   3$^+$    \\
              &            & 4994.21(10)      &     8(1)   & 3640.48  & (2,3)$^+$  \\
              &            & 4996.38(11)      &     5(1)   & 3638.33  &   1$^-$    \\
              &            & 5006.24(9)       &    26(2)   & 3628.50  &   3$^+$    \\
              &            & 5134.37(15)      &     6.5(5) & 3500.04  &   2$^+$    \\
              &            & 5162.43(12)      &     9(1)   & 3472.29  &   1$^+$    \\
              &            & 5171.70(15) *    &     0.9(2) & 3463.1   &  (4)$^+$   \\
              &            & 5182.46(12)      &    11(2)   & 3452.47  &  (3$^-$)   \\
              &            & 5226.54(31) *    &     1.0(2) & 3407.94  &   2$^-$    \\
              &            & 5263.29(9)       &    37(3)   & 3371.43  &   1$^-$    \\
              &            & 5346.0(2) *      &     1.2(3) & 3289.0   &   3$^+$    \\
              &            & 5347.12(11)      &     2.3(4) & 3287.17  &  (2$^+$)   \\
              &            & 5358.72(11) *    &     4.8(4) & 3276.00  &  (3$^+$)   \\
              &            & 5371.60(14)      &     6.0(5) & 3262.82  &   2$^+$    \\
              &            & 5443.55(17) *    &     1.7(3) & 3191.07  &  (4$^+$)   \\
              &            & 5455.90(22) *    &     1.4(3) & 3178.50  &   4$^+$    \\
              &            & 5576.74(18) *    &     2.4(2) & 3057.56  &   2$^+$    \\
              &            & 5595.17(16)      &     4.6(3) & 3039.83  &   3$^-$    \\
              &            & 5725.20(14) *    &     3.7(4) & 2909.65  &   3$^+$    \\
              &            & 5770.01(24) *    &     1.0(2) & 2864.82  &   4$^+$    \\
              &            & 5814.93(14)      &     3.1(3) & 2819.71  &   2$^+$    \\
              &            & 6236.44(13)      &     2.6(2) & 2398.42  &   4$^+$    \\
              &            & 6294.80(8)       &   226(7)   & 2339.80  &   3$^-$    \\
              &            & 6567.72(13)      &     3.0(4) & 2066.74  &   2$^+$    \\
              &            & 6787.01(25) *    &     0.8(2) & 1847.37  &   2$^+$    \\
              &            & 7139.03(14)      &    1.0(2)  & 1495.52  &   4$^+$    \\
              &            & 7251.64(16)      &    1.1(2)  & 1382.76  &   0$^+$    \\
              &            & 7699.82(10)      &    8(1)    &  934.52  &   2$^+$    \\
              &            & 8634.29(11)      &    14(1)   &    0.00  &   0$^+$    \\
\hline
\end{tabular}
\end{center}
\label{Zr92_table_ng_5}
\end{table}

Only 10 levels do not receive direct feeding from the capture level at 8634.81 keV. For seven
of them this is consistent with their spin-parity of 4$^-$, 5$^{\pm}$ or 6$^+$ (see next
section). The 3236.8-keV level reported with spin-parity 4$^+$ \cite{NDS_Zr92} is not populated
by a primary transition. In contrast, the 3191.07-keV level has a clear feeding  from the
capture level, in spite of (4$^-$) spin-parity proposed in the compilation \cite{NDS_Zr92}.
We note that the observed feeding and decay branchings are consistent with spin-parity 4$^+$
for the 3191.07-keV level and spin 4$^-$ for the 3236.8-keV level. No primary decays to the
two remaining levels, 0$^+$ at 2904.19 keV and 1$^{(+)}$ at 3124.93 keV may indicate their
special character.

In previous (n,$\gamma$) studies \cite{NDS_Zr92} the summed intensity of primary $\gamma$
transitions amounts to 33$\%$ of the total $\gamma$ intensity feeding the ground state in
$^{92}$Zr. The present work increases this figure to 54$\%$. Thus, nearly half of the primary
$\gamma$ intensity is still not identified. It is probably ``hidden'' in an unresolved
continuum between 2 and 4 MeV, signs of which can be seen in Fig. \ref{Zr_even_light_fig6}(b).

\subsubsection{Neutron binding energy of $^{92}$Zr}

The (n,$\gamma$) reaction with cold neutrons produces a neutron-capture level situated some
milli-electronvolts off the neutron binding energy, allowing to determine very precisely this
energy in measurements of $\gamma$ decays from the neutron-capture level using Ge arrays  \cite{Urb13,Rza18}. The present (n,$\gamma$) data provided a neutron binding energy in $^{92}$Zr
of $S(n)$=8634.75(4) \cite{Rza18}, more precise than the $S(n)$=8634.79(11)-keV value in the
compilation \cite{NDS_Zr92}. The neutron binding energy of $^{28}$Al, $S(n)$=7725.174(11) keV,
obtained from the same data set \cite{Rza18}, compared against the literature value
$S(n)$=7725.17(1) \cite{AME20}, shows that statistical uncertainties and the systematic error
for this technique are not higher than 0.01 keV.

The value reported in Ref. \cite{Rza18} was determined as an average of summed energies in a
few strongest $\gamma\gamma$ cascades linking the capture level with the ground state.
In the present work we have determined the neutron binding energy of $^{92}$Zr as an average
of 78 sums of a primary-$transion$ energy and the excitation energy of the level populated
by this transition. Such an approach is more accurate than summing energies in $\gamma-\gamma$
cascades, as an excited level energy may be more precise than the energy of the second
transition in a $\gamma-\gamma$ cascade. The weighted average of the 78 sums, S$_n$=8434.81(2)
keV, determines the new neutron binding energy of $^{92}$Zr, which agrees with previously published
values and is more accurate.

\subsubsection{Measurement of $\beta^-$ decay of $^{92}$Y to levels in $^{92}$Zr.}

The present, 4 hour $\beta^-$ measurement provided more precise energies of $\gamma$ lines
and excited levels compared to those reported in Ref. \cite{Tal70}. The results of the
measurement are listed in Table \ref{Zr92_table_beta}. One notes good agreement with
energies determined in the (n,$\gamma$) measurement.

\begin{table}[h]
\caption{Energies of excited states and their $\gamma$ decays with relative $\gamma$
intensities in $^{92}$Zr, as obtained in the present work from the measurement of $\beta^-$
decay of $^{92}$Y.}
\begin{center}
\begin{tabular}{l c c r r r}
\hline
& & & & & \\
~E$_i$(keV)~~&~J$^{\pi}_i$~&E$_{\gamma}$(keV&I$_{\gamma}$ (rel.)&~E$_f$(keV)&~~J$^{\pi}_f$\\
\hline
 ~934.53(3)  &   2$^+$    &  934.52(3)       & 1000(30)   &    0.00  &   0$^+$    \\
 1382.78(4)  &   0$^+$    &  448.25(3)       &   171(6)   &  934.53  &   2$^+$    \\
 1495.58(4)  &   4$^+$    &  561.02(3)       &   189(7)   &  934.53  &   2$^+$    \\
 1847.32(4)  &   2$^+$    &  912.79(4)       &    44(2)   &  934.53  &   2$^+$    \\
             &            & 1847.31(7)       &    21(1)   &    0.00  &   0$^+$    \\
 2066.82(9)  &   2$^+$    & 1132.28(9)       &    18(1)   &  934.53  &   2$^+$    \\
 2339.84(4)  &   3$^-$    &  492.43(5)       &    30(2)   & 1847.32  &   2$^+$    \\
             &            &  844.29(4)       &    93(3)   & 1495.58  &   4$^+$    \\
             &            & 1405.32(3)       &   316(12)  &  934.53  &   2$^+$    \\
 2819.4(3)   &   2$^+$    &  972.2(3)        &     3.3(5) & 1847.32  &   2$^+$    \\
             &            & 1884.5(4)        &     1.4(3) &  934.52  &   2$^+$    \\
 3039.3(4)   &   3        &  699.4(5)        &     1.0(2) & 2339.84  &   3$^-$    \\
             &            & 2104.7(4)        &     1.5(3) &  934.53  &   2$^+$    \\
 3372.2(5)   & 1$^{(-)}$  & 1989.4(5)        &    0.4(2)  & 1382.78  &   0$^+$    \\
 3407.5(4) * &2$^-$,3$^-$ & 2473.0(4) *      &    0.5(2)  &  934.53  &   2$^+$    \\
 3452.8(4) * &    3$^-$   & 1113.0(4) *      &    0.5(2)  & 2339.84  &   3$^-$    \\
\hline
\end{tabular}
\end{center}
\label{Zr92_table_beta}
\end{table}

\begin{figure*}
\centering
\scalebox{0.45}{\includegraphics{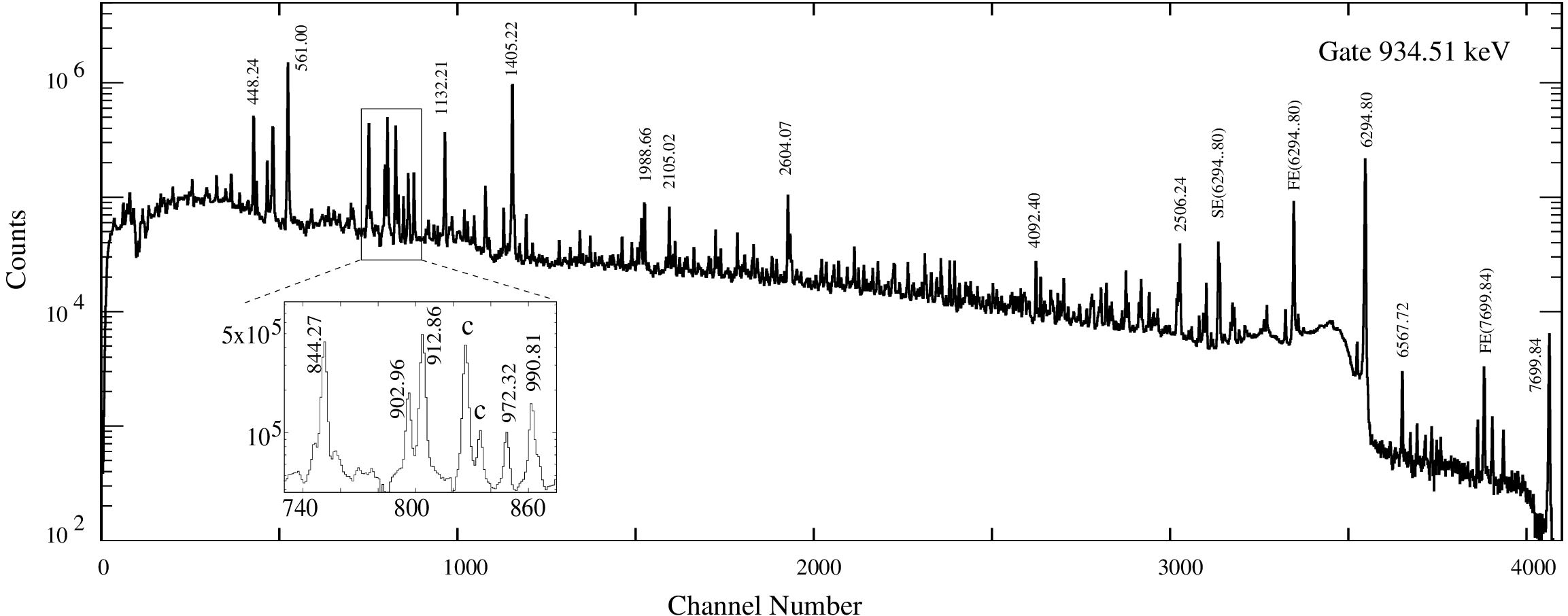}}
\caption{$\gamma$ spectrum gated on the 934.61-keV line in the data from the $^{235}$U+n
EXILL run. Constant-peak-width energy calibration was applied (see text for details).
Lines are labeled with their $\gamma$ energies in keV. Label ``c'' marks contaminating
lines. See text for more comments.}
\label{Zr_even_light_fig10}
\end{figure*}

We confirm the levels reported in Ref. \cite{Tal70}, except the tentative 2473.4-keV level.
Our coincidence data indicate that the 2473.0-keV transition (2473.4-keV in \cite{Tal70})
feeds the 934.53-keV level, defining a new level at 3407.5 keV. Another new level has been
introduced at 3452.8 keV. The two new levels populated in $\beta^-$ decay were observed in
the (n,$\gamma$) measurement at 3407.94 and 3452.47 keV.

There is good agreement of $\gamma$ intensities listed in Table \ref{Zr92_table_beta} with
those reported in Ref. \cite{Tal70}. Taking the ground state feeding of $^{92}$Zr in
$\beta^-$ decay from Ref. \cite{Tal70} we estimated {\it log ft} values for the new, 3407.5
and 3452.8 keV levels to be 5.4(2) and 5.1(2), respectively. These values are compatible
with an allowed character of $\beta^-$ decay to both levels. This indicates their negative
parity, considering the I$^{\pi}$=2$^-$ of the ground state of $^{92}$Y.

\subsubsection{Spin-parity assignments to levels in $^{92}$Zr}

To determine spins and parities of excited levels in $^{92}$Zr we used data from the 21-day
measurement of cold-neutron-induced fission of $^{235}$U, the observed decay branchings and
$log ft$ values from $\beta^-$ decay of $^{92}$Y.

A $\gamma$ spectrum doubly-gated on the 934.51-keV line and on the prompt DT peak in the
$\gamma\gamma$DT histogram is shown in Fig. \ref{Zr_even_light_fig10}. Comparing this
spectrum with that in Fig. \ref{Zr_even_light_fig6}(b) we estimated yields of the
$^{91}$Zr($n,\gamma )^{92}$Zr reaction in the $^{235}$U data to be factor five higher
than in the 21-hour run on the $^{96}$Zr enriched target. This data has been used to
determine more precise energies of $\gamma$ lines and decay branchings for excited levels
and most importantly to measure angular and directional-polarization correlations in
$^{92}$Zr, using EXOGAM clover detectors as Compton polarimeters \cite{Urb23} (see Refs.
\cite{Urb13,Urb16,Jen17} for more information on the correlations technique used). The
obtained correlation results for $\gamma\gamma$ cascades in $^{92}$Zr are listed in
Tables \ref{Zr92_angcorr_tab8}, \ref{Zr92_angcorr_tab9} and \ref{Zr92_polcorr_tab10}.

Our work confirms most of spin-parity assignments reported in the compilation
\cite{NDS_Zr92} and shown in Tables \ref{Zr92_table_ng_1} - \ref{Zr92_table_ng_5}, except
a few cases discussed below, and adds 61 new spin-parity assignments in $^{92}$Zr.

Mixing ratios, $\delta$ are in good agreement with $\delta$ values reported in Ref.
\cite{NDS_Zr92}. In particular, $\delta$=0.003(10) for the 1405.22-keV transition
confirms the evaluators choice \cite{NDS_Zr92}. The very small mixing ratio $\delta$=
-0.004(16) of the 912.7-keV decay from the 1847.37-keV level confirms its pure M1 character
whereas large $\delta$=-2.9(2) of the 1132.1-keV transition, compatible with the -3.2(+5,-4)
adopted value \cite{NDS_Zr92}, points to an E2 nature of the 1132.1-keV transition. However,
for the 972.32-keV transition we find $\delta$=1.1(+31,-7), suggesting its collective
character in contrast to an enhanced M1 nature reported in \cite{NDS_Zr92}.

\begin{table}[]
\caption{Experimental $a_k^{exp}$ angular correlation coefficients for $\gamma\gamma$
cascades and mixing ratios, $\delta$ for $\gamma$ transitions in $^{92}$Zr populated in the
$^{91}$Zr($n,\gamma$) reaction, as obtained in the present work. $E\gamma_1$ is the upper
transition in a cascade. Index $\delta$ marks transitions for which the $\delta$ value was
determined. Indices $u$ or $f$ mark an unmixed or fixed-$\delta$ transition, respectively.
The last column shows either the $\delta$ value or $\chi^2$ value, the latter for a cascade of
two transitions with known $\delta$.}
\begin{center}
\begin{tabular}{ l c c c c }
\hline
$E\gamma_1-E\gamma_2$~~~~      &Spins in&             & &$\delta(\gamma^{\delta})$ \\
(keV) ~ (keV)~~                &cascade  & $a_2^{exp}$&$a_4^{exp}$ & or $\chi^2$   \\
\hline
& & & & \\
 257.70$^{\delta}$-990.49$^f$  & 4-5-4   &  0.03(6)   & -0.11(14)  & -0.12(8)      \\
 379.6 $^{\delta}$-1795.14$^f$ & 3-2-0   & -0.15(14)  & -0.01(32)  & -0.11(21)     \\
 448.24$^u$-934.51$^u$         & 0-2-0   &  0.356(21) &  1.120(41) & $\chi^2$=0.3  \\
 492.42$^{\delta}$-912.86$^f$  & 3-4-2   & -0.012(31) & -0.132(71) &  0.07(6)      \\
 561.00$^{\delta}$-934.51$^u$  & 4-2-0   &  0.109(12) & -0.008(30) &  0.012(21)    \\
 700.21$^{\delta}$-1405.22$^f$ & 3-3-2   &  0.137(9)  & -0.002(19) &  0.7(4)       \\
 780.06$^{\delta}$-902.96$^f$  & 4-4-4   &  0.36(11)  & -0.04(27)  & -0.4(4)       \\
 842.93$^{\delta}$-1132.21$^f$ & 3-2-2   & -0.270(84) &  0.26(20)  & -0.40(19)     \\
                               &         &            &            & -5(-9,+2)     \\
 844.27$^{\delta}$-561.00$^u$  & 3-4-2   & -0.097(11) & -0.013(26) & -0.051(14)    \\
 902.96$^{\delta}$-561.00$^u$  & 4-4-2   &  0.208(28) & -0.037(60) & -0.04(9)      \\
 912.86$^{\delta}$-934.51$^u$  & 2-2-0   &  0.253(10) &  0.006(24) & -0.004(15)    \\
 972.32$^{\delta}$-912.86$^f$  & 2-2-2   &  0.096(44) &  0.038(98) &  1.1(8)       \\
 990.49$^{\delta}$-561.00$^u$  & 5-4-2   & -0.023(22) & -0.026(46) & -0.07(3)      \\
 990.81$^{\delta}$-1132.21$^f$ & 2-2-2   & -0.233(72) &  0.10(15)  & -0.09(15)     \\
                               &         &            &            & -1.8(6)       \\
 995.56$^{\delta}$-990.49$^f$  & 5-5-4   & -0.05(11)  & -0.11(23)  & -0.22(15)     \\
                               &         &            &            & -15(12)       \\
1132.21$^{\delta}$-934.51$^u$  & 2-2-0   &  0.189(19) &  0.306(41) & -2.9(2)       \\
1209.18$^{\delta}$-1132.21$^f$ & 2-2-2   &  0.051(74) &  0.6(17)   &  0.02(15)     \\
                               &         &            &            &  6.5(35)      \\
1222.1 $^{\delta}$-1132.21$^f$ & 3-2-2   & -0.203(97) &  0.03(21)  &  0.75(15)     \\
1248.22$^{\delta}$-561.00$^u$  & 4-4-2   &  0.133(52) &  0.07(12)  &  0.19(14)     \\
1369.31$^{\delta}$-561.00$^u$  & 4-4-2   &  0.291(42) & -0.065(99) & -0.35(23)     \\
1405.22$^{\delta}$-934.51$^u$  & 3-2-0   & -0.068(9)  &  0.008(18) &  0.003(10)    \\
1414.05$^{\delta}$-561.00$^u$  & 3-4-2   &  0.280(54) & -0.065(99) & -2.7(8)       \\
1463.79$^u$-934.51$^u$         & 4-2-0   &  0.143(41) & -0.06(9)   & $\chi^2$=1.5  \\
1652.90$^{\delta}$-912.86$^f$  & 2-2-2   &  0.32(14)  &  0.23(31)  & -0.4(3)       \\
1695.53$^{\delta}$-561.00$^u$  & 4-4-2   &  0.150(63) &  0.01(14)  &  0.14(18)     \\
1741.1 $^{\delta}$-561.00$^u$  & 4-4-2   & -0.09(14)  &  0.08(30)  &  0.9(5)       \\
1885.19$^{\delta}$-934.51$^u$  & 2-2-0   &  0.201(57) &  0.062(99) &  0.07(7)      \\
1967.6 $^{\delta}$-561.00$^u$  & 4-4-2   &  0.105(90) & -0.05(21)  &  0.25(0.24)   \\
                               &         &            &            & -1.5(8)       \\
1975.14$^{\delta}$-934.51$^u$  & 3-2-0   & -0.014(39) &  0.048(83) &  0.07(6)      \\
2105.20$^{\delta}$-934.51$^u$  & 3-2-0   & -0.105(31) & -0.041(72) & -0.04(4)      \\
                               &         &            &            &  7.9(21)      \\
2153.56$^{\delta}$-561.00$^u$  & 3-4-2   &  0.03(12)  &  0.39(26)  & -0.19(15)     \\
2328.30$^{\delta}$-934.51$^u$  & 2-2-0   &  0.251(47) & -0.06(12)  & -0.002(66)    \\
2341.24$^{\delta}$-934.51$^u$  & 3-2-0   &  0.09(13)  & -0.19(31)  &  0.23(19)     \\
                               &         &            &            &  2.4(12)      \\
2354.1 $^{\delta}$-934.51$^u$  & 3-2-0   & -0.088(62) &  0.04(32)  & -0.02(8)      \\
2436.84$^{\delta}$-934.51$^u$  & 1-2-0   & -0.276(37) & -0.063(77) &  0.02(4)      \\
2473.53$^{\delta}$-934.51$^u$  & 2-2-0   &  0.15(8)   & -0.12(20)  &  0.12(10)     \\
                               & 3-2-0   &            &            &  0.34(15)     \\
2517.90$^{\delta}$-934.51$^u$  & 2-2-0   & -0.078(43) &  0.072(99) &  0.45(7)      \\
                               & 3-2-0   &            &            & -0.01(6)      \\
2537.88$^{\delta}$-934.51$^u$  & 1-2-0   & -0.455(81) & -0.08(16)  &  0.21(10)     \\
\hline
\end{tabular}
\end{center}
\label{Zr92_angcorr_tab8}
\end{table}

\begin{table}[]
\caption{Continuation of Table \ref{Zr92_angcorr_tab8}.}
\begin{center}
\begin{tabular}{ l c c c c }
\hline
$E\gamma_1-E\gamma_2$~~~~      &Spins in&           & &$\delta(\gamma^{\delta})$   \\
(keV) ~ (keV)~~                &cascade  & $a_2^{exp}$&$a_4^{exp}$ & or $\chi^2$   \\
\hline
& & & & \\
2608.1$^{\delta}$-561.00$^u$   & 2-4-0   &  0.15(22)  & -0.01(43)  & $\chi^2$=0.1  \\
                               & 3-2-0   &            &            & -0.4(3)       \\
2612$^{\delta}$-561.00$^u$     & 3-4-0   & -0.02(9)   & -0.21(17)  &  22(15)       \\
2694.07$^{\delta}$-934.51$^u$  & 3-2-0   & -0.042(17) & -0.068(37) &  0.01(2)      \\
                               &         &            &            &  4.8(4)       \\
2705.90$^{\delta}$-934.51$^u$  & 2-2-0   &  0.141(32) & -0.023(79) &  0.14(4)      \\
                               & 3-2-0   &            &            &  0.32(7)      \\
                               &         &            &            &  2.0(3)       \\
2896.0$^{\delta}$-934.51$^u$   & 1-2-0   &  0.089(70) & -0.06(16)  & -0.29(7)      \\
                               & 2-2-0   &            &            &  0.21(9)      \\
3045.97$^{\delta}$-934.51$^u$  & 1-2-0   &  0.141(31) &  0.033(75) & -0.34(3)      \\
                               & 2-2-0   &            &            &  0.15(4)      \\
                               & 3-2-0   &            &            &  0.31(6)      \\
3145.97$^{\delta}$-1132.21$^f$ & 3-2-2   &  0.45(18)  &  0.0(5)    & -1.2(9)       \\
3148.6$^{\delta}$-934.51$^u$   & 1-2-0   &  0.14(7)   &  0.15(14)  & -0.09(6)      \\
                               & 2-2-0   &            &            &  0.6(2)       \\
3169.3$^{\delta}$-934.51$^u$   & 2-2-0   &  0.18(15)  &  0.5(4)    & -3.1(16)      \\
                               & 3-2-0   &            &            &  0.4(2)       \\
3295.1$^{\delta}$-934.51$^u$   & 2-2-0   &  0.14(12)  &  0.22(24)  &  0.6(3)       \\
                               &         &            &            & 11(8)         \\
3474.81$^{\delta}$-934.51$^u$  & 2-2-0   &  0.043(67) & -0.07(15)  &  0.27(9)      \\
                               & 3-2-0   &            &            &  3.0(9)       \\
3579.42$^{\delta}$-934.51$^u$  & 2-2-0   &  0.047(40) & -0.07(9)   &  0.26(6)      \\
                               & 3-2-0   &            &            &  2.9(6)       \\
3608.2$^{\delta}$-934.51$^u$   & 3-2-0   & -0.051(38) & -0.11(9)   &  5.1(15)      \\
4027.91$^{\delta}$-4606.69$^u$ & 2-2-0   & -0.032(40) & -0.07(9)   &  0.37(6)      \\
4091.90$^{\delta}$-3608.2$^f$  & 2-3-2   &  0.159(39) &  0.02(9)   & -0.37(6)      \\
4104.40$^{\delta}$-4530.28$^u$ & 2-2-0   &  0.115(30) & -0.15(7)   &  0.17(4)      \\
4120.72$^{\delta}$-3579.42$^f$ & 2-2-2   &  0.182(40) &  0.02(9)   &  0.12(5)      \\
                               & 2-3-0   &            &            & -1.3(8)       \\
4530.6$^{\delta}$-4103.8$^u$   & 2-2-0   &  0.093(22) & -0.08(6)   &  0.21(3)     \\
4654.39$^{\delta}$-3045.97$^f$ & 2-2-2   &  0.146(53) &  0.07(13)  &  0.05(16)     \\
                               &         &            &            & -3.3(8)       \\
4804.03$^{\delta}$-2896.0$^f$  & 2-2-2   & -0.084(75) & -0.001(9)  &  0.13(7)      \\
5994.21$^{\delta}$-2705.90$^f$ & 2-2-2   &  0.211(38) & -0.001(9)  &  0.04(5)      \\
5006.24$^{\delta}$-2694.07$^f$ & 2-3-2   &  0.092(19) & -0.065(42) &  0.28(3)      \\
5134.37$^{\delta}$-3499.97$^u$ & 2-2-0   &  0.025(60) &  0.03(13)  &  0.30(8)      \\
5162.43$^{\delta}$-2537.88$^f$ & 2-1-2   &  0.112(44) & -0.03(9)   &  1.2(2)       \\
5162.43$^f$-3472.09$^{\delta}$ & 2-1-0   & -0.164(60) & -0.04(12)  &  0.1(1)       \\
5182.46$^{\delta}$-2517.90$^f$ & 2-3-2   &  0.103(45) & -0.01(8)   & -0.06(7)      \\
                               &         &            &            &  5.2(15)      \\
5263.29$^{\delta}$-2436.84$^f$ & 2-1-2   & -0.006(32) &  0.005(69) &  0.15(40)     \\
5263.29$^f$-3371.42$^{\delta}$ & 2-1-0   &  0.104(35) &  0.004(78) &  0.08(5)      \\
5347.12$^{\delta}$-3287.13$^u$ & 2-2-0   &  0.212(80) & -0.06(18)  &  0.05(11)     \\
5371.60$^{\delta}$-2328.30$^f$ & 2-2-2   &  0.183(47) & -0.10(11)  &  3.3(24)      \\
5595.17$^{\delta}$-2105.20$^f$ & 2-3-2   & -0.117(59) & -0.09(12)  & -0.34(9)      \\
                               &         &            &            & -2.7(8)       \\
5814.93$^{\delta}$ -972.32$^f$ & 2-2-2   &  0.238(89) & -0.11(21)  &  0.07(10)     \\
                               &         &            &            & -2.7(8)       \\
6236.44$^{\delta}$ -902.96$^f$ & 2-4-4   &  0.137(98) & -0.15(22)  & $\chi^2$=0.6  \\
6294.80$^{\delta}$-1405.22$^f$ & 2-3-2   &  0.137(9)  & -0.002(19) &  0.03(2)      \\
                               &         &            &            &  3.6(2)       \\
6567.72$^{\delta}$-1132.21$^f$ & 2-2-2   & -0.163(62) & -0.11(12)  &  0.04(11)     \\
                               &         &            &            & -2.5(8)       \\
7699.82$^{\delta}$-934.51$^u$  & 2-2-0   & -0.116(24) & -0.11(5)   &  0.52(5)      \\
\hline
\end{tabular}
\end{center}
\label{Zr92_angcorr_tab9}
\end{table}

\begin{table}[]
\caption{Experimental, $P_{exp}$ and theoretical, $P_{th}$ values of linear polarization for
         $\gamma_1$ transitions of $^{92}$Zr populated in the $^{91}$Zr($n,\gamma$) reaction,
         as obtained in this work. Theoretical $P_{th}(\gamma_1)$ values are calculated for
         spin-parity hypotheses and $\delta_{\gamma_1}$ values shown in the table. The
         correlating transitions, ${\gamma}_2$ are stretched quadrupole or dipole transitions.}
\begin{center}
\begin{tabular}{r c c l l}
\hline
$E\gamma_1-E\gamma_2$~~&Spins in&${\delta}_{\gamma 1}$&$P_{th}(\gamma_1)$&$P_{exp}(\gamma_1)$\\
(keV) ~ (keV)& cascade & & & \\
\hline
  448.24- 934.51 &0$^+$-$2^+$-$0^+$  &  0.0       & +1.000    & +1.02(9)  \\
  561.00- 934.51 &0$^+$-$2^+$-$0^+$  &  0.0       & +0.167    & +0.14(2)  \\
  844.27- 561.00 &3$^-$-$4^+$-$2^+$  & -0.051(14) & +0.221(7) & +0.29(7)  \\
  902.96- 934.51 &2$^+$-$2^+$-$0^+$  & -0.04(9)   & +0.32(2)  & +0.1(2)   \\
  912.86- 934.51 &2$^+$-$2^+$-$0^+$  & -0.004(15) & +0.428(4) & +0.33(8)  \\
 1132.21- 934.51 &2$^+$-$2^+$-$0^+$  & -2.9(2)    & +0.007(2) & +0.02(5)  \\
 1405.22- 934.51 &3$^-$-$2^+$-$0^+$  &  0.003(10) & +0.105(4) & +0.11(4)  \\
 2105.20- 934.51 &3$^+$-$2^+$-$0^+$  & -0.04(4)   & +0.09(2)  & +0.39(22) \\
                 &                   &  7.9(21)   & +0.39(2)  &           \\
 2436.84- 934.51 &1$^-$-$2^+$-$0^+$  &  0.02(4)   & +0.32(2)  & +0.12(26) \\
                 &1$^+$-$2^+$-$0^+$  &            & - 0.32(2) &           \\
 2694.07- 934.51 &3$^+$-$2^+$-$0^+$  &  4.8(4)    & -0.428(8) & -0.53(9)  \\
 6294.80-1405.22 &2$^+$-$3^-$-$2^+$  &  0.003(2)  & -0.180(6) & -0.22(9)  \\
                 &                   &  3.6(2)    & +0.131(3) &           \\
\hline
\end{tabular}
\end{center}
\label{Zr92_polcorr_tab10}
\end{table}

Angular correlations for the 448.24-934.51-, 561.00-934.51- and 1405.22-934.51-keV cascades
illustrate high accuracy of the results and confirm spin I=0, I=4 and I=3 assignments or the
1382.76-, 1495.52- and 2339.80-keV levels, respectively. The directional-polarization
correlations for the three cascades uniquely confirm negative parity of the 2339.80-keV level
and positive parity of the 1382.76- and 1495.52-keV levels.

For the 3039.83-keV level the correlation data provide unique spin-parity assignment of
3$^-$, indicating rather large $\delta$ value for the 2105.20-keV transition. The assignment
is supported by the low-energy decay to the 4$^-$ levels at 2743.72 keV.

The (4$^+$) and  (4$^-$) spin-parity assignments for the 3191.07- and 3236.8-keV levels,
proposed above due to their feeding pattern from the neutron-capture level, are consistent
with angular correlations and branchings.

Linear polarization supports negative parity of the 3371.43-keV level. For the 3407.94-keV
level we propose spin-parity 2$^-$ considering mixing ratios obtained presently and reported
in the compilation \cite{NDS_Zr92}.

Our angular correlations reject the previous (2$^+$) assignment to the 3452.47-keV level.
Tentative spin-parity (2$^-$) is proposed instead considering also a new, 709.1-keV decay to
the 4$^-$ level.

Spin-parity of the 3628.50-keV level is 3$^+$ instead of (4$^+$) reported in the compilation
\cite{NDS_Zr92}, as uniquely determined by angular and directional-polarization correlations
for the 2694.07-934.5-keV cascade with a large $\delta$=4.8 of the 2694.07-keV transition.

Angular correlations for the 7698.9-934.5-keV cascade provide unique, I=2 spin assignment
to the capture level in $^{92}$Zr. The 7698.9-keV, primary $\gamma$ line is too weak for
directional-polarization correlation measurement but the positive parity of the capture
level is ensured by the positive parity of ground state of $^{91}$Zr. Therefore, the
6294.9-keV transition populating the 3$^-$ level at 2339.6 keV in $^{92}$Zr is an E1+M2
mixture with $\delta$=0.030(16) determined by correlations in the 6294.9-1405.22-keV,
which indicates its nearly pure E1 character.

\subsection{Excitations in $^{94}$Zr}

New information on $^{94}$Zr has been obtained in this work from $\beta^-$ decay of the
2$^-$, ground state of $^{94}$Y, strongly produced in neutron-induced fission of $^{235}$U
and measured with EXILL.

\begin{figure}
\centering
\scalebox{.42}{\includegraphics{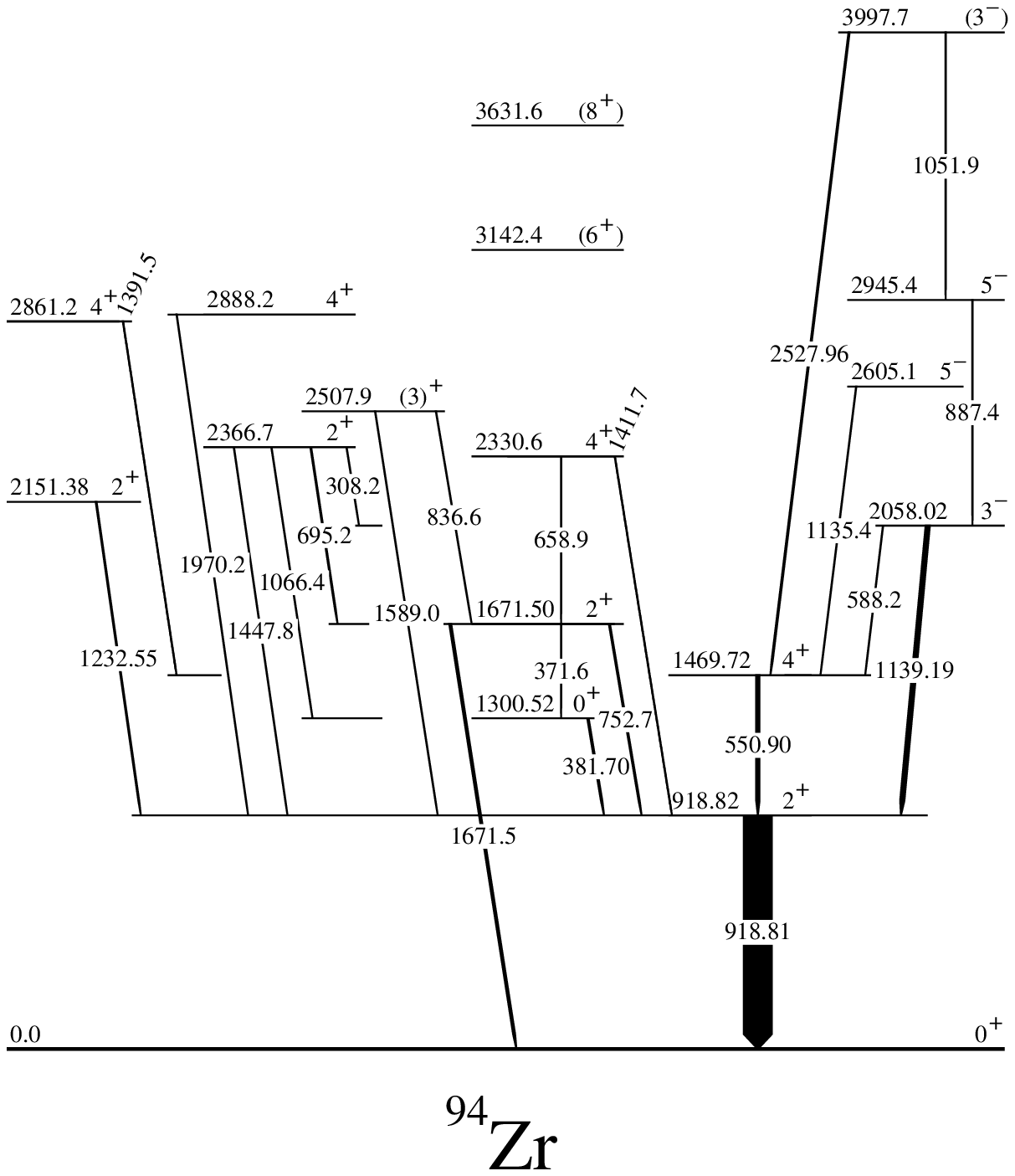}}
\caption{Partial level scheme of $^{94}$Zr populated in $\beta^-$ decay of $^{94}$Y, as
obtained in this work. Arrow width is proportional to the observed $\gamma$ intensity.
More comments in the text.}
\label{Zr94_scheme.eps}
\end{figure}

Properties of excited states in $^{94}$Zr observed in the present work are listed in Table
\ref{Zr94_energies}. The partial excitation scheme in Fig. \ref{Zr94_scheme.eps} shows levels
discussed in the text. To help the discussion also
shown are levels at 3142.4- and 3631.6 keV reported in the compilation \cite{NDS94}.

The results of a previous $\beta^-$ decay study \cite{Sin76} are mostly confirmed. We add 10
levels and 15 transitions, marked with asterisks in Table \ref{Zr94_energies}, which are
either new or now confirmed tentative data of Ref. \cite{Sin76}. The intense 2527.96-keV
transition with uncertain placement in Ref. \cite{Sin76} depopulates the 3997.7-keV level.
We do not confirm the 4002.2-, 4052.4- and 4098.5-keV levels reported in Ref. \cite{Sin76}.
The 3961.8-, 4198.8-, 4237.6- and 4637.9-keV levels proposed tentatively in Ref. \cite{Sin76}
are also not confirmed. The 2492.45-keV transition depopulates the 3411.2-keV level, the
intense 2527.96-keV transition depopulates the 3997.7-keV level and the 2899.2-keV
transition depopulates the 3818.0-keV level. The 966.6-keV transition reported in Ref.
\cite{Sin76} in not seen in our data.

\begin{table}[]
\caption{Experimental properties of excited levels in $^{94}$Zr populated in $\beta^-$
decay of $^{94}$Y, as observed in the present work. Levels, transitions and spins, which
are new compared to Ref. \cite{Sin76} are marked with asterisks. Data with superscript
``a'' are taken from Refs. \cite{Zr94NDS,Cha13}. See text for more comments.}
\begin{center}
\begin{tabular}{l c c c c c}
\hline
\
~~~~E$_i$    &I$^{\pi}_i$&E$_{\gamma}$  &I$_{\gamma}$& E$_f$   &I$^{\pi}_f$\\
  ~~(keV)    &           &   (keV)      &  (rel.)    &  (keV)  &           \\
\hline
~918.82(4)   & 2$^+$     &  918.81(4)   &  1000$^a$  &    0.00 &  0$^+$    \\
1300.52(7)   & 0$^+$     &  381.70(5)   &    27(3)   &  918.82 &  2$^+$    \\
1469.72(7)   & 4$^+$     &  550.90(5)   &    90(8)   &  918.82 &  2$^+$    \\
1671.50(8)   & 2$^+$     &  371.6(4)    &     0.4(2) & 1300.52 &  0$^+$    \\
             &           &  752.7(1)    &    25(4)   &  918.82 &  2$^+$    \\
             &           & 1671.5(1)    &    67(6)   &    0.00 &  0$^+$    \\
2058.02(6)   & 3$^-$     &  588.2(4)    &    2.5(8)  & 1469.72 &  4$^+$    \\
             &           & 1139.19(4)   & 107(7)$^a$ &  918.82 &  2$^+$    \\
2151.38(2)   & 2$^+$     & 1232.55(14)  &    7(1)    &  918.82 &  2$^+$    \\
2330.6(2)    & 4$^+$     &  658.9(4)    & $\le$0.8   & 1671.50 &  2$^+$    \\
             &           & 1411.7(2)    &    1.8(6)  &  918.82 &  2$^+$    \\
2366.7(1)    & 2$^+$     &  308.2(3)    &  0.9(3)    & 2058.02 &  3$^-$    \\
             &           &  695.2(2)    & 3.4(6)$^a$ & 1671.50 &  2$^+$    \\
             &           & 1066.4(2)    &    1.2(4)  & 1300.52 &  0$^+$    \\
             &           & 1447.8(1)    &    4(1)    &  918.82 &  2$^+$    \\
2507.9(3)*   & (3)$^+$   &  836.6(4)*   &    0.6(3)  & 1671.50 &  2$^+$    \\
             &           & 1589.0(3)*   &    2.2(5)  &  918.82 &  2$^+$    \\
2605.1(4)    &  5$^-$    & 1135.4(3)    &    0.8(4)  & 1469.72 &  4$^+$    \\
2846.9(3)    & (1$^-$)   & 1928.1(3)    &    1.3(4)  &  918.82 &  2$^+$    \\
2861.2(5)*   &  4$^+$*   & 1391.5(4)*   &    1.3(4)  & 1469.72 &  2$^+$    \\
2889.0(5)*   &  4$^+$*   & 1970.2(4)*   &    0.5(2)  & 1469.72 &  2$^+$    \\
2907.9(1)    & (2$^+$)   & 1236.4(1)    &    2.2(4)  & 1671.50 &  2$^+$    \\
             &           & 1989.0(5)    &    0.6(3)  &  918.82 &  2$^+$    \\
2945.4(3)    & 5$^-$     &  887.4(2)    &    1.4(4)  & 2058.02 &  3$^-$    \\
3060.0(2)    & (3$^+$) * & 1002.2(4)    &    0.5(2)  & 2058.02 &  3$^-$    \\
             &           & 2141.2(1)    &   17(2)    &  918.82 &  2$^+$    \\
3173.8(3)*   & (2,3) *   & 2254.5(3)*   &    1.2(3)  &  918.82 &  2$^+$    \\
3219.80(7)   & (1,2,3)   & 1162.0(1)    &    5(1)    & 2058.02 &  3$^-$    \\
             &           & 2301.3(2)    &    2.9(4)  &  918.82 &  2$^+$    \\
3361.3(2)    &  (2,3) *  & 1891.6(2)    & 6.9(8)$^a$ & 1469.72 &  4$^+$    \\
             &           & 2442.5(2)    &    2.4(3)  &  918.82 &  2$^+$    \\
3411.2(2)*   &  (2,3) *  & 1941.0(5)*   &    0.7(2)  & 1469.72 &  4$^+$    \\
             &           & 2492.45(15)* &    4.8(5)  &  918.82 &  2$^+$    \\
3724.5(2)*   &(2,3,4)$^+$& 2254.5(3)*   &    1.2(3)  & 1469.72 &  4$^+$    \\
             &           & 2805.8(2)*   &    0.8(3)  &  918.82 &  2$^+$    \\
3818.0(2)*   &  (2,3) *  & 2348.2(4)*   &    0.4(2)  & 1469.72 &  4$^+$    \\
             &           & 2899.2(2)*   &    1.9(3)  &  918.82 &  2$^+$    \\
3997.7(2)*   & (3$^-$) * & 1051.9(4)*   &    0.5(2)  & 2945.4  &  5$^-$    \\
             &           & 2527.96(14)* &    3.4(4)  & 1469.72 &  4$^+$    \\
4107.4(5)*   &  (2,3) *  & 2637.7(5)*   &    0.4(2)  & 1469.72 &  4$^+$    \\
4670.3(5)*   & (3$^-$) * & 3751.5(5)*   &    0.3(1)  &  918.82 &  2$^+$    \\
\hline
\end{tabular}
\end{center}
\label{Zr94_energies}
\end{table}

\begin{table}
\caption{Angular correlation coefficients for $\gamma\gamma$ cascades in $^{94}$Zr following
$\beta^-$ decay of $^{94}$Y produced in neutron-induced fission of $^{235}$U. The correlating,
550.90- and 918.82-keV $\gamma_2$ are assumed to be stretched, E2 transition with $\delta$=0.}
\begin{center} 
\begin{tabular}{ r r l l c }
\hline
E$_{\gamma 1}$-E$_{\gamma 2}$&~$A_2/A_0$~&$A_4/A_0$&Spins in&$\delta_{exp}$(E$_{\gamma 1}$)\\
   cascade                   &   exp.    &  ~~exp.     &cascade~~~~~&    or  $\chi^2$      \\
\hline
   381.70-918.82             &  0.32(4)  &  1.09(8)    & 0 - 2 - 0  &   $\chi^2$=1.1       \\
   550.90-918.82             &  0.11(2)  & -0.02(4)    & 4 - 2 - 0  &   $\chi^2$=0.6       \\
   752.7 -918.82             &  0.20(4)  &  0.11(8)    & 2 - 2 - 0  &   0.07(5) or         \\
                             &           &             &            &  -2.6(4)             \\
  1139.19-918.82             & -0.09(2)  &  0.01(4)    & 3 - 2 - 0  &  -0.02(2)            \\
  1232.55-918.82             &  0.36(6)  &  0.11(16)   & 2 - 2 - 0  &  -0.16(11) or        \\
                             &           &             &            &  -1.5(4)             \\
  1447.8 -918.82             & -0.18(11) &  0.06(23)   & 2 - 2 - 0  &   0.6(3)             \\
  2141.2 -918.82             &  0.37(25) &  0.17(57)   & 1 - 2 - 0  &  -0.51(6)            \\
                             &           &             & 2 - 2 - 0  &  -0.08(8)            \\
                             &           &             & 3 - 2 - 0  &   0.8(3)             \\
  2492.45-918.82             &  0.05(12) & -0.33(27)   &            &                      \\
  2527.96-550.90             & -0.28(11) &  0.08(23)   & 3 - 4 - 2  &   0.03(10)           \\
                             &           &             & 4 - 4 - 2  &   0.9(-3,+18)        \\
\hline
\end{tabular}
\end{center}
\label{Sr94_angcorr}
\end{table}

\begin{table}
\caption{Experimental, P$_{exp}(\gamma_1)$ and theoretical, P$_{th}(\gamma_1)$ values of
linear polarization for the $\gamma_1$ (upper) transition in a $\gamma_1-\gamma_2$ cascade of
$^{94}$Zr, as obtained in this work. The correlating  918.82-keV $\gamma_2$ is assumed to be
stretched, E2 transition.}
\begin{center} 
\begin{tabular}{c c c c c}
\hline
E$_{\gamma 1}$-E$_{\gamma 2}$&P$_{exp}(\gamma 1)$& Spin-parity&$\delta_{exp}(\gamma 1)$&
                                                                    P$_{th}(\gamma 1)$\\
\hline
  381.70-918.82            & +0.93(12)   & 4$^+$-2$^+$-0$^+$ & 0.0   & +1.0;     E2 \\
  550.90-918.82            & +0.14(3)    & 4$^+$-2$^+$-0$^+$ & 0.0   & +0.1667;  E2 \\
 1139.19-918.82            & +0.12(7)    & 3$^-$-2$^+$-0$^+$ & 0.0   & +0.1034;  E1 \\
 2141.2 -918.82            & -0.5(4)     & 1$^-$-2$^+$-0$^+$ &-0.5(1)& -0.70(7); M1 \\
                           &             & 2$^-$-2$^+$-0$^+$ &-0.1(1)& -0.40(4); E1 \\
                           &             & 3$^-$-2$^+$-0$^+$ & 0.8(3)& -0.46(9); M1 \\
 2527.96-918.82            & -0.1(2)     & 3$^-$-2$^+$-0$^+$ & 0.0   & +0.1034;  E1 \\
                           &             & 2$^-$-2$^+$-0$^+$ & 0.9   &  0.19(17);M1 \\
\hline
\end{tabular}
\end{center}
\label{Sr94_polcorr}
\end{table}

\begin{table}
\caption{$log ft$ values calculate in this work for newly-observed levels in $^{94}$Zr.
See text for further explanation.}
\begin{center}
\begin{tabular}{|c c | c c | c c |}
\hline
E$_{exc}$(keV)& $log ft$  &E$_{exc}$(keV)& $log ft$& E$_{exc}$(keV)& $log ft$ \\
\hline
2507.9   &  8.4(1)        & 3411.2       &  7.3(1) & 4107.4        &  7.4(3)  \\
2861.2   &  9.5(2)$^{1u}$ & 3724.5       &  7.3(1) & 4670.3        &  5.9(2)  \\
2889.0   &  9.9(2)$^{1u}$ & 3818.0       &  7.1(1) &               &          \\
3173.8   &  8.2(1)        & 3997.7       &  6.6(3) &               &          \\
\hline
\end{tabular}
\end{center}
\label{Sr94_logft}
\end{table}

Because the singles $\gamma$ spectrum was too complex in the present measurement, we used
$\gamma\gamma$ coincidences to determine intensities of $\gamma$ lines. The four intensities
marked with superscript ``a'' in Table \ref{Zr94_energies} are taken from Ref. \cite{Sin76}
and used to normalize coincidence $\gamma$ intensities to singles $\gamma$ intensities. One
notes overall good agreement with singles intensities of Ref. \cite{Sin76}.

For the strongest $\gamma\gamma$ cascades angular correlations and directional-polarization
correlations have been determined, as described in Ref. \cite{Urb23}. The results are listed
in Table \ref{Sr94_angcorr} and Table \ref{Sr94_polcorr}, respectively.

We have calculated $log ft$ values for new levels in $^{94}$Zr. In calculations we used the
LOGFT Analysis Program available at \cite{ENSDF} and assumed $\beta^-$  feeding to the ground
state and the 2$^+_1$ levels in $^{94}$Zr as reported in the Ref. \cite{NDS94}. The results
are shown in Table \ref{Sr94_logft}.

Spins in $^{94}$Zr populated in $\beta^-$ decay of the 2$^-$ ground state of $^{94}$Y are
limited to 0 - 4 values. Spin-parity assignments shown in Table \ref{Zr94_energies}
are taken from the compilation \cite{NDS94} except those marked with asterisks, which are
proposed in this work based on angular correlations, directional-polarization correlations,
the observed decay branchings and $log ft$ values obtained in this work.

Accurate angular correlations and directional-polarization correlations for the
381.70-918.82-keV, 550.90-918.82-keV and 1139.19-918.82-keV cascades confirm spins and
parities of the 0$^+$, 1300.52-, 4$^+$, 1469.72- and 3$^-$, 2058.02-keV levels.

For the 3060.0-keV level we propose (3$^+$) spin-parity, considering the present correlations
and no decay to the ground state. To the 3997.7-keV level we assign (3$^-$) spin-parity as
the most likely solution. The I=4 option with $\delta$=0.9 would implicate positive parity
for this level, which is less compatible with its decay branching and linear polarization of
the 2527.96-keV transition.

The $log ft > 7$ values in $^{94}$Zr \cite{NDS94} point to first-forbidden transitions
from the 2$^-$ ground state of $^{94}$Y to levels in $^{94}$Zr. These transitions may be due
to the $\nu d_{5/2}\rightarrow \pi p_{1/2}$ or $\nu d_{5/2}\rightarrow \pi p_{3/2}$ decays.

The 2$^+$ levels at 2151.38 and 2366.7 keV and the 2507.9-keV level, with $log ft$ of 8.6,
8.2 \cite{NDS94} and 8.4, respectively, are populated in $\beta^-$ decay much weaker than
the 2$^+_1$ level at 918.82 keV with $log ft$ of 7.2 \cite{NDS94}, which is proposed to be a
member of the $\pi (g_{9/2})^2$ s.p. multiplet \cite{Pan05}. We propose that the 2151.38 and
2366.60 keV levels are of a collective nature.

The (3$^-$), 4670.3-keV level with $log ft$ value of 5.8, as calculated in this work,
may be populated in part by the $\nu g_{7/2}\rightarrow \pi g_{9/2}$, Gamow-Teller transition
from the $(\nu g_{7/2}, \pi p_{3/2})_{2^-}$ contribution to the wave function of the 2$^-$
ground state in $^{94}$Y. This produces the $\pi (g_{9/2}, p_{3/2})_{3^-}$, s.p. contribution
to the wave function of this level. The (3$^-$) spin-parity of the 3997.7-keV level also fits
this scenario, though its $log ft$ of 6.6 is larger.

\subsection{Excitations in $^{98}$Zr}

Excited levels in $^{98}$Zr were studied before in many experiments listed in the compilation
\cite{NDS98}, including $\beta$-decay of $^{98}$Y \cite{Kaw82,Urb17}, transfer reaction
\cite{Mey86}, spontaneous fission of $^{248}$Cm \cite{Urb01} and $^{252}$Cf \cite{Hwa12,Urb17}
and $\alpha$-induced fission of $^{238}$U \cite{Wu004}. Level live-times were reported in a
number of works \cite{Sim06,Bet10,Ans17,Sin18,Wit18,Kar20,Pas23}. In the present work the
experimental information on $^{98}$Zr has been updated in order to address open questions
left by previous studies and to search for new excitations in this nucleus.

We used high-statistics data from the measurement of $\gamma$ rays following neutron-induced
fission of $^{235}$U, performed at the ILL, Grenoble during the EXILL campaign \cite{Jen17}.
Partial results of this analysis, concerning levels populated in $\beta$-decay of $^{98}$Y
were reported in Ref. \cite{Urb17}. In this work we report results for levels which are
populated directly in fission, as assured by coincidences with $\gamma$ rays in the
complementary, $^{134}$Te fission fragment.

\begin{figure*}
\centering
\scalebox{.73}{\includegraphics{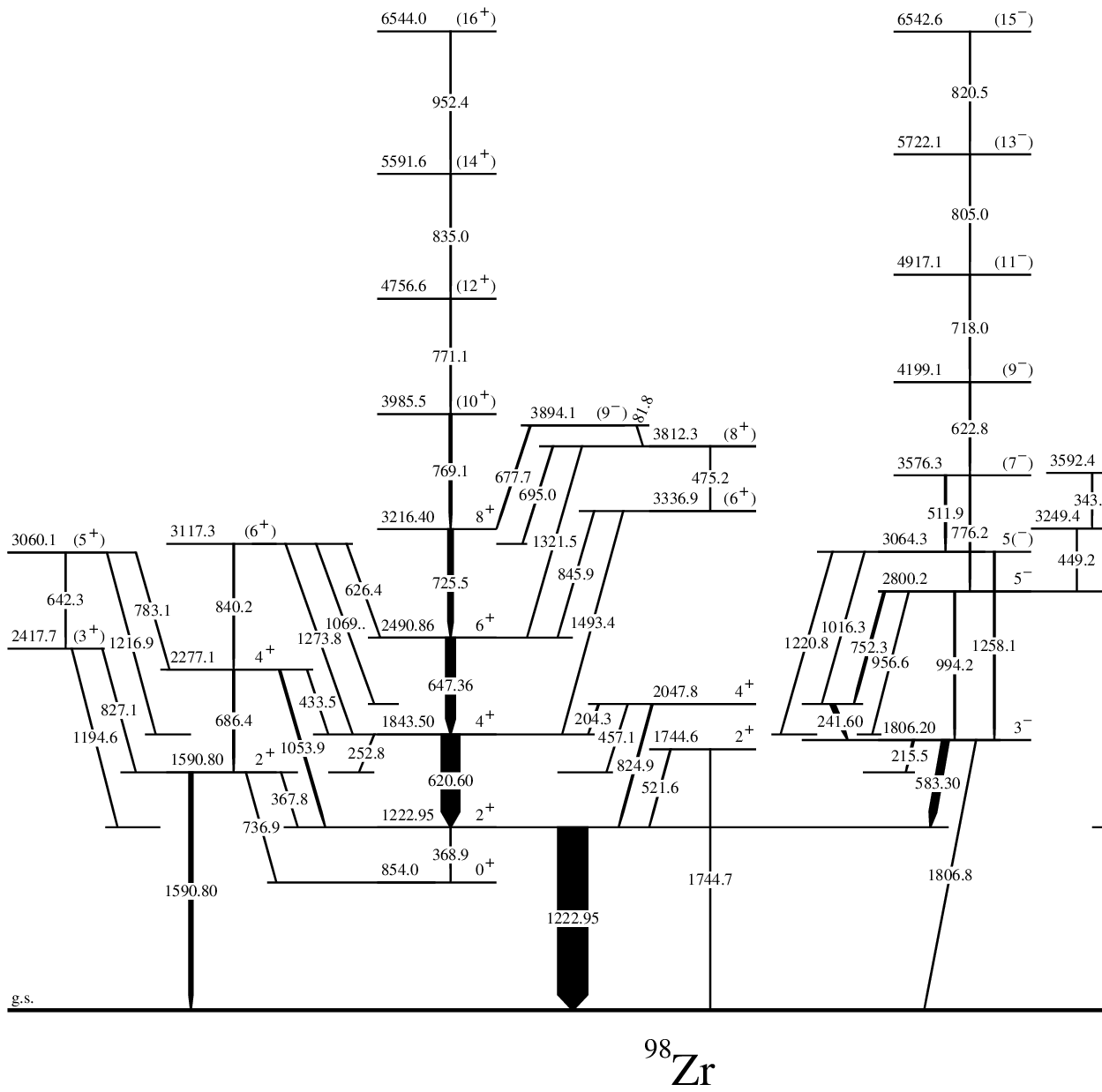}}
\caption{Partial level scheme of $^{98}$Zr obtained in this work in the measurement of
$\gamma$ rays from neutron-induced fission of $^{235}$U.}
\label{Zr98_scheme}
\end{figure*}

The excitation scheme of $^{98}$Zr, obtained from triple-$\gamma$ coincidences is shown in
Fig. \ref{Zr98_scheme}. Properties of excited levels and their $\gamma$ decays are listed in
Table \ref{table_Zr98levels}. Three levels and nine transitions, which are new compared to Ref.
\cite{NDS98}, are marked by asterisks. We note significantly different energies of high-energy
levels compared to \cite{NDS98}. The T$_{dbe}$ half-life estimates are shown in Table
\ref{table_Zr98levels} at those $\gamma$ transitions for which they were obtained.

The present work confirms results obtained in $\beta^-$ the decay study of $^{98}$Y
\cite{Urb17} (not shown in Table \ref{table_Zr98levels} and Fig. \ref{Zr98_scheme}). In
particular we confirm the 253.7(2)-1801.50(5)-keV from the (7$^+$) level at 4546.1(3)-keV
level, populated in $\beta^-$ decay of the 2.36~s isomer in $^{98}$Y.

\begin{table}[]
\caption{Excited levels and their $\gamma$ decays in $^{98}$Zr as observed in this work
following neutron-induced fission of $^{235}$U.}
\begin{center}
\begin{tabular}{c c c r c l}
\hline
E$_{ex}$(keV&I$^{\pi}$& E$_{\gamma}$(keV)&I$_{\gamma}$(rel)&T$_{dbe}$(ps)&B(E$\lambda$)(W.u.)\\
\hline
 1222.95(5)    &  2$^+$   &    368.9(2)       &   21(3)   & 4.5(15) & B(E2)$\ge$14           \\
               &          &   1222.95(5)      & 1000.0    &   6(2)  & B(E2)$\ge$1.3          \\
 1436.1(2)     &  0$^+$   &    213.3(2)       &   10(5)   & &                                \\
 1590.80(5)    &  2$^+$   &    367.8(2)       &   14(4)   & &                                \\
               &          &    736.9(2)       &    8(3)   & &                                \\
               &          &   1590.80(5)      &  116(8)   &   6(2)  & B(E2)$\ge$0.3          \\
 1744.6(1)     &  2$^+$   &    521.6(1)       &   17(4)   & &                                \\
               &          &   1744.7(2)       &   12(3)   & &                                \\
 1806.20(7)    &  3$^-$   &    215.5(1)       &   13(3)   & &                                \\
               &          &    583.30(5)      &  240(9)   &  $>$15  & B(E1)$<$0.00008        \\
               &          &    1806.8(4)      &    0.4(1) &         &                        \\
 1843.50(7)    &  4$^+$   &    252.6(2)       &    8(3)   & &                                \\
               &          &    620.55(6)      &  605(35)  &  5.1(8) & B(E2)$\ge$46           \\
 2047.8(1)     &  4$^+$   &    204.3(1)       &   26(2)   & &                                \\
               &          &    241.60(7)      &  144(7)   & &                                \\
               &          &    457.1(3)       &   10(3)   & &                                \\
               &          &    824.9(1)       &   38(5)   & &                                \\
 2225.4(2)     &  2$^+$   &    789.3(3)       &    9(4)   & &                                \\
               &          &   1002.3(3)       &   25(7)   & &                                \\
               &          &   2225.6(3)       &   12(4)   & &                                \\
 2277.1(1)*    & 4$^+$ *  &    433.5(2)*      &    6(1)   & &                                \\
               &          &    686.4(1)*      &   40(8)   &   7(2)  & B(E2)$\ge$14           \\
               &          &   1053.9(1)*      &   44(5)   &  3.2(6) & B(E2)$\ge$15            \\
 2417.7(2)*    & (3$^+$)* &    827.1(3)*      &   11(4)   & &                                \\
               &          &   1194.6(3)*      &    8(3)   & &                                \\
 2490.86(10)   &  6$^+$   &    647.36(7)      &  310(11)  &  4.5(8) & B(E2)$\ge$41           \\
 2800.2(2)     &  5$^-$   &    752.3(1)       &   58(4)   &   6(2)  & B(E1)$\ge$0.00008      \\
               &          &    956.6(2)       &    8(2)   & &                                \\
               &          &    994.2(2)       &   34(3)   &  3.5(7) & B(E2)$\ge$2            \\
 3060.1(3)*    & (5$^+$)* &    642.3(2)*      &   10(3)   & &                                \\
               &          &    783.1(3*       &    3(1)   & &                                \\
               &          &   1216.9(4)       &    5(2)   & &                                \\
 3064.3(2)     & (5$^-$)  &   1016.3(2)       &    8(2)   &  3.0(6) & B(E1)$\ge$0.00002      \\
               &          &   1220.8(2)       &    7(2)   & &                                \\
               &          &   1258.1(1)       &   41(5)   &   5(1)  & B(E2)$\ge$1.3          \\
 3117.3(2)     & (6$^+$)  &    626.4(3)       &    4(1)   & &                                \\
               &          &    840.2(1)       &   30(5)   &  2.9(6) & B(E2)$\ge$14           \\
               &          &   1273.8(2)       &    6(2)   & &                                \\
 3216.40(15)   &  8$^+$   &    725.5(1)       &  158(9)   &  3.0(6) & B(E2)$\ge$35           \\
 3249.4(3)     & (6$^-$)* &    449.2(2)       &   13(3)   & &                                \\
 3336.9(3)     & (6$^+$)  &    845.9(3)       &    4(1)   & &                                \\
               &          &   1493.4(2)       &    9(1)   & &                                \\
 3576.3(2)     & (7$^-$)  &    511.9(2)       &   35(6)   & &                                \\
               &          &    776.2(1)       &   29(5)   &  5.2(12) & B(E2)$\ge$7          \\
 3592.4(3)     & (7$^-$)  &    343.4(4)       &    2(1)   & &                                \\
               &          &    792.0(2)       &   22(5)   & &                                \\
 3812.3(3)     & (8$^+$)  &    475.2(3)*      &    2(1)   & &                                \\
               &          &    695.0(2)       &   14(3)   & &                                \\
               &          &   1321.5(2)       &   10(2)   & &                                \\
 3894.1(2)     & (9$^-$)* &     81.8(1)*      &    5(1)   & &                                \\
               &          &    677.7(1)       &   25(2)   &  $>$15   &                       \\
 3985.5(2)     & (10$^+$) &    769.1(1)       &   65(5)   &  2.2(6)  & B(E2)$\ge$36          \\
 4199.1(3)     &  (9$^-$) &    622.8(2)       &   23(5)   &  4.0(12) & B(E2)$\ge$56          \\
 4756.6(3)     & (12$^+$) &    771.1(2)       &   20(5)   &  2.2(8)  & B(E2)$\ge$35          \\
 4917.1(4)     & (11$^-$) &    718.0(3)       &   17(4)   &  4.5(15) & B(E2)$\ge$24          \\
 5591.6(4)     & (14$^+$) &    835.0(2)       &   17(3)   &  2.1(5)  & B(E2)$\ge$25          \\
 5722.1(6)     & (13$^-$) &    805.0(4)       &   10(3)   &  6(2)    & B(E2)$\ge$10          \\
 6542.6(7)     & (15$^-$) &    820.5(4)       &    6(2)   &  $>$15   & B(E2)$<$4             \\
 6544.0(5)     & (16$^+$) &    952.4(3)       &    8(2)   &  3.5(12) & B(E2)$\ge$8           \\
\hline
\end{tabular}
\end{center}
\label{table_Zr98levels}
\end{table}

Table \ref{table_Zr98angcorr} lists angular correlations coefficients for $\gamma$-$\gamma$
cascades in $^{98}$Zr. They complement angular correlations for levels populated in $\beta$
decay of $^{98}$Zr \cite{Urb17}. Column ``Spins'' shows spins in $\gamma_1 - \gamma_2$
cascades.

The directional-polarization correlations technique developed for EXILL \cite{Urb23} provided
linear polarization values for two strong transitions in coincidence with the 1222.95-keV,
stretched E2 transition. For the 583.30- and 620.60-keV transitions we obtained P$_{exp}$=
+0.14(3) and P$_{exp}$= +0.07(2), respectively, confirming their E1 and E2 multipolarities,
respectively.

\begin{table}[]
\caption{Normalized, experimental angular-correlation coefficients, $a_{k}/a_{0}$ and the 
corresponding mixing ratios, $\delta$, for $\gamma$ transitions in $^{98}$Zr, as observed
in neutron-induced fission of $^{235}$U. The 1222.9- and 1590.9-keV transitions are the
reference, stretched quadrupole transitions \cite{NDS98} and ``sum'' denotes summed gates
on quadrupole transitions below $\gamma_1$.}
\begin{center}
\begin{tabular}{c c c c c}

\hline
Cascade            &$a_{2}/a_{0}$& $a_{4}/a_{0}$& Spins           & $\delta$(${\gamma}_1$) \\
$\gamma_1 - \gamma_2$& (exp.)    & (exp.)       & $I_1, I_2, I_3$ &                        \\
\hline
&&&& \\
241.60  - 583.30  & ~0.027(15)  &  -0.008(35)  & 4$\rightarrow$3$\rightarrow$2 & 0.05(3)  \\
725.5~ - ~sum~    & ~0.111(16)  &  -0.037(36)  & 8$\rightarrow$6$\rightarrow$4 &          \\
~620.60 - 1222.95 & ~0.102(12)  &  -0.037(28)  & 4$\rightarrow$2$\rightarrow$0 &          \\
1002.3~ - 1222.95 & -0.069(57)  &  ~0.100(99)  & 2$\rightarrow$2$\rightarrow$0 & 0.44(9)  \\
1053.9~ - 1222.95 & ~0.108(29)  &  -0.051(70)  & 4$\rightarrow$2$\rightarrow$0 &          \\
1258.1~ - 583.30~ & -0.083(57)  &  -0.009(95)  & 5$\rightarrow$3$\rightarrow$2 &          \\
1321.5~ - ~sum~~  & ~0.142(62)  & ~~0.135(105) & 8$\rightarrow$6$\rightarrow$4 &          \\
1493.4~ - ~sum~~  & ~0.125(57)  & ~-0.055(125) & 6$\rightarrow$4$\rightarrow$2 &          \\
\hline
\end{tabular}
\end{center}
\label{table_Zr98angcorr}
\end{table}

Spins reported in the yrast cascade \cite{NDS98} are confirmed up to I=8. The (10$^+$)
assignment for the 3985.5-keV level remains tentative because of the 769.1+771.1-keV doublet.
We confirm, based on the presently observed $\gamma$ intensities, that both members of the
doublet belong to the yrast cascade, as reported in Refs. \cite{Wu004,Sim06}, rather than
forming a bifurcation suggested in Ref. \cite{Urb01}. Spins and parities of levels proposed
in the band above the 3985.5-keV level are tentative but supported by short T$_{dbe}$ values
supporting positive-parity, $\Delta$I=2 transitions.

The negative-parity band reported in Ref. \cite{Sim06} up to spin (15$^-$) is confirmed in
the present work. We could not observe directly the 1.9 $\mu$s isomer reported in Ref.
\cite{Sim06} because of its low population but its presence is supported by the T$_{dbe}$
for the 952.4- and 820.5-keV transitions, higher than the T$_{dbe}$ for the respective,
835.0- and 805.0-keV transitions below; feeding of the 6544.0- and 6542.6-keV levels by
an isomer would lead to higher T$_{dbe}$ values for these levels. We note the different
energies of the two levels, indicating two different isomer decay branches, one
of the options considered in Ref. \cite{Sim06}.

The 2277.1-, 3117.3-, 3336.9- and 3812.3-keV levels, reported in Ref. \cite{Urb01} are
confirmed and the 686.4- and 845.9-keV transitions, reported in \cite{Urb01} as tentative,
are now firmly observed. To this structure we add two new levels at 2417.7 and 3060.1 keV
and ten new decays, in total. Levels in this structure decay to the positive-parity, yrast
band, suggesting their positive parity.

The 3249.4-keV level, reported with spin-parity (6$^+$) in Ref. \cite{Urb17}, and the
3592.4-keV level are proposed wit tentative spin parity (6$^-$) and (7$^-$), respectively,
based on their decay properties, including the upper intensity limit of the unobserved,
1202-keV decay of the 3249.4-keV level to the 4$^+$ at 2047.8 keV, which amounts to 0.1 in
the relative units of Table \ref{table_Zr98levels}. This is two orders of magnitude less
than the intensity of the 449.2-keV decay, making the 6$^+$ spin-parity assignment to the
3249.4-keV level unlikely.

Angular correlations for the 1493.4- and 1321.5-keV transitions are consistent with
spin-parity of 6$^+$ and 8$^+$ for the 3336.9- and 3812.3-keV levels, respectively.

The spin of the 3894.1-keV level is limited to 8 or 9 due to the observed decay branches.
Spin I=9 is favoured by the rather strong population of the 3894.1-keV level suggesting
its yrast character. Negative parity for this level is suggested by the T$_{dbe}>$20 ps
value for the 677.7-keV transition. The high T$_{dbe}$ value points to an isomeric nature
of the 3894.1-keV level. One notes, that the new, 81.8-keV decay of the 3894.1-keV level
is much faster than the 677.7-keV decay branch. Assuming B(E1) of 0.0001 W.u. for
the 81.8-keV decay, similar to other E1 decays in $^{98}$Zr, one obtains a half-life
estimate for the 3894.1-keV level of about 5 ns.

To help further discussions we estimated upper limits for $\gamma$ intensities of two
unobserved decays. The upper limit for the 154.7-keV, unobserved branch from the
1590.80-keV level to the 1436.1-keV level is 4$\times$10$^{-4}$ fraction of $\gamma$
intensity of the 1590.80-keV transition. This is significantly lower than the
1.9$\times$10$^{-2}$ fraction reported in the compilation \cite{NDS98}. For the 308.5-keV,
unobserved branch from the 1744.6-keV level to the 1436.1-keV level the upper limit is
0.0014 with respect to the I$_{\gamma}$(1744.7).

\subsection{T$_{dbe}$ estimates in $^{94}$Sr, $^{96}$Sr and $^{96}$Zr}

\begin{table}[]
\caption{T$_{dbe}$ half-life estimates in $^{94}$Sr, $^{96}$Sr and $^{96}$Zr obtained
in the present work. Level and $\gamma$ energies, spins and branchings are taken
from Refs. \cite{Urb21,Wis23} and the present work.}
\begin{center}
\begin{tabular}{l c c c l l}
\hline
& & & & & \\
E$_{ex}$(keV)&I$^{\pi}$&E$_{\gamma}$(keV)&I$_{\gamma}$ (rel.)&T$_{dbe}$(ps)&B($\pi \lambda$)(W.u.)\\
\hline
& & & & & \\
              &            &         &$^{94}$Sr&          &                               \\
& & & & & \\
 1926.40      &  3$^-$     & 1089.45  & 100(3) &  6.4(7)  & B(E1)$\ge$0.00004             \\
              &            & 1926.7(3)&  1.1(2)&          & B(E3)$\ge$40                  \\
 2146.00      &  4$^+$     & 1309.05  &        &  3.5(5)  & B(E2)$\ge$1.7                 \\
 2414.50      &  3$^+$     & 1577.55  &        &  6(1)    &                               \\
 2604.1       &  4$^-$     &  677.70  &        &  9(2)    &                               \\
 2649.8       &  4$^+$     & 1812.7   &        &  5(1)    & B(E2)$\ge$0.2                 \\
 2739.4       &  4$^{(-)}$ &  813.0   &        &  5(1)    &                               \\
 2856.9       &  5$^+$     &  710.90  &        &  6(1)    &                               \\
 2972.0       &  5$^-$     &  826.00  & 100(8) &  4.2(7)  & B(E1)$\ge$0.00007             \\
              &            & 1045.60  &  82(6) &  4.0(7)  & B(E2)$\ge$2                   \\
 3155.9       &  6$^+$     & 1009.90  &        &  8(2)    & B(E2)$\ge$2.5                 \\
 3310.6       &  5$^-$     &  661.0   &  75(5) &  6(1)    & B(E1)$\ge$0.00008             \\
              &            & 1384.3   & 100(5) &  5.5(9)  & B(E2)$\ge$0.4                 \\
 3705.8       &  6$^{(+)}$ & 1559.8   &        &  5(1)    & B(E2)$\ge$0.3                 \\
 3923.3       &  7$^-$     &  130.20  &  35(2) &          &                               \\
              &            &  767.35  & 100(3) &  7(2)    & B(E1)$\ge$0.00007             \\
              &            &  951.15  &  26(2) &  9(2)    & B(E2)$\ge$0.8                 \\
              &            & 1066.5   &   9(2) &          &                               \\
 4034.6       & (7$^+$)    & 1177.9   &        &  3.5(5)  & B(E2)$\ge$1.2                 \\
 4952.9       & (8$^+$)    & 1000.8   &100(15) &  6(1)    & B(E2)$\ge$2.8                 \\
              &            & 1247.1   & 30(10) &          & B(E2)$\ge$0.3                 \\
\hline
& & & & & \\
              &            &          &$^{96}$Sr&         &                               \\
  815.00      &  2$^+$     &  815.00  &        &  4.5(5)  & B(E2)$\ge$13.4                \\
 1507.0       &  2$^+$     &  277.7   & 2.8(5) &          & B(E2)$\ge$40                  \\
              &            &  692.00  &100(3)  &          &                               \\
              &            & 1506.95  & 53(2)  &   6(1)   & B(E2)$\ge$0.2                 \\
 1975.9       &  4$^+$     & 347.3    & 18(1)  &          &                               \\
              &            & 468.9    &100(20) &          &                               \\
              &            & 1160.8   & 54(2)  &  5(1)    & B(E2)$\ge$0.7                 \\
 2120.1       &  4$^+$     & 1305.10  &        &  3(1)    & B(E2)$\ge$1.8                 \\
 2466.8       &  6$^+$     &  491.00  &  82(3) &          & B(E2)$\ge$55                  \\
              &            &  674.00  & 100(3) &  6(1)    & B(E2)$\ge$14                  \\
 2786.0       &  6$^+$     &  810.3   &  50(3) &          & B(E2)$\ge$4                   \\
              &            &  993.20  & 100(3) &  5(1)    & B(E2)$\ge$3                   \\
 2899.8       & (6$^+$)    &  779.6   &100(20) &  6(2)    & B(E2)$\ge$6                   \\
              &            & 1107.0   & 66(12) &          & B(E2)$\ge$1                   \\
 3126.1       &  8$^+$     &  659.30  &        &  3.0(4)  & B(E2)$\ge$60                  \\
 3708.5       & (8$^+$)    &  922.5   &        &  3(1)    & B(E2)$\ge$11                  \\
 3887.3       & (10$^+$)   &  761.2   &        &  1.8(3)  & B(E2)$\ge$47                  \\
 4725.6       & (12$^+$)   &  838.3   &        &  1.0(2)  & B(E2)$\ge$52                  \\
\hline
& & & & & \\
              &            &          &$^{96}$Zr&         &                               \\
 3082.50      &     4$^+$  & 1185.2   &        &  7(1)    & B(E1)$\ge$0.00003             \\
 3176.5       &     4$^+$  & 1279.2   &        &  4(1)    & B(E1)$\ge$0.00004             \\
 4126.5       &    (6,7)   & 1006.0   &        &  7(1)    &                               \\
 4234.5       &     7$^-$  &  750.8   &  20(10)&          & B(E1)$\ge$0.00003             \\
              &            & 1114.6   & 100(9) &  4(1)    & B(E2)$\ge$3                   \\
 4751.3       &  (7,8)$^+$ &  979.1   &        &  7(1)    &                               \\
 4907.2       &    (9$^+$) &  517.6   &        &  7(2)    & B(E2)$\ge$12                  \\
 5484.3       &   (10$^+$) & 1094.7   &        &  4(1)    & B(E2)$\ge$3                   \\
 5738.3       &   (11$^+$) &  831.1   &        &  5(1)    & B(E2)$\ge$9                   \\
\hline
\end{tabular}
\end{center}
\label{table_halflives}
\end{table}

In the present work we estimated half-lives, T$_{dbe}$, for several levels in
$^{94}$Sr, $^{96}$Sr and $^{96}$Zr nuclei, complementing our studies of these nuclei reported
in Refs. \cite{Urb21,Wis23}. This information is used in the following sections to discuss
development of collectivity in the A$\approx$100 region.

The obtained T$_{dbe}$ values, listed in Table \ref{table_halflives} are mostly consistent
with T$_{1/2}$ values reported for $^{94}$Sr, $^{96}$Sr and $^{96}$Zr in the compilations
available at ENSDF base \cite{ENSDF} and in Ref. \cite{Reg17}. For low-spin levels in
$^{96}$Zr the DBE analysis was obstructed  by feeding from the strongly populated, 8$^+$,
4389.60-keV isomer with T$_{1/2}$=127 ps \cite{NDS96}.

\section{Discussion} 

The dominating mode of low-energy excitations in spherical nuclei are single-particle
excitations, though there may also be low-energy, collective modes due to various vibrations
(phonons), believed to be virtual excitations between valence orbitals enhanced by coupling
to the corresponding giant vibrations \cite{Ber19,Bor16,Wal11}.

In deformed nuclei another low-energy collectivity emerges, the nuclear rotation, a quantum
effect restoring broken spherical symmetry. There one may expect also the, so-called $\beta$
vibrations, whereas particle excitations will appear at higher energies.

The structure of transitional nuclei, positioned in between, can be seen as a ``skeleton''
of particle excitations in a nuclear potential ``dressed'' with various collective modes as
described theoretically in Refs. \cite{Kur72,Kur75,Paa73,Mat16,Kis25} and observed in
experiments \cite{Rza13,Wis19,Urb25}. With the increasing number of valence nucleons these
``dressed'' excitations evolve towards deformed structures \cite{Cak08}.

Various orbitals of high-$j$ valence shells populated in transitional nuclei act differently
depending on their K values, driving either towards a deformed (low K) or spherical (high K)
shape  (an example analysis for the $j$=9/2 shell can be found in Ref. \cite{Boy21}). In the
A$\sim$100 region the high-K, 9/2$^+$[404] extruder orbital can play a dual role, helping
also a transition towards a deformed shape, as will be argued in the following.

\begin{figure}
\centering
\scalebox{.35}{\includegraphics{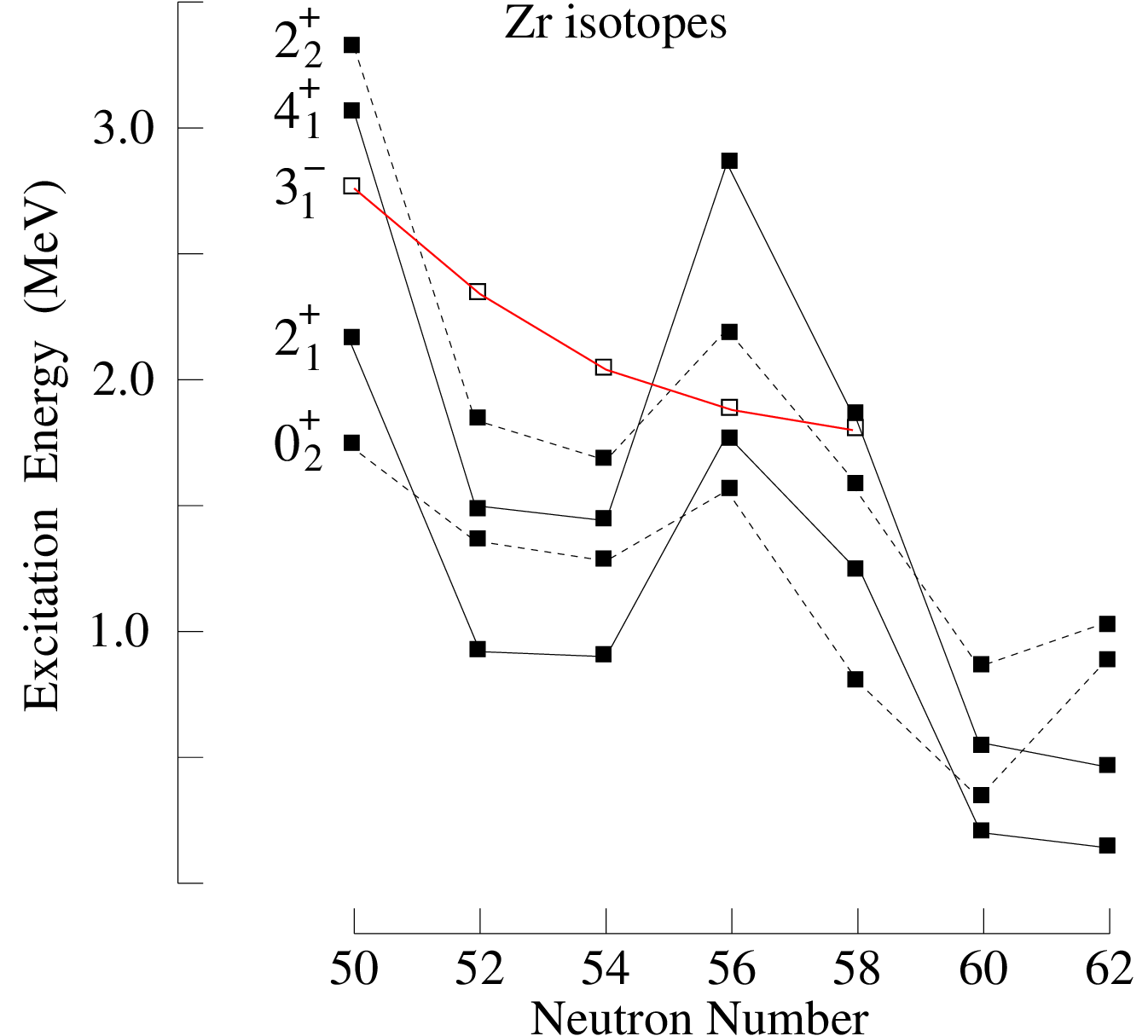}}
\caption{Low-energy excitations in Zr isotopes. The experimental data are taken from the
         ENSDF data base \cite{ENSDF}.}
\label{Zr_even_light_low-en-ex.eps}
\end{figure}

Figure \ref{Zr_even_light_low-en-ex.eps} shows that the lowest 0$^+$, 2$^+$ and 4$^+$ excited 
levels in $^{92,94,96,98}$Zr isotopes are strongly influenced by s.p. excitations whereas
above N=58 there is a  spectacular lowering of positive-parity levels, indicating a change in
the nature of these excitations. In contrast, the 3$^-$ levels, which are most likely due to
collective, octupole excitations at N$<$ 58, do not show such rapid variations. In the
following subsections we will discuss these effects in more detail.

\subsection{2$^+$ excitations}

\begin{figure}
\centering
\scalebox{.44}{\includegraphics{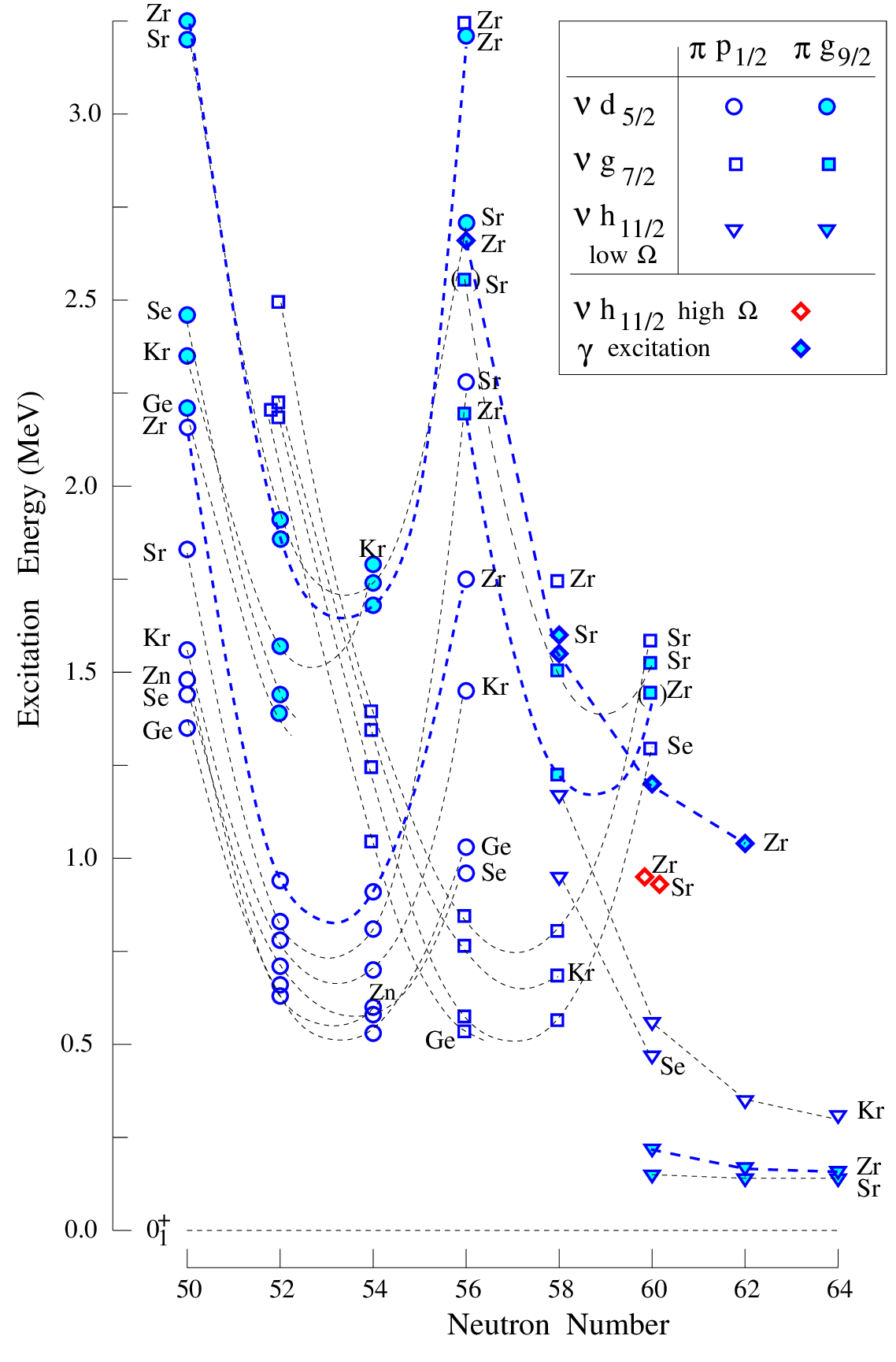}}
\caption{Excitation energies of 2$^+_1$, 2$^+_2$ and some of 2$^+_3$ and 2$^+_4$  levels in
even-even, Zn-to-Zr isotopes. The data are taken from the ENSDF base \cite{ENSDF} and from
the present work. Lines are drawn to guide the eye. Tentative spin assignments are in
parenthesis. See text for further explanation.}
\label{Zr_even_light_2plus}
\end{figure}

Energies of 2$^+$ levels in Zn-Sr even-even nuclei in the 50$\le$N$\le$60 neutron range
display regular systematic trends, as found in our recent works \cite{Urb21,Wis23}. As
noted there, the 2$^+$ excitations are not of a phonon nature. They are due to s.p.
excitations of valence protons and neutrons, ``dressed in collectivity'' \cite{Mat16}.
Figure \ref{Zr_even_light_2plus} is an extended version of Fig. 13 from Ref. \cite{Urb21},
updated with 2$^+$ levels in even-even Zr isotopes.

Over 60 data points shown in Fig. \ref{Zr_even_light_2plus}, corresponding to 2$^+_1$,
2$^+_2$ and some 2$^+_3$ and 2$^+_4$ levels could be arranged along ``parabolic'' curves.
There are four groups of levels linked by dashed  ``parabolas'' corresponding to specific
proton-neutron configurations shown in the legend.

Low-energy 2$^+_1$ excitations in the 50$\leq$N$\leq$54 neutron range (open circles) are due
to filling of the $\nu$d$_{5/2}$ shell with valence protons in the $\pi$p$_{1/2}$ shell. High
energy 2$^+_1$ levels at N=50, which follow ``parabolas'' shown in Fig.
\ref{Zr_even_light_2plus} may be solely due to excitations of a neutron pair in the
$\nu$d$_{5/2}$ shell promoted from the $\nu g_{9/2}$ shell. Their energy drops quickly from
Zr to Ge, coinciding with the weakening of the N=50 shell closure with the decreasing proton
number, as proposed in Ref. \cite{Rza07}.

Filling of the $\nu$g$_{7/2}$ shell at higher N produces 2$^+_1$ levels marked by open squares
(note that at N=56 the ($\nu$d$_{5/2}$$\pi$p$_{1/2}$) configuration corresponds to 2$^+_2$
levels (open circles) whereas at N=54 the 2$^+_2$ levels are assigned to the
($\nu$g$_{7/2}$$\pi$p$_{1/2}$) configuration).

Crossing of the two configurations and the increased density of 2$^+$ levels at N=54 and
N=56 were very well reproduced by the Large Scale Shell Model (LSSM) calculations \cite{Urb21}
for the Sr isotopic chain, where also the s.p. structure of leading configurations in these
configurations has been determined (see Table XVI in Ref. \cite{Urb21}).

\begin{figure}
\centering
\scalebox{.50}{\includegraphics{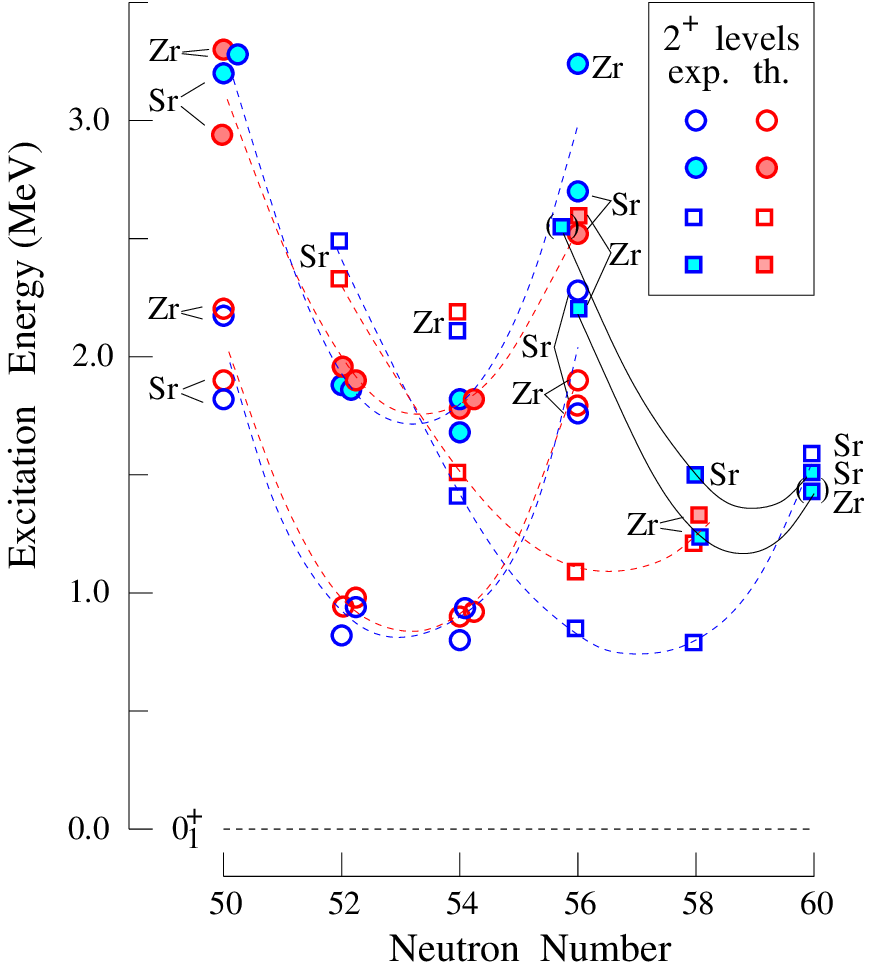}}
\caption{Comparison of excitation energies of 2$^+_1$, 2$^+_2$ and 2$^+_3$ levels in Sr and
Zr isotopes observed experimentally (blue symbols) and calculated with LSSM (red symbols).
The experimental data are taken from the present work and from Refs. \cite{Urb21,ENSDF} and
the LSSM  calculations are taken from  Refs. \cite{Sie09,Urb21,Sie22}. Lines are drawn to
guide the eye.}
\label{Zr_even_light_2plus_th}
\end{figure}

Figure \ref{Zr_even_light_2plus_th} is an extended version of Fig. 14 from Ref. \cite{Urb21},
now containing also data for Zr isotopes. It compares energies of experimental 2$^+$ levels
with energies of 2$^+$ levels obtained from LSSM calculations for Sr \cite{Urb21} and Zr
\cite{Sie09} isotopes. As in case of Sr isotopes \cite{Urb21} the LSSM calculations for Zr
\cite{Sie09} reproduce well shapes and positions of experimental ``parabolas'' as well as the
absence of ($\nu$g$_{7/2}$$\pi$p$_{1/2}$) configurations (open squares) in Zr isotopes at
N=56, 58 whereas such configurations are seen at low energies in $^{94}$Sr and $^{96}$Sr.

The ``parabolas'' linking higher energy points (filled circles and filled squares) show trends
for configurations with protons in the $\pi g_{9/2}$ shells. In the 50$\leq$N$\leq$56 range
they are dominated by the ($\nu d_{5/2},\pi g_{9/2}$) coupling (filled circles in Figs.
\ref{Zr_even_light_2plus} and \ref{Zr_even_light_2plus_th}), whereas in the 56$\leq$N$\leq$60
range they are dominated by the ($\nu g_{7/2}, \pi g_{9/2}$) coupling (filled squares in Figs.
\ref{Zr_even_light_2plus} and \ref{Zr_even_light_2plus_th}). At N=56 both configurations are
present, the latter being lower in energy due to the Fermi level approaching the $\nu g_{7/2}$
shell. The parabolas in 56$\leq$N$\leq$60 range are visibly shifted towards higher N, compared
to analogous parabolas shown by open squares. This may be due to an admixture of neutrons in
the $h_{11/2}$ shell, which starts to be populated around N=60.

Detailed reproduction in Fig. \ref{Zr_even_light_2plus_th} of experimental ``parabolas'' by
the LSSM calculations confirms that this phenomenological classification is a meaningful,
not an ad hoc idea.

One may wonder about the ($\nu d_{5/2},\pi g_{9/2}$) coupling in Ge and Se isotopes being
lower in energy than in Zr and Sr isotopes. The, somewhat surprising, low energy of the
$g_{9/2}$ protons there, confirmed experimentally in Br isotopes \cite{Nya21,Nya21b}, could
be due to the mentioned weakening of the N=50 shell closure \cite{Rza07}, which lowers the
energy of the $g_{9/2}$ proton shell due to the $\nu g_{9/2}-\pi g_{9/2}$ monopole interaction.

Levels shown in Fig. \ref{Zr_even_light_2plus} by blue triangles, blue diamonds and red
diamonds are due to collective 2$^+$ excitations, which are discussed in the following
sections.

\subsection{0$^+$ excitations}

\subsubsection{0$^+_2$ and 0$^+_3$ levels in $^{98}$Zr}

\begin{figure*}
\centering
\scalebox{.33}{\includegraphics{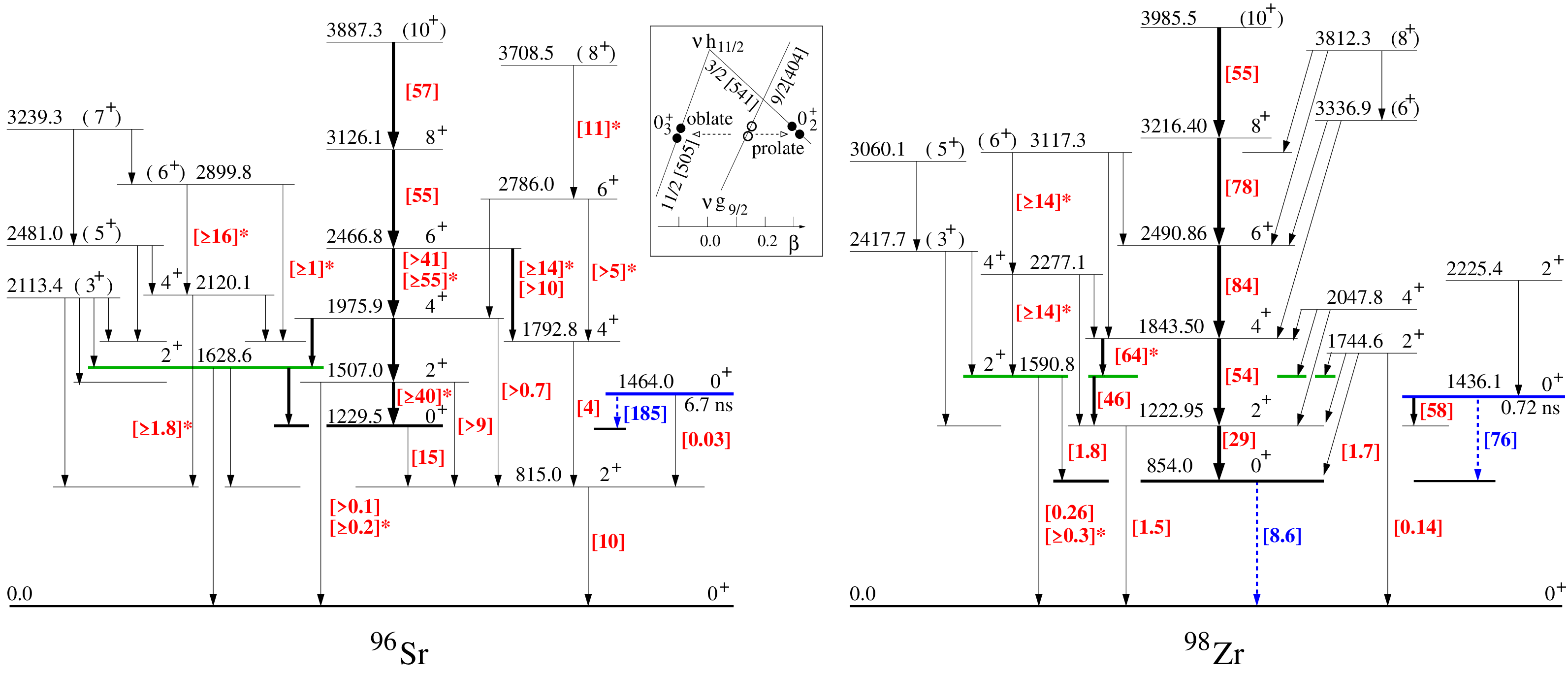}}
\caption{Excitation energies of selected levels in $^{96}$Sr and $^{98}$Zr nuclei shown to
help the discussion. Shown in red are B(E2) rates in W.u., whereas $(\rho)^2 \times 10^3$
rates of E0 transitions are shown in blue. Experimental data are taken from  Refs. \cite{Urb21,NDS98,NDS96,Sin18,Smi12,Reg17,Pas23,Toc25} and those marked by asterisk are
derived in this work, using either T$_{dbe}$ or present branching. See text for further
explanation.}
\label{Zr_even_light_2plus_gamma}
\end{figure*}

One of the most studied effects in the Zr chain is the sudden change of the ground-state
deformation around N=59 found long time ago \cite{Chi70}. The striking abruptness of the
change was first attributed to lowering of an excited 0$^+$, deformed configuration
when the neutron number grows \cite{She72,Mey86} (first-order phase transition
\cite{Cas07,Hey04}).

Later theoretical works \cite{Fed77} proposed adifferent explanation of the change undergoing
in ground states as due to rapid occupation of the $g_{7/2}$ proton shell caused by the
monopole-tensor attraction with protons in the $g_{9/2}$ shell, called Spin-Orbit-Partner
(SOP) mechanism (second order phase transition \cite{Cas07,Hey04}).

The original idea of crossing structures \cite{She72} was supported by the proposed strongly
deformed band on top of the $third$  0$^+$ level at 1436.1 keV in $^{98}$Zr \cite{Kaw82,Lhe94}. However, this band was not confirmed \cite{Ham95} and the 0$^+_3$ level remained unexplained \cite{WH1992}.

Moderately deformed bands in $^{96}$Sr and $^{98}$Zr were identified on top of the $second$
0$^+$ level in a measurement using a large Ge array \cite{Urb01}. It has been proposed that
the increasing deformation of 0$^+_2$ structures with increasing neutron number (due to
the SOP mechanism acting at {\it excited} 0$^+$ levels) induces an increase of the deformation
in ground states \cite{Urb01}, the process later described as intertwined quantum phase
transition \cite{Gav22} (compare Fig. 9 from \cite{Urb01} with Fig. 1 from \cite{Gav22}).

The idea of 0$^+_2$ as moderately deformed configuration was supported in recent works
\cite{Wit18,Sin18,Kar20,Pas23}, though the authors of Ref. \cite{Sin18} suggested that the
4$^+_1$ level at 1843.4 keV in $^{98}$Zr belongs to a strongly deformed band based on the
0$^+_3$, 1436.1-keV level, comprising the 0$^+_3$, 1436.1-, 2$^+_2$, 1590.8-, 4$^+_1$,
1843.4- and 6$^+_1$2490.5-keV levels.

Figure \ref{Zr_even_light_2plus_gamma} shows levels of interest in $^{98}$Zr and, for
comparison, analogous levels in the N=58 isotone, $^{96}$Sr. We will argue that there is
moderately deformed band on top of the 0$^+_2$, 854.0-keV level, as shown in the figure, and
there is no strongly deformed band on top of the 0$^+_3$, 1436.1-keV level in $^{98}$Zr.

The key element of the strongly deformed band on top of the 0$^+_3$ level in $^{98}$Zr is the
$2^+_2 \rightarrow 0^+_3$, 154.5-keV stretched E2 transition between the 1590.8-keV and
1436.1-keV levels. One notes immediate problems here:

- taking B(E2)=0.26 W.u. reported recently for the 2$^+_2\rightarrow 0^+_1$ transition
\cite{Kar20} and relative intensities, $I_{\gamma}$(154.5 keV)=1.9 and $I_{\gamma}$(1590.78
keV)=100(3) from the  compilation \cite{NDS98} one obtains unrealistic B(E2) of 30100 W.u.
for the 154.5-keV, 2$^+_2\rightarrow 0^+_3$ transition

- for a strongly deformed band one expects a nearly linear increase of in-band $\gamma$
energies. The 6$^+\rightarrow 4^+$ in-band transition should have energy of about 350 keV,
much lower than the observed, 647.1-keV

- the proposition of a strongly deformed $0^+_3$ level was based on the Monte Carlo Shell Model
(MCSM) calculation (see Fig. 3 in Ref. \cite{Sin18}). The 5DCH/DS1 model, also used in Ref.
\cite{Sin18}, does not predict any strongly deformed band on top of the $0^+_3$ level in $^{98}$Zr.

- the 2$^+_2\rightarrow 0^+_3$ transition in $^{98}$Z has not been observed neither in our
detailed $\beta^-$ decay study \cite{Urb17} nor in the present work. The upper limit for its
$\gamma$ intensity obtained in the two studies is 4$\times$10$^{-4}$ fraction of $\gamma$
intensity of the 1590.80-keV transition. This puts the upper limit of 12 W.u. for the
154.5-keV transition, a B(E2) rate too low for a strongly-deformed band member.

The lifetimes in $^{98}$Zr reported in Ref. \cite{Sin18} are, on average, over two times shorter
than analogous values reported in Ref. \cite{Kar20}. The most recent measurement \cite{Pas23}
reports values which are between the two. In Fig. \ref{Zr_even_light_2plus_gamma} B(E2) rates
(shown in red) are taken from Table 3 of Ref. \cite{Pas23} for 2$^+_1\rightarrow 0^+_1$,
4$^+_1\rightarrow 2^+_1$ and 6$^+_1\rightarrow 4^+_1$ transitions, from Table III of Ref.
\cite{Kar20} for 2$^+_2\rightarrow 0^+_1$, 2$^+_2\rightarrow 0^+_2$ and 2$^+_2\rightarrow 2^+_1$
transitions (for the 2$^+_2\rightarrow 2^+_1$ transition pure E2 multipolarity was recently
confirmed \cite{Mas25}) and from Ref. \cite{NDS98} for 0$^+_3\rightarrow 2^+_1$,
2$^+_4\rightarrow 0^+_1$ and 2$^+_4\rightarrow 2^+_1$  transitions. For 2$^+_1\rightarrow 0^+_2$
and 4$^+_1\rightarrow 2^+_2$ transitions we estimated B(E2) rates taking relevant lifetimes
from Ref. \cite{Pas23} and branchings obtained in the present work.

The moderately deformed band in $^{98}$Zr shown in Fig. \ref{Zr_even_light_2plus_gamma} on top
of the 0$^+_2$, 854.0-keV level differs from that shown in Fig. 4 of Ref. \cite{Urb01} in that
it comprises now the 1222.95-keV, 2$^+_1$ level instead of the 1590.8-keV,
2$^+_2$ level, as proposed in Ref. \cite{Mey86}. The B(E2) of 46 W.u. of the 367.8-keV,
2$^+_2\rightarrow 2^+_1$ transition indicates strong mixing between the 2$^+_2$ and $2^+_1$
levels, which repeal each other. With the energy of the $2^+_1$ level increased by about 70 keV
the 0$^+_2$ band becomes a regular, rotational cascade.

The 0$^+_2$ band in $^{98}$Zr is similar to the analogous band in $^{96}$Sr based on the
0$^+_2$, 1229.5-keV level, which shows similar collectivity to that in $^{98}$Zr. The strong
6$^+_1\rightarrow 4^+_1$ transition in $^{96}$Sr indicates mixing with the ground-state band
causing the $4^+_1$ and $4^+_2$ levels to repeal each other. With the energy of the $4^+_2$
level decreased by about 60 keV the 0$^+_2$ band becomes a regular, rotational cascade.

As in $^{98}$Zr there in no deformed band on top of the 0$^+_3$ isomer in $^{96}$Sr. Both
0$^+_3$ isomers have enhanced E0 decay rates to 0$^+_2$ levels (shown in blue in Fig. \ref{Zr_even_light_2plus_gamma}). This was the reason to propose large deformation for the
0$^+_3$, 1436.1-keV isomer in $^{98}$Zr in Ref. \cite{Kaw82}. In their model the large E0 rate
means a large difference in deformation of the two states linked by an E0 transition. In Ref.
\cite{Kaw82} the authors came to the conclusion that the ground state and the 0$^+_2$ state
in $^{98}$Zr are both spherical, because of a slow E0 link between them and, therefore the
1436.1-keV isomer in $^{98}$Zr must be strongly deformed. However, there is no
0$^+_3\rightarrow 0^+_1$, E0 decay in either $^{98}$Zr or $^{96}$Sr.

It has long been debated whether large E0 rates require large difference in deformation or
a large mixing of the two levels linked by the E0 transition \cite{Hey88,Mac90a,Hey90,Mac90b}.
In Ref. \cite{Woo99} Wood and coworkers proposed that both conditions have to be met. Thus,
in $^{98}$Zr and $^{96}$Sr close lying 0$^+_3$ and $0^+_2$ states mix whereas distant 0$^+_3$
and $0^+_1$ states do not and there is no 0$^+_3\rightarrow 0^+_1$, E0 decay.

The argument of Wood has been reiterated in the review \cite{Leo24}. However, as seen in Fig.
31 and Table VI of Ref. \cite{HW2011}, E0 decays of 0$^+_2$ levels to 0$^+_1$ ground states
in $^{98}$Sr and $^{100}$Zr are strong whereas mixing between 0$^+_2$ and 0$^+_1$ states in
the two nuclei is weak \cite{Urb19,Toc25}. Furthermore, in Ref. \cite{Sin18} the strong
$0^+_3\rightarrow 0^+_2$, E0 transition in $^{98}$Zr was invoked as a sign of strong mixing
of the two structures. However, as the authors of Ref. \cite{Sin18} stress, their MCSM
calculation does not predict any strong mixing between $0^+_3$ and $0^+_2$ levels. Thus,
the understanding of E0 transitions in the region may need further attention.

Looking for an interpretation of the discussed 0$^+_3$ isomers we note that in $^{98}$Zr the
2$^+$ excitation on top of the 0$^+_3$, 1436.1-keV band head has an energy of 789.3 keV,
suggesting spherical deformation for the 0$^+_3$ level. This is similar to cascades on top of
spherical 0$^+_2$ isomers in $^{98}$Sr and $^{100}$Zr \cite{Urb19}. Therefore, for the 0$^+_3$,
1436.1-keV isomer in $^{98}$Zr and the 0$^+_3$, 1464.0-keV isomer in $^{96}$Sr we propose the
same $\nu (11/2^-[505])^2$, dominating configuration, as proposed for 0$^+_2$ isomers in
$^{98}$Sr and $^{100}$Zr \cite{Urb19}. The inset in Fig. \ref{Zr_even_light_2plus_gamma},
analogous to Fig. 7 (b) of Ref. \cite{Urb19}, illustrates schematically the formation of
0$^+_2$ and 0$^+_3$ states in $^{96}$Sr and $^{98}$Zr via ``pair hopping'' \cite{Bra90,Bro94}
from the $\nu 9/2^+$[404] extruder orbital to either the $\nu 3/2^-$[550] orbital or to the
high-$\Omega$, $\nu 11/2^-$[505] orbital, respectively. The recent measurement of E0 strength
\cite{Toc25} confirms moderate deformation of the 0$^+_2$ level in $^{98}$Zr and spherical
shape of the 0$^+_2$ level in $^{100}$Zr.

The unexpectedly fast 0$^+_3\rightarrow 2^+_1$, transition in $^{98}$Zr with B(E2)=58 W.u. is
similar to the 0$^+_2\rightarrow 2^+_1$ transitions in $^{98}$Sr and $^{100}$Zr showing B(E2)
of 71 W.u. and 67 W.u., respectively \cite{ENSDF}. This observation as well as other properties
of unusual 0$^+_2$ levels in $^{98}$Sr and $^{100}$Zr present a challenging
chance for nuclear structure models.

It is of interest to measure the half-life of the $4^+_2$ level at 1975.9-keV in $^{96}$Sr,
to verify the proposed collectivity in band on top of the $0^+_2$ level there. We note the
significantly higher B(E2) limit for the 6$^+_1\rightarrow 4^+_1$, 674.00-keV transition in
this band obtained in the present work, compared to that reported in Ref. \cite{Reg17}.

\subsubsection{Evolution of 0$^+$ levels and deformation in Zr isotopes}

\begin{figure}
\centering
\scalebox{.34}{\includegraphics{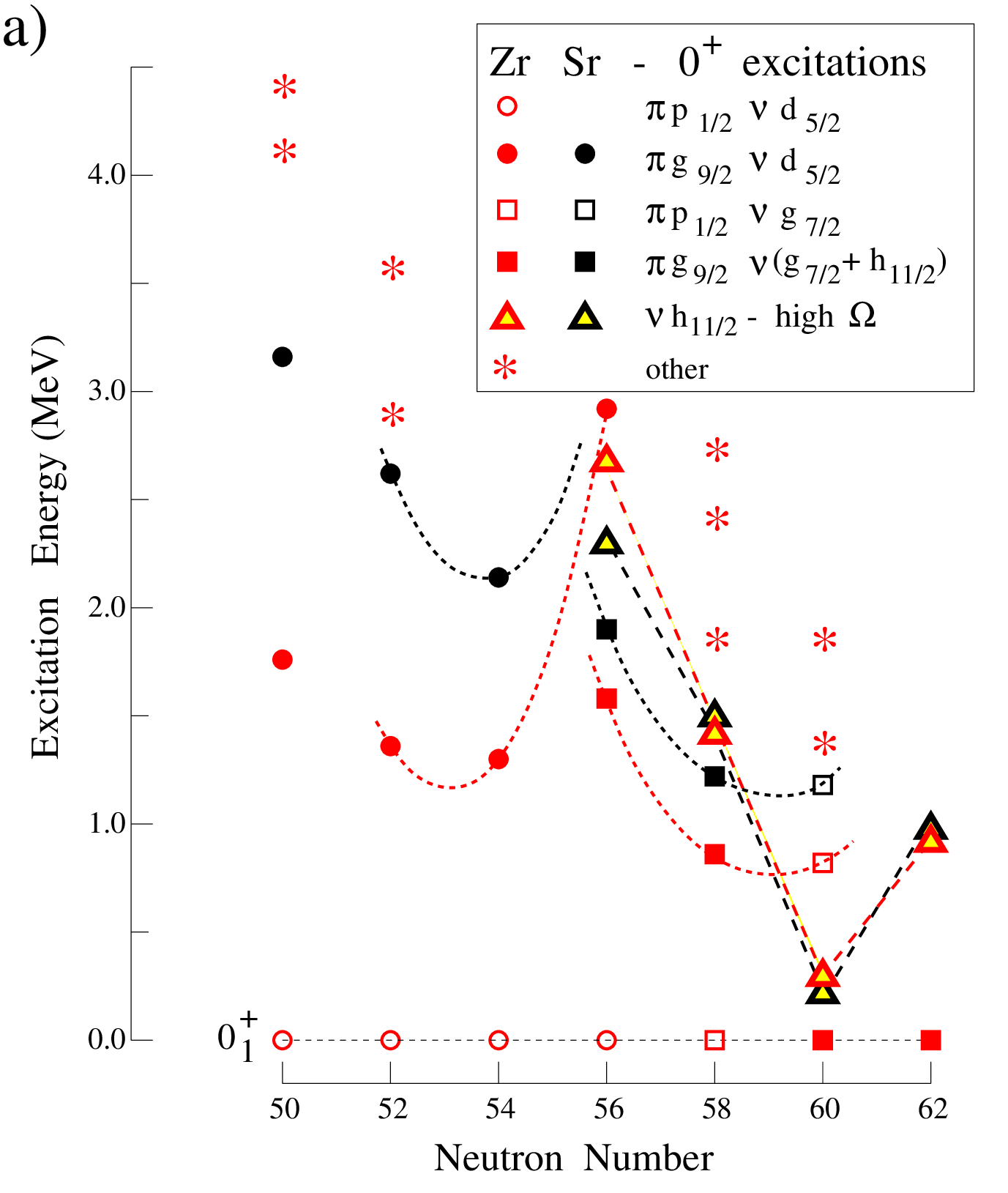}}
\scalebox{.34}{\includegraphics{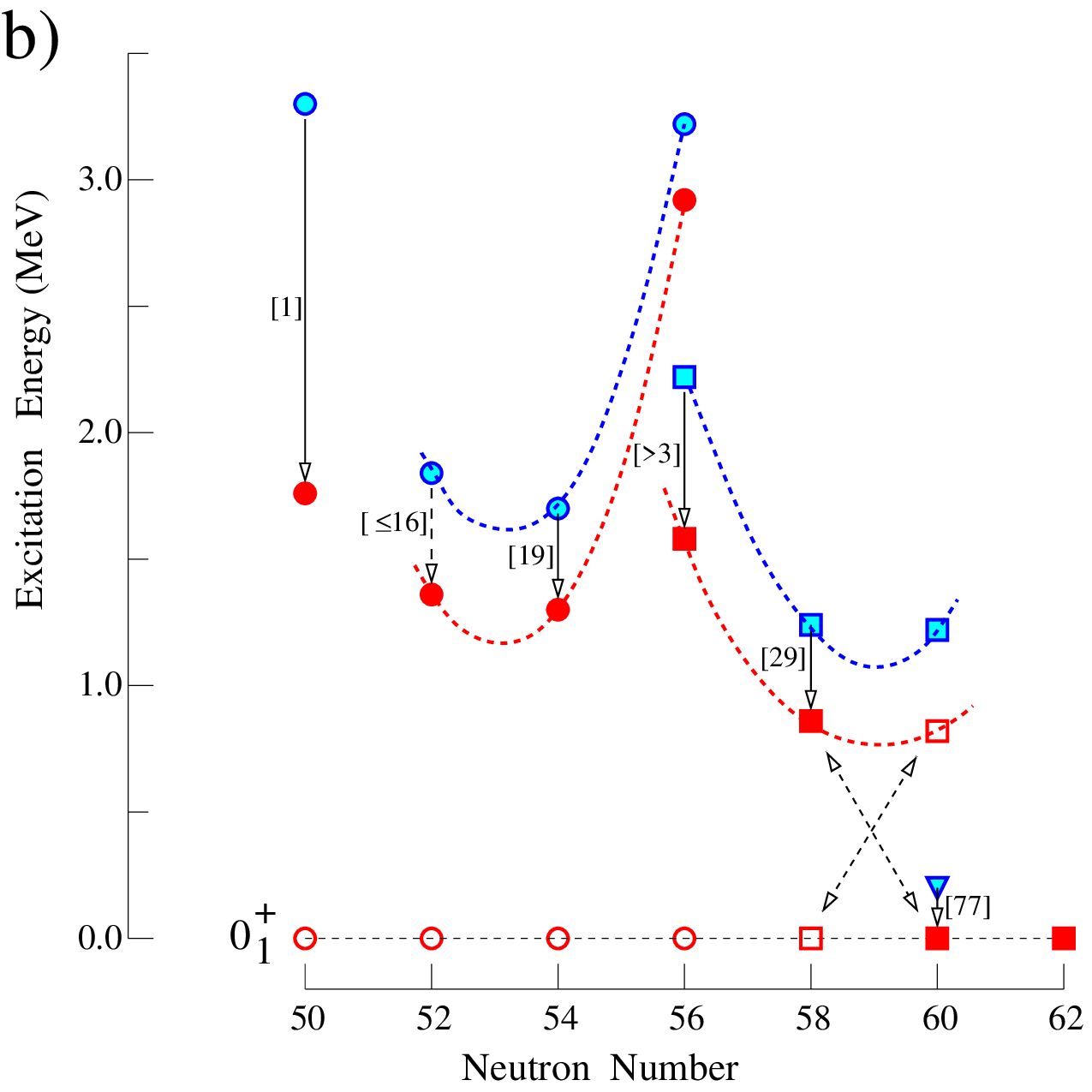}}
\caption{a) Excitation energies of known $0^+$ levels in Zr isotopes (red symbols). The
$0^+_2$ and some $0^+_3$ levels in Sr isotopes (black symbols) are shown for comparison.
b) Excitation energies of $0^+_2$, $0^+_3$ (red symbols) and 2$^+_2$, 2$^+_3$ (blue symbols)
levels in Zr isotopes. Numbers is square brackets next to arrows are B(E2) rates in W.u.
The experimental data are taken from \cite{Cru20,Cha13,NDS_Zr92,NDS96,NDS98,Wu024,Mas25,ENSDF}
and the present work. Further explanations are in the text.}
\label{Zr_even_light_0plus}
\end{figure}

Figure \ref{Zr_even_light_0plus}(a) shows known $0^+$ levels in even $^{90-102}$Zr nuclei
(red symbols). For comparison shown are $0^+_2$ and some $0^+_3$ levels in even $^{88-100}$Sr
nuclei (black symbols). Energies of $0^+$ excited levels generally decrease with the
increasing neutron number. The trend is not as smooth as for Mo and Ru isotopes in Fig.
\ref{Zr_even_light_fig2} but several $0^+$ levels can be arranged along two ``parabolas'',
one linking $0^+_2$ levels in $^{92,94}$Zr and the $0^+_4$ level in $^{96}$Zr and the other
linking the $0^+_2$ level in $^{96}$Zr, the $0^+_2$ level in $^{98}$Zr and the $0^+_3$ level
in $^{100}$Zr. Analogous ``parabolas'' can be drawn for Sr isotopes.

The ``parabolas'' comprising $0^+$ levels are analogous to those linking $2^+$ levels. Figure
\ref{Zr_even_light_0plus}(b) highlights their similarity. The increased collectivity of
excited $0^+$ configurations in Sr and Zr isotopes stems from the proton-neutron interaction
\cite{Mey86,Wer94}, which shows ``parabolic'' variations with increasing population within
neutron shells.

In $0^+_1$ ground states up to N=58, protons occupy the $\pi p_{1/2}$ shell. Excited $0^+$
levels are created by elevating a pair of protons to the $\pi g_{9/2}$ shell:

- excited $0^+$ levels on the first ``parabola'' are due to the $\pi g_{9/2} -\nu d_{5/2}$
interaction, as the $2^+_2$ levels shown by filled circles in Fig. \ref{Zr_even_light_2plus}
and in Fig. \ref{Zr_even_light_0plus}(b) as quadrupole excitations on top of $0^+$ levels,
with B(E2) of 19 W.u. in $^{94}$Zr. The $2^+_2$, 1847.27 keV level in $^{92}$Zr, was seen
as a mixed-symmetry excitation \cite{Wer02,Yat05} though this was questioned in \cite{Sie09}.
It is of high interest to verify the tentative 464.4-keV, $2^+_2 \rightarrow 0^+_2$
transition in $^{92}$Zr with its B(E2)$\le$16 W.u. limit obtained in the present work

- excited $0^+$ levels on the second ``parabola'' are due to the $\pi g_{9/2} -\nu g_{7/2}$
interaction, as the 2$^+$ levels shown by filled squares in Fig. \ref{Zr_even_light_2plus}
and in Fig. \ref{Zr_even_light_0plus}(b) as quadrupole excitations on top of $0^+$ levels,
with B(E2) of 29 W.u. in $^{98}$Zr. One should verify the B(E2) limit in $^{96}$Zr and find
the B(E2) rate for the $2^+_3 \rightarrow 0^+_3$ transition in $^{100}$Zr.

Around N=60 the $0^+_2$, ($\pi g_{9/2},\nu g_{7/2}$) configuration undergoes an avoided
crossing with the  $0^+_1$, ($\pi p_{1/2},\nu g_{7/2}$) configuration, as marked by dashed
arrows in Fig. \ref{Zr_even_light_0plus}(b). Furthermore, the $\nu h_{11/2}$ shell starts
to be populated and to interacts with the $\pi g_{9/2}$ shell already there, increasing
collectivity in $^{100}$Zr higher than expected from the avoided crossing alone. As
proposed in Ref. \cite{Urb19}, at N=60 the $\nu 9/2^+[404]$ extruder orbital delivers an
extra pair of neutrons, which is passed to the down sloping $\nu 3/2^-[541]$ orbital of
the $\nu h_{11/2}$ shell (see inset in Fig. \ref{Zr_even_light_2plus_gamma}).

The inset in Fig. \ref{Zr_even_light_2plus_gamma} illustrates also the creation of an
oblate 0$^+$ excitation when a pair of neutrons is passed from the $\nu 9/2^+[404]$ extruder
to the $\nu 11/2^-$[505] orbital. This orbital differs from other orbitals present at the
Fermi surface and does not interact with them. Therefore, the excitation energy of the
resulting $0^+$ level varies nearly linearly with the neutron number. This is the case for
$0^+_3$ levels in $^{94}$Sr and $^{96}$Zr, $0^+_3$ levels in $^{96}$Sr and $^{98}$Zr and
$0^+_2$ levels in $^{98}$Sr, shown in Fig. \ref{Zr_even_light_0plus}(a) by triangles.

If the linear trend holds, then at N=62 an analogous oblate configuration is expected abut 0.9
MeV below the Fermi surface and it would take this much to excite it. This could explain the
mysterious $0^+_2$ levels in $^{100}$Sr and $^{102}$Zr around 0.9 MeV, not explained to date.

The $0^+_3$ level in $^{94}$Sr, called ``strange'' in Ref. \cite{Wit18}, is different from
the $0^+_2$ level there. However, the two levels produced by the extruder action share
common $\nu (h_{11/2})^2\otimes \nu (g_{9/2}[404]^{-2}$ structure, which could explain
their apparent mixing \cite{Cru20}. Analogous mixing between $0^+_2$ and $0^+_3$ levels in
$^{96}$Sr and $^{98}$Zr is reported in Ref. \cite{Cru18}. Interestingly, mixing between
$0^+_1$ and 0$^+_2$ levels in both, $^{98}$Sr and $^{100}$Zr, is weak, though bands on top
of these two levels are strongly linked \cite{Urb19}. The weak mixing could be due to a large
difference in deformation whereas electromagnetic decays between $0^+_1$ and 0$^+_2$ levels
could be due to the same $\nu (h_{11/2})^2\otimes \nu (g_{9/2}[404])^{-2}$ components in
their wave functions.

\begin{figure*}
\centering
\scalebox{.33}{\includegraphics{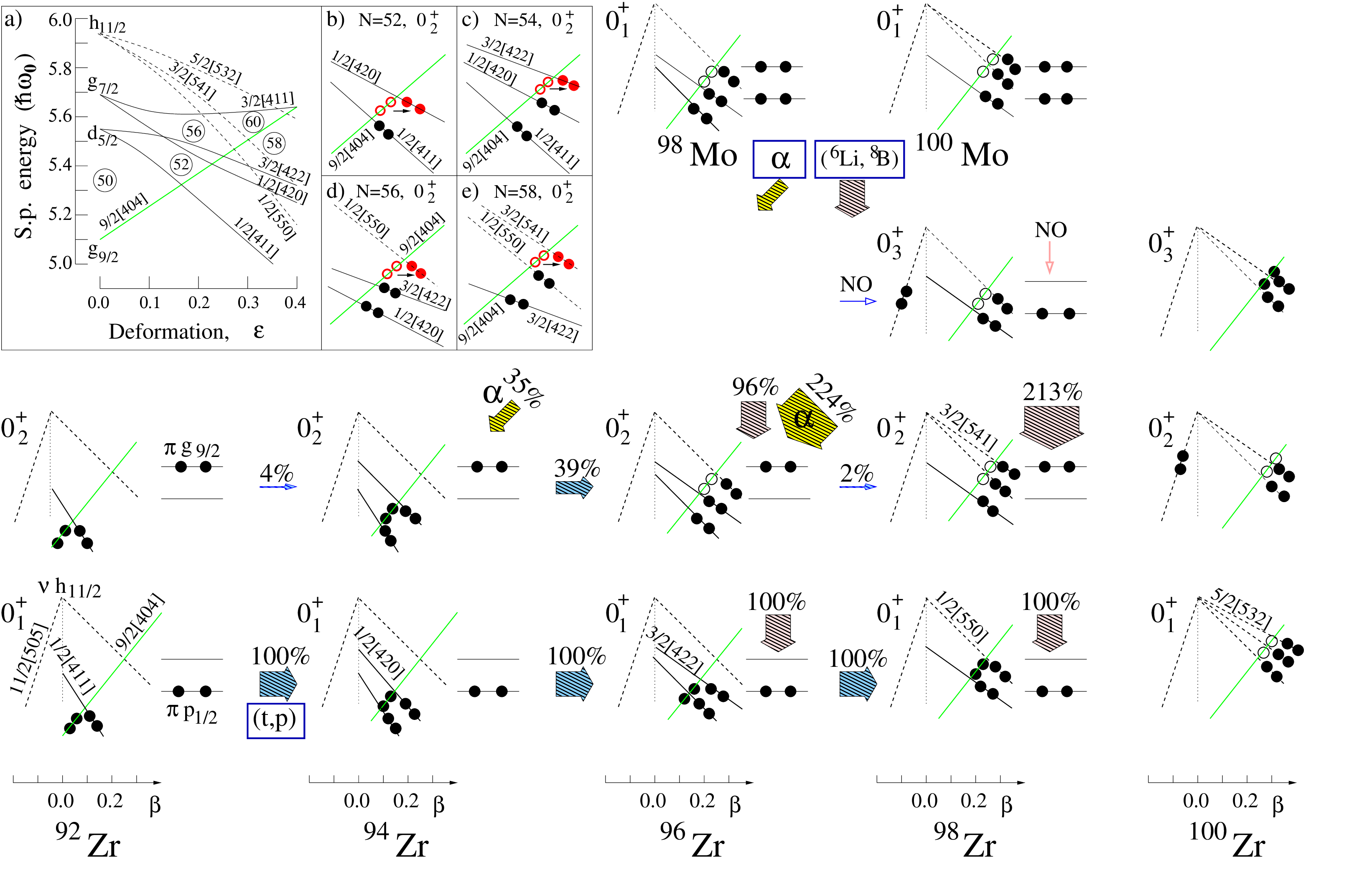}}
\caption{Schematic presentation of dominating neutron and proton configurations in
transitional Zr isotopes and their relation to cross section of transfer reactions in the
region. Experimental data are from Refs. \cite{ENSDF,HW2011}. Further explanation in the
text.}
\label{A100_Zr_transfer}
\end{figure*}

The picture presented above supports the avoided-crossing scenario proposed in Fig.
\ref{Zr_even_light_0plus}(b) in contrast to the configuration evolution shown in Fig. 18 of
Ref. \cite{Kib22} linking the 0$^+_2$ level in $^{98}$Zr with the 0$^+_2$ level in $^{100}$Zr.

Out of 63 0$^+_2$ and 0$^+_3$ levels observed at N$>$90 in the A$\approx$100 region, which
are shown in Fig. 6 of Ref. \cite{Urb19}, only eight deviate from regular trends. Five of
them belong to Zr isotopes, highlighting the special nature of Zr nuclei. The new
classification proposed in Fig. \ref{Zr_even_light_0plus}(a) offers an explanation for the
deviating 0$^+$ levels in even Zr nuclei.

Finally, let us comment on the 0$^+$ levels at N=50. The 0$^+_1$ and 0$^+_2$ levels in
$^{90}$Zr have been interpreted in Ref. \cite{HW2011} as due to a particular mix of protons
in the $\pi p_{1/2}$ and $\pi g_{9/2}$ shells, both active at the Fermi surface. The 0$^+_2$
level in $^{90}$Zr has rather low excitation energy, comparable to that of the 0$^+$ level in
$^{92}$Zr. The same is seen in $^{88}$Sr. It is of interest to check whether the
$\nu d_{5/2}$ shell gets some population at N=50 by exciting neutrons from the $\nu g_{9/2}$
shell, elevated by the $\nu g_{9/2}-\pi g_{9/2}$ monopole attraction.

\subsubsection{0$^+$ levels and transfer reactions in A$\approx$100 region}

In the study of the phase transition in A$\approx$150 region \cite{Urb26} we have proposed
explanation of various structure effects associated with this transition employing the
11/2$^-$[505] neutron extruder. In particular we could explain cross sections for (t,p) and
(p,t) transfer reactions in the region, rapidly changing around neutron number N=89, where
the shape transition takes place (see Figs. 16(b) and 17 in Ref. \cite{Urb26}).

The discussion of transfer cross sections for $^{98}$Zr reported in Ref. \cite{Kaw82} can
now be updated and extended to more nuclei in the A$\approx$100 region, using explanations
analogous to those in the A$\approx$150 region.

Figure \ref{A100_Zr_transfer}, similar to Fig. 17 of Ref. \cite{Urb26}, shows the
proposed dominating neutron and proton configurations in 0$^+$ levels of $^{92}$Zr,
$^{94}$Zr, $^{96}$Zr, $^{98}$Zr, $^{100}$Zr, $^{98}$Mo, and $^{100}$Mo nuclei. The inset (a)
shows part of the  Nilsson scheme for neutrons in the A$\approx$100 region \cite{Mey85} and
in insets (b)-(e) the excitation of 0$^+_2$ states in $^{92-98}$Zr isotopes is schematically
illustrated  (we note that in Fig. 10 in Ref. \cite{Urb21}, which shows analogous
information, labels N=58 and N=60 in parts (d) and (e) should be corrected to N=56 and N=58,
respectively). Blue arrows represent cross sections for the (t,p) two-neutron transfer
between Zr isotopes. Pink arrows represent cross sections for ($^6$Li,$^8$B) two-proton
transfer between Mo and Zr isotopes and yellow arrows represent cross sections for $\alpha$
transfer between Mo and Zr isotopes. Cross sections are normalized to 100$\%$ for transfers
between ground states.

As discussed in Ref. \cite{Wis23} ground states in Mo isotopes are more deformed than in
their Zr isotones due to the increased population of the $\pi g_{9/2}$ shell helped by the
``extruder-like'' action of 3/2$^+$[301] and 5/2$^+$[303] up sloping proton orbitals (see
Fig. 8 in Ref. \cite{Wis23}). We propose that the 0$^+_1$ ground state in $^{100}$Mo has
the same dominating neutron configuration as the 0$^+_2$ level in $^{98}$Zr.

The configurations proposed in Fig. \ref{A100_Zr_transfer}, which are consistent with those
in Fig. \ref{Zr_even_light_0plus}(a) explain the particularly high ($^6$Li,$^8$B) transfer
cross section to the 0$^+_2$ level in $^{98}$Zr since the 0$^+_1$ ground state in $^{100}$Mo
has the same dominating neutron configuration as the 0$^+_2$ level in $^{98}$Zr. On the other
hand the (t,p) transfer to 0$^+_2$ and 0$^+_3$ in $^{98}$Zr is negligible because of rather
different neutron compositions of these states compared to the 0$^+_1$ level in $^{96}$Zr.

The observed transfers to 0$^+_2$ level in $^{96}$Zr may need more attention. Very strong
$\alpha$ transfer to this state indicates similar dominating neutron configuration as in
the ground state of $^{100}$Mo. This is supported by the strong ($^6$Li,$^8$B) transfer
indicating the same neutron configuration as in the ground state of $^{98}$Mo. The moderate
(t,p) cross section to this 0$^+_2$ level is somewhat high, considering the difference in
the proposed leading configurations of the ground state of $^{94}$Zr and of the 0$^+_2$
level in $^{96}$Zr.

We propose that, in accord with the crossing shown in Fig. \ref{Zr_even_light_0plus}(b)
and the decrease of the oblate configurations shown in Fig. \ref{Zr_even_light_0plus}(a),
the 0$^+_2$ and 0$^+_3$ levels in $^{98}$Zr have the same configurations as the 0$^+_1$
and 0$^+_2$ levels in $^{100}$Zr. However, whereas there are plausible arguments for the
proton structures shown in Fig. \ref{A100_Zr_transfer} in $^{92,94,96,98}$Zr isotopes, in
$^{100}$Zr the proton structure of 0$^+$ levels is not determined. It is of high interest
to measure two proton transfer to 0$^+_1$ and 0$^+_2$ levels in $^{100}$Zr using
radioactive beams. The two proton transfer to the 0$^+_2$ level in $^{100}$Zr on the
$^{102}$Mo target is expected to be negligible, in sharp contrast to that in $^{98}$Zr.

\subsubsection{0$^+$ levels and S$_{2n}$ separation energies in A$\approx$100 region}

Variations of the two neutron-separation energy, S$_{2n}$, have been analysed vs. various
structural changes in nuclei for a long time (see e.g. Refs. \cite{Ang09,Spa25}). In a
recent study of transitional Nd isotopes \cite{Urb26} we have proposed an approach to this
subject allowing the observation of specific, $\Delta$S$_{2n}$ changes of S$_{2n}$ over a
wide neutron range such as a span of the $\nu i_{13/2}$ intruder shell. A similar analysis
can be employed to trace $\Delta$S$_{2n}$ changes in transitional nuclei of the
A$\sim$100 region.

\begin{figure}
\centering
\scalebox{.35}{\includegraphics{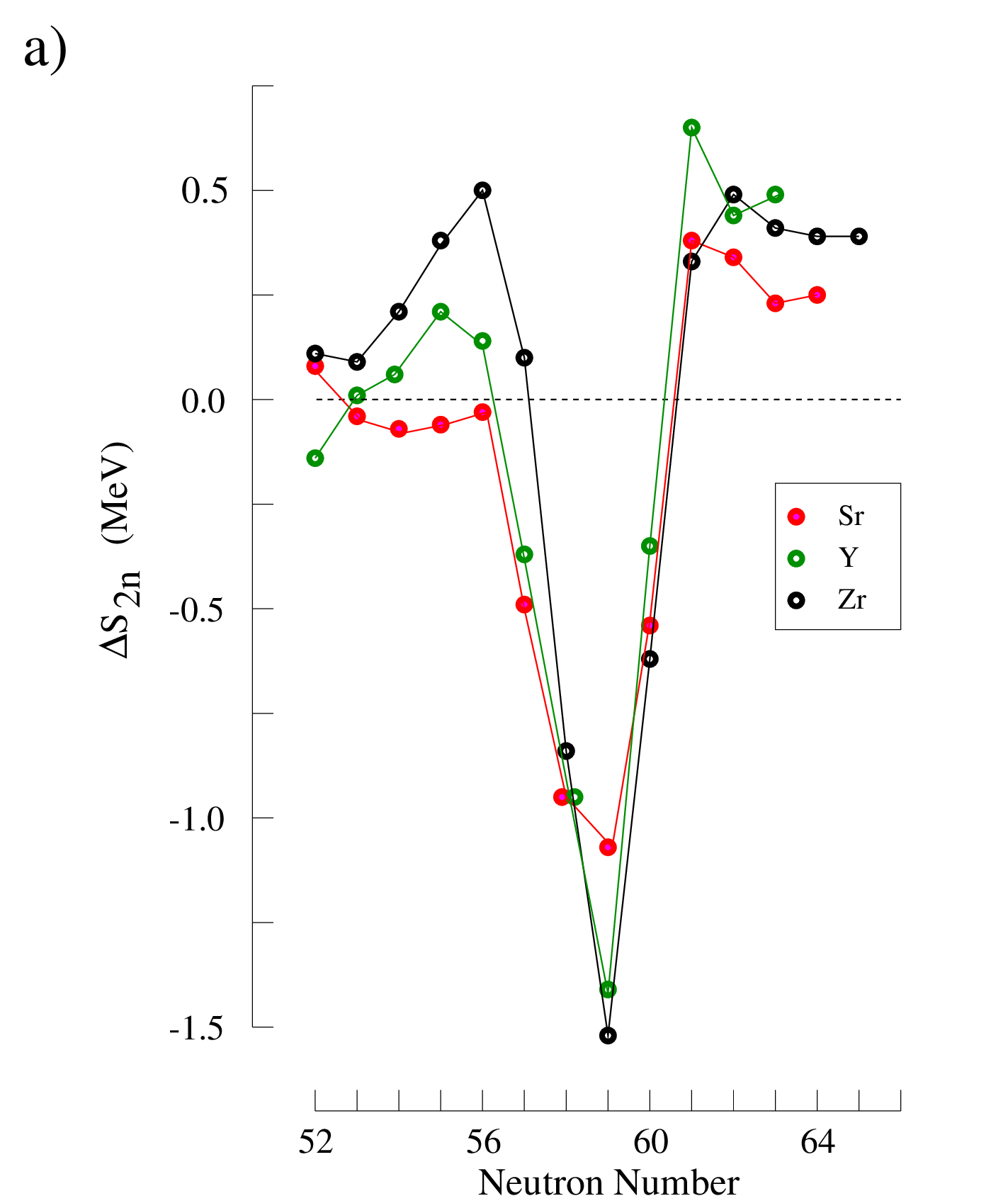}}
\scalebox{.35}{\includegraphics{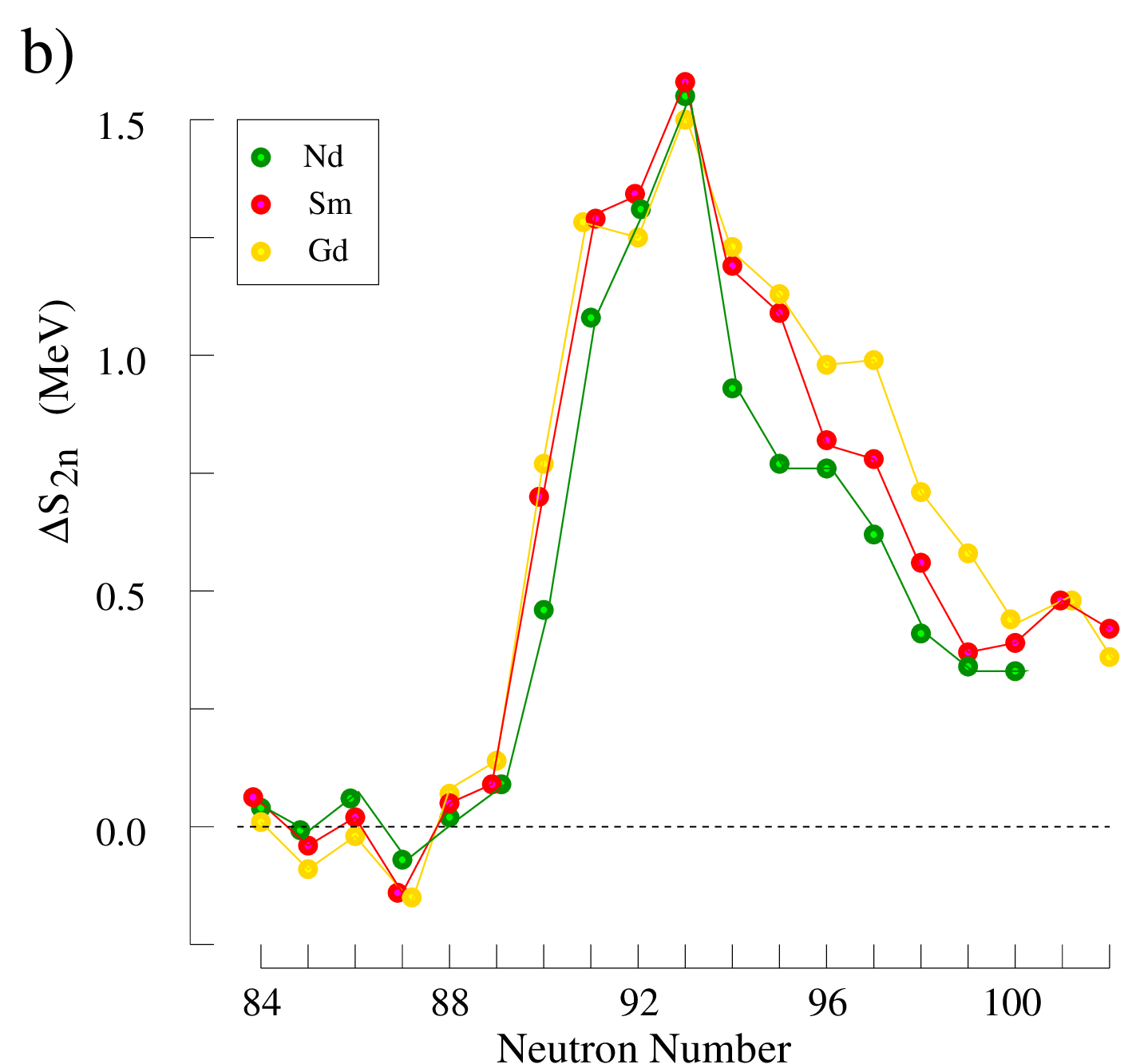}}
\caption{$\Delta$S$_{2n}$ excess in the A$\sim$100 region. Data points are obtained using
the S$_{2n}$ values from the NuDat base \cite{NUDAT}. See text for further explanation.}
\label{S2n_excess_A100}
\end{figure}

As seen in the NuDat systematics \cite{NUDAT}, except for local deviations, the S$_{2n}$
energy changes almost linearly in many regions of the nuclear chart (see also formula (6) in
Ref. \cite{Gar05}). This allows to define {\it local} reference plane, S$_{2n}^{ref}$(N,Z)
= Z$\times$A$_n$ + N$\times$B$_n$ + C$_n$, for a studied region, and use it to show
$\Delta$S$_{2n}$(N,Z) = S$_{2n}$(Z,N) - S$_{2n}^{ref}$(N,Z) deviations from this reference
in a range of interest.

Figure \ref{S2n_excess_A100}(a) displays $\Delta$S$_{2n}$(N,Z) deviations in the
A$\approx$100 region. The coefficient A$_n$=0.830 MeV/Z has been obtained from fitting the
increase rate of the S$_{2n}$ separation energy along the N=50, 52, 54 and 56 isotonic
lines in the 38$\leq$Z$\leq$48 proton range. Nuclei in these proton and neutron ranges are
expected to be spherical. The coefficient B$_n$= -0.490 MeV/N has been obtained from fitting
the decrease rate of the S$_{2n}$ separation energy along the line of spherical, semi
magic Sn isotopes in the neutron range 52$\leq$N$\leq$64 and along lines of Sr isotopes in
the 52$\leq$N$\leq$56 neutron range where nuclei are expected to have spherical shapes.
The $local$ normalization coefficient C$_n$=8.05 MeV has been adjusted to reproduce the
average S$_{2n}$ value in $^{90,92,94}$Sr nuclei.

Figure \ref{S2n_excess_A100}(a) shows a rapid decrease of the $\Delta$S$_{2n}$(N,Z) above
N=56 with the maximum drop of 1.5 MeV at N=59, exactly where the sudden change of nuclear
deformation occurs in the region. Later the $\Delta$S$_{2n}$(N,Z) grows and stabilizes at
about 0.3 MeV. One notes high similarity of the trends in the Sr, Y and Zr isotopic chains.

The behaviour of $\Delta$S$_{2n}$(N,Z) in the  A$\approx$100 region sharply contrasts
with that of the A$\approx$150 region shown in Fig. \ref{S2n_excess_A100}(b), which is an
extended version of Fig. 16(a) from Ref. \cite{Urb26}. The sudden increase of the
$\Delta$S$_{2n}$(N,Z) value above N=88 corresponds to filling of low-$\Omega$, deformation
driving orbitals of the $\nu i_{13/2}$ shell in the prolate-deformed potential. After
these orbitals are filled the $\Delta$S$_{2n}$(N,Z) drops to about 0.4 MeV.

In Sr, Y and Zr isotopes with 56$<$N$<$62 neutrons the 9/2$^+$[404], high-$\Omega$,
oblate-shaped orbital of the $\nu g_{9/2}$ shell, appears at the Fermi surface
\cite{Urb03,Urb04}. The pair of neutrons on this extruder is weakly bound in the
prolate nuclear potential of these nuclei which may explain lowering of neutron binding
energies in Fig. \ref{S2n_excess_A100}(a).

One expects that a pair of neutrons on the high-$\Omega$, 11/2$^-[505]$ oblate orbital
will be strongly bound in an oblate nuclear potential. The 0$^+$, oblate configurations
have been predicted in Zr nuclei in works of Petrovici \cite{Pet12,Pet20,Pet25} and
others \cite{Ska97,Gor07,Sie09}. Figure 1 in the most recent of such calculations
\cite{Kos26} predicts close lying 0$^+_1$ and  0$^+_2$ levels in $^{94-106}$ even-even Zr
nuclei. As discussed in Ref. \cite{Urb19} the 0$^+_2$ states in $^{98}$Sr and $^{100}$Zr
nuclei have two neutrons in the high-$\Omega$, oblate 11/2$^-$[505] orbital. Such a
configuration is lowered in energy relative to the ground state. This may explain the
exceptionally low values of 0$^+_2$ states $^{98}$Sr and $^{100}$Zr.

\subsection{Emerging $\gamma$ collectivity in $^{92-98}$Zr isotopes}

An intriguing observation in $^{96}$Sr and $^{98}$Zr is the presence of $two$ $2^+$
levels closely associated with the 0$^+_2$ band, as seen in Fig.
\ref{Zr_even_light_2plus_gamma}.

In Ref. \cite{Urb21} the 2120.1-, 2481.0-, 2899.8- and 3239.3-keV levels in $^{96}$Sr,
were assigned spins as shown in Fig. \ref{Zr_even_light_2plus_gamma}. To this set we
add the 2113.4-keV level \cite{NDS_Sr96} with spin-parity (3$^+$) suggested by its
decay branchings and the fact that it is populated in fission, which excludes spins
lower than 3. An analogous set comprising 2277.1-, 2417.7-, 3060.1- and 3117.3-keV
levels is observed in $^{98}$Zr, as shown in Fig. \ref{Zr_even_light_2plus_gamma}.

We propose that in addition to weakly deformed bands based on the 0$^+_2$ levels in
$^{96}$Sr and $^{98}$Zr there are ``proto-$\gamma$'' bands in both nuclei. The 2$^+_3$
level in $^{96}$Sr and 2$^+_2$ level in $^{98}$Zr may be due to ``$\gamma$ vibrations''
on top of ground states and bands above
them are built around the $\pi (g_{9/2}^2)_{0,2,4,6,8}$ multiplet ``dressed'' with this
``$\gamma$ collectivity'', as suggested by the B(E2) lower limits of several W.u. for
the 4$^+_3\rightarrow 2^+_2$ transition and 6$^+_3\rightarrow 4^+_3$ transitions in
both nuclei, estimated from T$_{dbe}$ values. The 5DCH/DS1 model calculations reported
in Ref. \cite{Sin18}, support triaxiality in $^{98}$Zr.

Analogous picture has been proposed at N=56 in $^{96}$Zr, with 0$^+_2$ band comprising
1581-, 2225-, 2857- and 3761-keV levels (see Fig. 3 (a) in Ref. \cite{Mey86}) . In our
study of $^{96}$Zr \cite{Wis23} the 2857.40- and 3772.20-keV levels were assigned to
the $(g_{9/2}^2)_{0,2,4,6,8}$, ground state cascade, whereas levels at 3082.50- and
3483.55-keV have been proposed as 4$^+$ and 6$^+$ members of the ``proto-$\gamma$''
band. This band, comprising also the 2$^+_3$ level at 2668.8 keV, 3$^+_1$ level at
2438.8 keV, 4$^+_3$ level at 3176.5 keV and (5$^+_1$) level at 3309.3 keV, is linked
to the 0$^+_2$ band and to the $(g_{9/2}^2)_j$ cascade.

A ``proto-$\gamma$'' band has also been proposed in $^{94}$Sr \cite{Urb21}. This early
collectivity is supported by B(E2) estimates in $^{94}$Sr and $^{96}$Zr shown in Table
\ref{table_halflives}. Thus, there are at N=56 structures similar to those at N=58 with
``proto-$\gamma$'' bands mixed with 0$^+_2$ bands and $(g_{9/2}^2)_j$ multiplets.
Admixture of $\gamma$ collectivity to the 0$^+_2$ band has been predicted by
calculations in Ref. \cite{Mar20}.

At N=54 the ground-state $\pi (g_{9/2}^2)_j$ multiplet in $^{94}$Zr is mixed with
the cascade on top of the 0$^+_2$ level at 1300.53 keV \cite{Pan05,Fot02}. The rather
low energies of the 2$^+_1$ and 4$^+_1$ levels are due to the contribution of the
$\nu (d_{5/2}^2)_j$ multiplet \cite{Jak99}. The 2$^+_2$ level at 1671.60 keV, first
seen as a candidate for a mixed-symmetry excitation \cite{Elh08}, was later interpreted
as member of the 0$^+_2$ collective band \cite{Cha13}. The 2$^+$ level at 2366.60-keV
shows decays similar to those of 2$^+$, $\gamma$-band heads at N=56. This level and
the (3)$^+$, 2507.7- and 4$^+$, 2888.2-keV, levels may belong to a ``proto-$\gamma$''
structure in $^{94}$Zr.

At N=52 the present study has revealed levels in $^{92}$Zr which indicate collectivity
in this nucleus. The large $\delta$=-2.86(23) of the 1132.1-keV decay of the
2066.74-keV 2$^+_2$, level points to a collective character of this transition and a
possible $\gamma$-vibrational character of the 2$^+_3$, 2066.8-keV level. The 3$^+$,
2909.65-, 4$^+$, 2864.84-, (5$^+$) 3675.3- and the new (6$^+$), 3719.5-keV level are
proposed as members of a ``proto-$\gamma$'' on top of the 2066.8-keV bad head. This is
supported by the newly observed, 798.1-, 810.4- and 855.0-keV transitions in this band.

\begin{figure}
\centering
\scalebox{.34}{\includegraphics{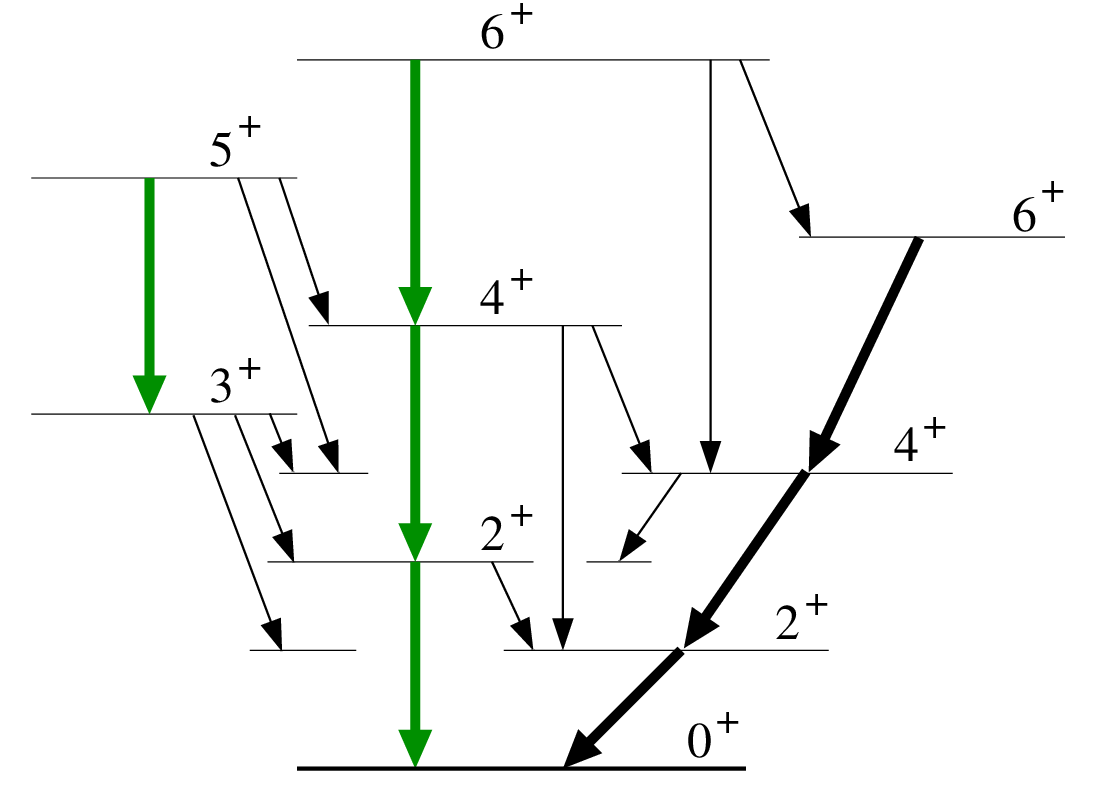}}
\caption{2D pattern of rotational excitations on top of a non-axial 0$^+$ state in
an even-even nucleus. See text for more comments.}
\label{2D_band_MTV}
\end{figure}

The common presence of ``$\gamma$ excitations'' in $^{92-98}$Zr isotopes suggests their
universal character, yet there are indications that the, so called, $\beta$ or $\gamma$
$vibrations$ are not present in nuclei \cite{Gar01,Sha11,Sha16} in contrast to
non-axial configurations, which are well documented and widely prevalent over the
nuclear chart.

To describe complex, rotational patterns in odd-A, triaxial nuclei a simple idea of
simultaneous rotation around $two$ principal axes has been proposed long time ago
\cite{MTV75}. Such two-dimensional pattern (2D band) has been applied recently to
describe rotations of Ag nuclei \cite{Urb25}.

An analogous idea applied to even-even nuclei results in a 2D band on top of a 0$^+$
configuration, as schematically sketched in Fig. \ref{2D_band_MTV}. Here thick, black
arrows connect levels corresponding to a rotation (R$_1$) with the largest moment of
inertia whereas thick, green arrows correspond to a rotation (R$_2$) around an axis
with smaller moment of inertia. The spin of the R$_1$ rotation serves as a non-zero,
K vector enabling levels with odd spins for the rotation R$_2$. All levels in this
2D band have positive parity and are interconnected by numerous transitions, shown by
thin black arrows, which are there because all the levels have same underlying
configuration of the 0$^+$ band head. The 0$^+$ head can be the ground state or an
excited state, provided its configuration is non-axial.

One notes that 0$^+_2$ bands with the associated ``proto-$\gamma$'' bands in $^{96}$Zr
and $^{98}$Zr shown in Fig. \ref{Zr_even_light_2plus_gamma} form structures similar to
the 2D band sketched in Fig. \ref{2D_band_MTV}.

\subsection{Negative-parity excitations in Zr isotopes}

\subsubsection{Octupole excitations}

Octupole collectivity in Zr isotopes is due to octupole coupling between neutrons in the
$d_{5/2}$, $g_{f/2}$, and $h_{11/2}$ shells and between protons in the $p_{3/2}$ and
$g_{9/2}$ shells. In addition to 3$^-$ octupole vibrations one may expect here coupling
between octupole and quadrupole phonons, producing the usual 3$^-\otimes2^+_1$ symmetric
multiplet as well as the more intriguing multiplet of the 3$^-\otimes2^+_{ms}$ asymmetric
states, which have been traced in the neighboring $^{96}$Mo \cite{Gre19} using EXILL.

\begin{figure}
\centering
\scalebox{.34}{\includegraphics{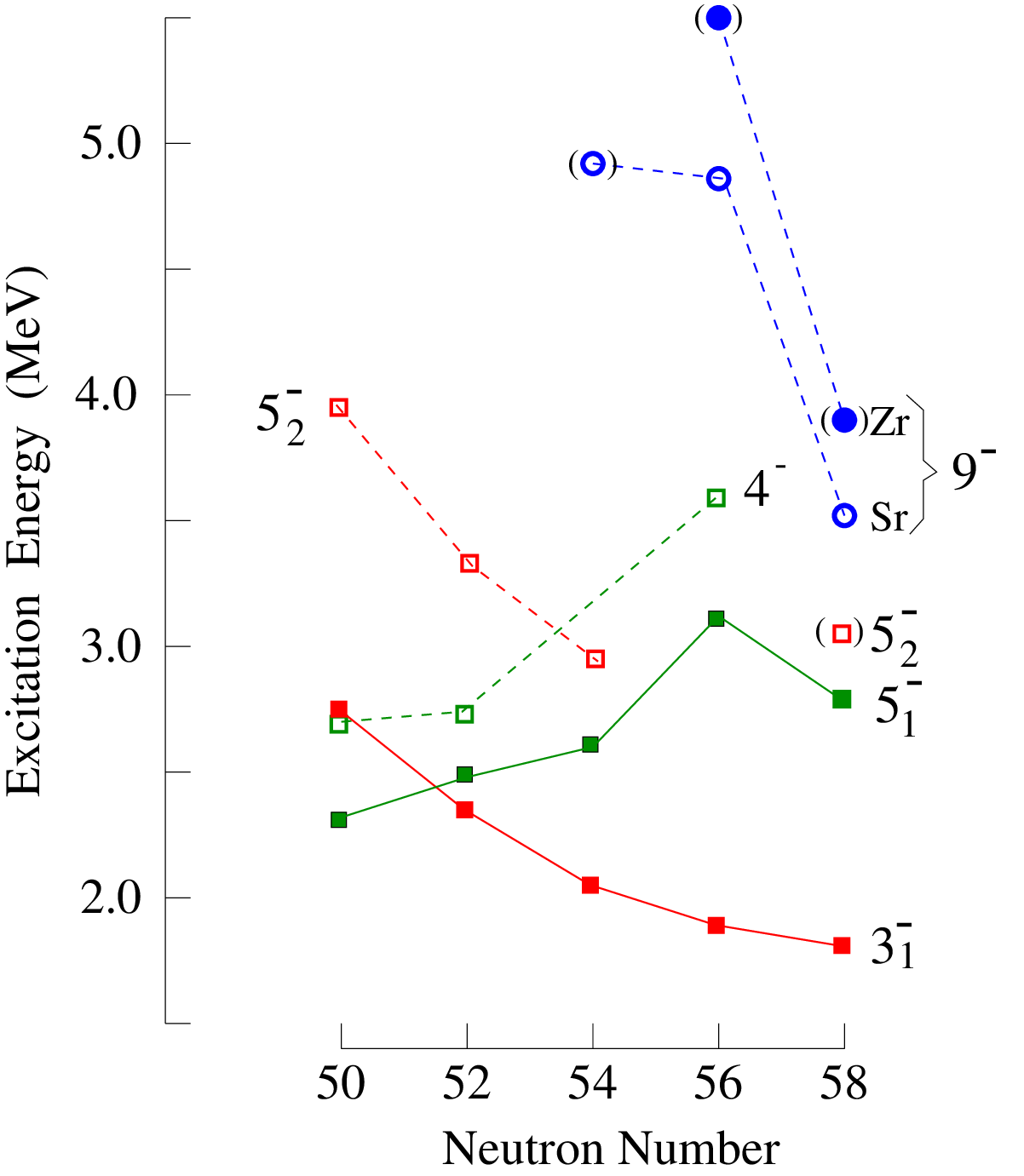}}
\caption{Excitation energies of selected negative-parity levels in Zr and Sr isotopes.
Experimental data are taken from Refs. \cite{Urb21,ENSDF,Sug17} and the present work.
Points in parentheses are tentative. Lines are drawn to guide the eye. See text for more
comments.}
\label{Zr_even_light_neg-exc}
\end{figure}

The $\pi (p_{3/2},g_{9/2})_{3^-}$ coupling, independent on the neutron number, results in
small changes of the 3$^-$ excitation energy. The 3$^-$ neutron coupling, which increases
with N due to the population of the $d_{5/2}$ and $g_{7/2}$ shells contributes to a steady
lowering of the 3$^-_1$ octupole excitation, shown in Fig. \ref{Zr_even_light_neg-exc}.

Octupole collectivity in Zr isotopes is very high, with B(E3) rate in $^{96}$Zr among the
highest observed in all nuclei. In the review work \cite{Mey88} Meyer have proposed
double-octupole-phonon excitations in $^{96}$Zr, though strongly mixed with the 0$^+_2$
band and the g.s. cascade. The $\gamma$ collectivity proposed in the previous section ads
to this mixed structure, recognized as a rather complex in calculations \cite{Saz19}.

This mixing causes irregularities in the 0$^+_2$ band and fragmentation of E1 strength
over many transitions \cite{Mey88}. This is confirmed by large B(E1) rates observed
previously \cite{Mey88} and in the present work for many transitions in $^{96}$Zr.
The B(E1)/B(E2) ratios in $^{94}$Sr, $^{96}$Zr and $^{98}$Zr are as high as in octupole
bands in the lanthanides (see Fig. 9 and 10 in Ref. \cite{Urb97}).

High octupole collectivity around $^{96}$Zr is supported by the B(E3) rate in $^{94}$Sr
determined in the present work. The T$_{dbe}$ value for the 1926.7-keV, E3 transitions
in $^{94}$Sr provide a high limit of B(E3)$\ge$40 W.u this transition.
For the 2340.0-keV transition in $^{92}$Zr a B(E3)=28(14) W.u. is determined in the
present work. The large uncertainty results from the weak E3 branching and a rather high
correction for summation effect in EXILL estimated at 0.019(1).

The B(E1) rates observed at N=56 suggests an important role of the $\nu d_{5/2}$ orbital
in producing octupole collectivity in the region. The empirical particle-vibration
description of the E3 decay in  $^{96}$Zr \cite{Stu04} depicts the 3$^-_1$ level as a
s.p. configuration ``dressed'' in octupole collectivity, a structure indescribable
within the shell model, alone.

\subsubsection{Single-particle excitations}

Figure \ref{Zr_even_light_neg-exc} shows two 5$^-$ excitations in $^{90-98}$Zr isotopes.
In $^{90-94}$Zr the 5$^-_2$ level corresponds to a quadrupole excitation coupled to the
3$^-$ octupole level, whereas the 5$^-_1$ level together with the 4$^-_1$ level may
both be due to the $\pi (p_{1/2},g_{9/2})_j$ s.p. coupling, (cf. Ref. \cite{Isk24}).
At N=56 the smooth trend of 4$^-_1$ and 5$^-_2$ levels is distorted, probably due to
the $\nu d_{5/2}$ shell closure and no 5$^-_2$ level is observed here. The very high
B(E1)/B(E2) ratio of 11$\times 10^{-6} fm^{-2}$ obtained in this work for the 5$^-_2$
level in $^{94}$Sr suggests a s.p. coupling for this level.

The excitation energy of the 3894.1-keV level in $^{98}$Zr with tentative spin-parity
(9$^-$) shown in Fig. \ref{Zr_even_light_neg-exc} fits that expected for the
$\nu(h_{11/2}\otimes g_{7/2})_{9^-}$ coupling at N=58, as seen in Fig.
\ref{Zr_even_light_neg-exc}. This configuration is observed systematically in the Sr
isotopes \cite{Rza09} and is well reproduced by the LSSM calculations  \cite{Urb21}.

The 1.9 $\mu$s, high-spin isomer at 6603.3 keV in $^{98}$Zr \cite{Sim06}, was interpreted
as the $[\pi~(g_{9/2})^2\otimes~\nu~(g_{7/2},h_{9/2})]_{17^-}$ configuration. Its decay
scenario is not obvious because only one isomeric transition of 63.0 keV was observed
whereas two different cascades are populated from the isomer. It was proposed that the
16$^+$ and 15$^-$ top levels in the two cascades have identical excitation energies and
the E1 and E2 isomeric transitions have identical energies. In the present work the
16$^+$ and 15$^-$ levels are established at 6544.0(5) and 6542.6(7) keV, respectively,
and the isomer energy is determined at 6607.0(5) keV taking identical energies of the
E1 and E2 isomeric transitions.

The data from Ref. \cite{Sim06} show that the E2 decay channel dominates. It is strongly
converted and, therefore, rather prompt. Thus the weak 63.0-keV $\gamma$ line corresponds
mostly to the more delayed, E1 decay branch. This is consistent with the present T$_{dbe}$
values which increase with spin much faster in the negative-parity band than in the
positive-parity band.

\section{Summary}

Exited states in $^{92}$Zr have been studied in cold-neutron capture reacttion and excited
states in $^{94}$Zr and $^{98}$Zr nuclei have been studied following neutron-induced
fission of $^{235}$U. The (n,$\gamma$) and fision data have been complemented by
measurements of $\beta$ decay of yttrium isotopes produced in fission of $^{235}$U. To
build level schemes of these nuclei we used triple-$\gamma$ coincidences measured with the
EXILL Ge array. Total of 54 new levels, 180 new $\gamma$ transitions and 70 new or improved
spin-parity assignments have been obtained in the present work for the three nuclei.

A new technique of estimating half-lives in the picosecond range, developed in this work,
provided 61 half-life estimates and 60 resulting B($\pi ,\lambda$)-rate limits in $^{94}$Sr,
$^{96}$Sr, $^{96}$Zr and $^{98}$Zr. The new technique uses triple-$\gamma$ coincidence and
provides half-lives even for weak transitions seen in doubly-gated spectra extracted from
complex $\gamma$ data measured in fission reaction. Considering details of the technique,
the conservative uncertainties of the half-life estimates obtained this work and their good
agreement with half-lives reported in the literature one may conclude that the obtained
T$_{dbe}$ half-life estimates are equal to T$_{1/2}$ half-lives with high credibility.

The precise neutron binding energy of 8634.81(2) keV has been  determined fot $^{92}$Zr.
New transitions found in this nucleus suggest band structures emerging on top of the
0$^+_2$ and 2$^+_3$ levels with B(E2) rates of several W.u.

In $^{94}$Zr a Gamow-Teller transition to the 4670.3-keV level has been proposed. The
(3$^+$) level at 2507.9 keV is proposed to belong to a collective structure due to an
emerging $\gamma$ instability in this nucleus.

In $^{98}$Zr a new, E3 decay from the 3$^-$ 1806.20-keV level has been found. The (9$^-$),
3894.1-keV level in $^{98}$Zr was proposed to be a few ns isomer and the decay pattern
of the 17$^-$, 6607.0-keV isomer has been clarified. A $\gamma$ structure has been found
on top of the 1590.80-keV level with B(E2) rates of in-band transitions of several W.u.
The deformed band on top of the 0$^+_3$, 1436.1-keV level, proposed in previous works is
not confirmed. The upper $\gamma$ intensity limit of the unobserved, 154.7-keV decay of
the 1590.80-keV level is 4$\times$10$^{-4}$ relative to the $\gamma$ intensity of the
1590.80-keV transitions.

New-type, phenomenological systematics, backed by LSSM calculations have been used to
classify 2$^+$ excitations in the Zn - Zr, even-even nuclei of the 50$\le$N$\le$60 neutron
range. This and the new phenomenological systematics of 0$^+$ excitations in Sr and Zr
isotopes explain the evolution of collectivity in zirconium isotopes, showing the important
role of various single particle excitations in the phase transition and coexistence in the
region. In particular, the role of the $\nu$9/2$^+$[404] extruder orbital as a catalyst in
creating prolate-deformed, 0$^+$ excitations and inducing sudden deformation change in the
region has been discussed. Special type of configurations, with pair of neutrons excited
from the $\nu$ 9/2$^+$[404] extruder to the $oblate$, 11/2$^-$[505] intruder orbital, has
been proposed for 0$^+_3$ levels in $^{94}$Sr, $^{96}$Zr, $^{96}$Sr, $^{98}$Zr and 0$^+_2$
levels in $^{98}$Sr, $^{100}$Zr, $^{100}$Sr and $^{102}$Zr.

The dominating proton and neutron configurations in the studied nuclei, based on and
consistent with the proposed phenomenological systematics of 0$^+$ excitations in Sr and Zr
isotopes provide very good explanation of rapidly varying, relative cross sections for the
$^A$Zr(t,p), $^A$Mo($^6$Li,$^8$B)$^{A-2}$Zr and $^A$Mo($\alpha$)$^{A-4}$Zr transfer
reactions to 0$^+_2$ and 0$^+_3$ states in $^{94-98}$Zr.

The same dominating configurations, particularly those containing the $oblate$,
11/2$^-$[505] intruder orbital, provide an explanation of two-neutron separation energies,
S$_{2n}$, rapidly varying in the region of the deformation change with the amplitude of
the change in excess of 1.5 MeV at N=59, as revealed by the new visualisation of S$_{2n}$
variation proposed in the present work.

Collective structures with characteristic 3$^+$ excitations, related to $\gamma$
collectivity, have been proposed in $^{92,94,96,98}$Zr nuclei. They are strongly mixed
with collective structures based on 0$^+_2$ levels. Both structures are built around the
two-proton, $\pi (g_{9/2})^2_{2^+,4^+,6^+,8^+}$  multiplet. The $\gamma$ collectivity
``dressing'' the single-particle $(g_{9/2})^2_j$ cascade is likely due to non-axiality
of the 0$^+_2$ levels, which is induced by the the high-$\Omega$, oblate-shaped
$\nu 9/2^+[404]$ extruder present near the Fermi surface on one hand and the population
of low-$\Omega$, prolate-driving orbitals of the $\nu g_{7/2}$ shell, on the other hand.

In this work we suggested a number of high-interest measurements aimed at verification
of the proposed structures and effects associated with the development of collectivity
in  transitional nuclei of the A$\sim$ 100 region.

More generally, we would like to stress the importance of studies of nuclei near the line
of stability for advancing our understanding of nuclear structure. Such nuclei can be
accessed with various, complementary experimental techniques. For example, high-resolution
measurements using large Ge arrays, which provide high-statistics data, help finding new
effects, whereas transfer-reaction measurements can probe the contents of wave functions,
especially of crucial, low-energy 0$^+$ and 2$^+$ levels.

\section{Acknowledgments}

The authors thank the technical services of the ILL Grenoble and GANIL, Caen for supporting
the EXILL campaign. The EXOGAM collaboration and the INFN Legnaro are acknowledged for the
loan of Ge detectors.


\begin{thebibliography}{}


\bibitem{Urb13} W. Urban, M.~Jentschel, R.F. Casten,  J.~Jolie, Ch. Bernards, B. Maerkisch, 
                Th. Materna, P. Mutti, L. Pr\'ochniak, T.~Rz\c{a}ca-Urban, G.S. Simpson, 
                V. Werner,and  S. Ahmed, Phys.~Rev. C {\bf 87}, 031304(R) (2013).

\bibitem{Urb21} W. Urban, K. Sieja, T. Rz\c{a}ca-Urban, J. Wi\'sniewski, A. Blanc, 
                M. Jentschel, P. Mutti,  U. K\"oster, T. Soldner, G. de France, 
                G. Simpson, C.A. Ur, A.G. Smith, and J.P. Greene, 
                Phys. Rev {\bf C} 104, 064309 (2021).

\bibitem{Lan81} L.D. Landau, E.M. Lifshitz, ``Quantum Mechanics: Non-Relativistic Theory'', 
                3-rd Edition, Elsevier, 1981.

\bibitem{Urb19} W. Urban, T. Rz\c{a}ca-Urban, J. Wi\'sniewski, I. Ahmad, A.G. Smith, 
                and G.S. Simpson, Phys.~Rev. C, {\bf 99}, 064325 (2019).
                
\bibitem{ENSDF} Evaluated Nuclear Structure Data File (ENSDF), {\it www.nndc.bnl.gov/ensdf/}.

\bibitem{Cru20} S. Cruz, K. Wimmer, S.S. Bhattacharjee, P.C. Bender, G. Hackman, R. Kr\"ucken,
                F. Ames, C. Andreoiu, R.A.E. Austin, C.S. Bancroft, {\it et al.},
                Rev. C {\bf 102}, 024335 (2020).

 \bibitem{Wis23} J. Wisniewski, W.Urban, T. Rz\c{a}ca-Urban, K. Sieja, A. Blanc, M. Jentschel,
                C. Micheagnoli, P. Mutti, U. K\"oster, T. Soldner, G. de France, G. S. Simpson, 
                and C.A. Ur, Phys.~Rev. C, {\bf 108}, 024302 (2023).              

\bibitem{Urb13b} W.~Urban, M.~Jentschel, B.~M\"arkisch, Th.~Materna, Ch.~Bernards, 
                C.~Drescher, Ch.~Fransen, J.~Jolie, U.~K\"oster, P.~Mutti, 
                T.~Rz\c{a}ca-Urban, and G. S.~Simpson, 
                JINST {\bf 8}, P03014 (2013).

\bibitem{Jen17} M.~Jentschel, A.~Blanc, G.~de~France, U.~Koster, S.~Leoni, P.~Mutti, 
                G.~Simpson, T. ~Soldner, C.~Ur, W.~Urban, and EXILL Collaboration,
                JINST {\bf 12}, 11003  (2017).

\bibitem{Tal70} W.L. Talbert Jr., F.K. Wohn, H.H., Hsu, and S.T. Hsue, 
                Nucl. Phys. A {\bf 146}, 149 (1970).
                
\bibitem{NGATLAS} Atlas Of Neutron Capture Cross Sections, www-nds.iaea.org/ngatlas/

\bibitem{Rza18} T. Rz\c{a}ca-Urban, W. Urban, M. Czerwi\'nski, J. Wi\'sniewski, A. Blanc,
                H. Faust, M. Jentschel, P. Mutti, U. K\"oster, T. Soldner, G. de France,
                G. S. Simpson, and C. A. Ur,
                Phys.~Rev.~C {\bf 98}, 064315 (2018).

 \bibitem{Cze15} M. Czerwi\'nski,  T.~Rz\c{a}ca-Urban, W.~Urban, P.B\c{a}czyk, K. Sieja,
                J. Timar, I. Kuti, T. Tornyi, B. Nyak\'o, L. Atanasova, A. Blanc,
                M. Jentschel, P. Mutti, U. K\"oster, T. Soldner, G. de France,
                G. Simpson, and C.A. Ur,
                Phys. Rev. C {\bf 92}, 014328  (2015).       

\bibitem{Urb16} W. Urban, K. Sieja, T. Materna, M. Czerwi\'nski,  T.~Rz\c{a}ca-Urban, 
                A. Blanc, M. Jentschel, P. Mutti, U. K\"oster, T. Soldner, G. de 
                France, G. Simpson, C.A. Ur,  C. Bernards, C. Fransen, J. Jolie,
                J.-M. Regis, T. Thomas, and N. Warr, 
                Phys. Rev. C {\bf 94}, 044328 (2016). 

\bibitem{Urb23} W.Urban, M. Czerwi\'nski, G. de France, M.~Jentschel, U.~K\"oster, P.~Mutti,
                T. Rz\c{a}ca-Urban, and J. Wi\'sniewski, JINST {\bf 18}, P11031 (2023).

\bibitem{Kra73} K.S. Krane and R.M. Steffen and R.M. Wheeler,
                Nuclear Data Tables {\bf  11}, 351 (1973).

\bibitem{Ham75} ``The Electromagnetic Interaction in Nuclear Spectroscopy'',
                W.D. Hamilton, Editor, North-Holland Publishing Company,
                Amsterdam (1975).

\bibitem{Nol94} P.J.~Nolan, ,F.A.~Beck, and D.B.~Fossan,
                Ann. Rev. Nuc. Part. Sci. {\bf 44}, 561 (1994).

\bibitem{Urb97} W. Urban, J.L. Durell, W.R. Phillips, A.G. Smith, M.A. Jones, I. Ahmad,
                A.R. Barnett, M. Bentaleb, S.J. Dorning, M.J. Leddy, E. Lubkiewicz,
                L.R. Morss, T. Rz\c{a}ca-Urban, R.A. Sareen, N. Schulz, and B.J. Varley,
                Z.Phys. A {\bf 358}, 145 (1997).

\bibitem{Smi94} A.G. Smith, W.R. Phillips, J.L. Durell, W. Urban, B.J. Varley, C.J. Pearson,
                J.A. Shannon, I . Ahmad, C.J. Lister,  L.R. Morss, K.L. Nash, C.W. Williams,
                M. Bentaleb, E. Lubkiewicz, and N. Schulz,
                Phys. Rev. Lett {\bf 73}, 2540 (1994).

\bibitem{Smi96} A.G. Smith, J.L. Durell, W.R. Phillips, M.A. Jones, M. Leddy,
                W. Urban, B.J. Varley, I. Ahmad, L.R. Morss, M. Bentaleb,
                A. Guessous, E. Lubkiewicz, N. Schulz, and R. Wyss,
                Phys. Rev. Lett {\bf 77}, 1711 (1996).

\bibitem{Urb01} W.~Urban, J.L. Durell, A.G. Smith, W.R. Phillips, M.A. Jones,
                B.J. Varley, T. Rz\c{c}a-Urban, I. Ahmad, L.R.Morss, M. Bentaleb,
                and N. Schulz, Nucl. Phys. A {\bf 689}, 605 (2001).


\bibitem{Smi12} A. G. Smith, J. L. Durell, W. R. Phillips, W. Urban, P. Sarriguren,
                and I. Ahmad, Phys. Rev. C {\bf 86}, 014321 (2012).

\bibitem{Reg17} J.-M. R\'egis, J. Jolie, N. Saed-Samii, N. Warr, M. Pfeiffer, A. Blanc,
                M. Jentschel, U. K\"oster, P. Mutti, T. Soldner,  {\it et al},
                Phys. Rev. C {\bf 95}, 054319 (2017).

\bibitem{Ans17} S. Ansari, J.-M. R\'egis, J. Jolie, N. Saed-Samii, N. Warr, W. Korten,
                M. Zieli\'nska, M.-D. Salsac, A. Blanc, M. Jentschel, {\it et al},
                Phys. Rev. C {\bf 96}, 054323 (2017).

\bibitem{Esm21} A. Esmaylzadeh, V. Karayonchev, K. Nomura, J. Jolie, M. Beckers, A. Blazhev,
                A. Dewald, C. Fransen, R.-B. Gerst, G. Häfner, A. Harter,  L. Knafla, M. Ley,
                L. M. Robledo, R. Rodríguez-Guzm\'an, and M. Rudigier,
                Phys. Rev. C {\bf 104}, 064314 (2021).

\bibitem{Ral17} D. Ralet,  S. Pietri,  T. Rodríguez,  M. Alaqeel,  T. Alexander,
                N. Alkhomashi,  F. Ameil,  T. Arici,  A. Ataç, R. Avigo, {\it et al},
                Phys. Rev. C {\bf 95}, 034320 (2017).

\bibitem{Kru01} R. Kr\"ucken, W.-T. Chou, J.R. Cooper, C.W. Beausang, C.J. Barton,
                M. A. Caprio,  R.F. Casten,  A.A. Hecht, J.R. Novak,  N. Pietralla,
                A. Wolf, and N.V. Zamfir, Phys. Rev. C {\bf 64}, 017305 (2001).

\bibitem{Yat05} S.W. Yates, Journal of Radioanalytical and Nuclear Chemistry, {\bf 265}, 
                291 (2005).
                
\bibitem{Wer02} V. Werner, D. Belic, P. von Brentano, C. Fransen, A. Gade, H. von Garrel, 
                J. Jolie, U. Kneissl, C. Kohstall, A. Linnemann, A.F. Lisetskiy, N. Pietralla, 
                H.H. Pitz, M. Sheck, K.-H. Speidel, F. Stedile, and S.W. Yates,
                Phys. Lett. {\bf B 550}, 140 (2002).

\bibitem{Sie09} K. Sieja, F. Nowacki, K. Langanke, and G. Martinez-Pinedo
                Phys. Rev. C {\bf 79}, 064310 (2009).

\bibitem{NDS_Zr92} Coral M. Baglin, Nucl. Data Sheets {\bf 113}, 2187 (2012). 

\bibitem{Mac90} H. Mach, E.K. Warburton, W. Krips, R.L. Gill, and M. Moszy\'nski,
                Phys.~Rev.~C {\bf 42}, 568 (1990).

\bibitem{Fot02} N. Fotiades, J.A. Cizewski, J.A. Becker, L.A. Bernstein, D.P. McNabb,
                W. Younes, R.M. Clark, P. Fallon, I.Y. Lee, A. Holt, and M. Hjorth-Jensen,
                Phys. Rev. C {\bf 65}, 044303 (2002).

\bibitem{Pan05} D. Pantelica, I.Gh. Stefan, N. Nica, M.-G. Porquet, G. Duchene, A. Astier, 
                S. Courtin, I. Deloncle, F. Hoellinger, A. Bauchet, {\it et al.}, 
                Phys. Rev. C {\bf 72}, 024304 (2005).
                
\bibitem{Sug17} M. Sugawara, Y. Toh, M. Koizumi, M. Oshima, A. Kimura, T. Kin, Y. Hatsukawa,
                and H. Kusakari, Phys. Rev. C {\bf 96}, 024314 (2017).

\bibitem{Pet13} E.E. Peters, A. Chakraborty, B.P. Crider, B.H. Davis, M.K. Gnanamani, 
                M.T. McEllistrem, F.M. Prados-Est\'evaz, J.R. Vanhoy, and S.W. Yates,
                Phys. Rev. C {\bf 88}, 024317 (2013).

\bibitem{NDS_Zr91} Coral M. Baglin, Nucl. Data Sheets {\bf 114}, 1293 (2013).

\bibitem{NDS_Zr93} Coral M. Baglin, Nucl. Data Sheets {\bf112}, 1163 (2011). 

\bibitem{NDS_Zr95} S.K. Basu, G. Mukherjee, and A.A. Sonzogni,
                Nucl. Data Sheets {\bf111}, 2555 (2010).

\bibitem{Rza23} T. Rz\c{a}ca-Urban, W. Urban, A. Blanc, M. Jentschel, P. Mutti, U. K\"oster, 
                G. de France, G.S. Simpson, and C.A. Ur,
                Phys.~Rev.~C {\bf 108}, 014324 (2023).

\bibitem{NDS28Al} M. Shamsuzzoha Basunia, Nucl. Data Sheets {\bf114}, 1189 (2013).

\bibitem{NDS2H} J. H. Kelley, and J. L. Godwin, ENSDF (2013).

\bibitem{Sch82} H.H. Schmidt, P. Hungerfort, H. Daniel, T. von Egidy, S.A. Kerr,
                R. Brissot, G. Barreau, H.G. B\"ornere, C. Hofmeyr, and K.P. Lieb, 
                Phys. Rev. C.  {\bf 25}, 2888 (1982).

\bibitem{Hon96} J. Honz\'atko, K. Kone\v{c}n\'y, I. Tomandl, J. Vacik, F. Be\v{c}v\'a\v{r},
                and P. Cejnar, Nucl. Instr. Meth. A {\bf 376}, 434 (1996).

\bibitem{AME20} M.~Wang, W.J. Huang, F.G. Kondev, G.~Audi,  and S. Naimi,
                Chin. Phys. C {\bf 45}, 030003 (2021).

\bibitem{NDS10} N. Nica, Nucl. Data Sheets {\bf 111}, 525 (2010).

\bibitem{NDS94} A. Negret, and A.A. Sonzogni, ENSDF (2011).

\bibitem{Sin76} B. Singh, H.W. Taylor, and P.J. Tivin,
                J. Phys. G: Nucl. Phys. {\bf 2}, 397 (1976).

\bibitem{Zr94NDS} D. Abriola, A.A. Sonzogni, Nucl. Data Sheets {\bf 107}, 2423 (2006).

\bibitem{Cha13} A. Chakraborty, E.E. Peters, B.P. Crider, C. Andreoiu, P.C. Bender, D.S. Cross,
                G.A. Demand A.B. Garnsworthy, P.E. Garrett, G. Hackman, {\it et al.},
                Phys. Rev. Lett. {\bf 110}, 022504 (2013).

\bibitem{NDS98} J. Chen, B. Singh, Nucl. Data Sheets {\bf 164}, 1 (2020).

\bibitem{Kaw82} K. Kawade, G. Battistuzzi, H. Lawin, H.A. Seli\v{c}, K. Systemich, F. Schussler
                E. Monnand, J.A. Pinston, B. Pfeiffer, and G. Jung,
                Z. Phys. A {\bf 304}, 293 (1982).

\bibitem{Urb17} W. Urban, M. Czerwi\'nski, J. Kurpeta, T. Rz\c{a}ca-Urban, J. Wi\'sniewski,
                T. Materna, \L.W. Iskra, A. G. Smith, I. Ahmad, A. Blanc, H. Faust,
                U. K\"oster, M. Jentschel, P. Mutti, T. Soldner, G. S. Simpson,
                J.A. Pinston, G. de France, C.A. Ur, V.-V. Elomaa, T. Eronen, J. Hakala,
                A. Jokinen, A. Kankainen, I. D. Moore, and J. Rissanen,
                Phys. Rev. C {\bf 96}, 044333 (2017).

\bibitem{Mey86} R.A. Meyer, E.A. Henry, L.G. Mann, and K. Heyde,
                Phys.~Lett.~B {\bf 177}, 271 (1986).

\bibitem{Hwa12} J.K. Hwang, J.H. Hamilton, A.V. Ramayya, N.T. Brewer, E.H. Wang, Y.X. Luo,
                and S.J. Zhu, Int. J. of Mod. Phys. E {\bf 21}, 1250080 (2012).

\bibitem{Wu004} C.Y. Wu, H. Hua, D. Cline, A.B. Hayes, R. Teng, R.M. Clark, P. Fallon,
                A. Goergen, A.O. Macchiavelli, and K. Vetter,
                Phys. Rev. C {\bf 70}, 064312 (2004).

\bibitem{Sim06} G.S. Simpson, J.A. Pinston, D. Balabanski, J. Genevey, G. Georgiev, J. Jolie,
                D.S. Judson, R. Orlandi, A. Scherillo, I. Tsekhanovich, W. Urban, and N. Warr,
                Phys. Rev. C {\bf 74}, 064308 (2006).

\bibitem{Bet10} L. Bettermann, J.-M. Regis, T. Materna, J. Jolie, U. K\"oster, K. Moschner,
                and D. Radeck, Phys. Rev. C {\bf 82}, 044310 (2010).

\bibitem{Sin18} P. Singh, W. Korten, T.W. Hagen, A. G\"orgen, L. Grente, M.-D. Salsac,
                F. Farget, E. Cl\'ement, G. de France, T. Braunroth, {\it et al},
                Phys.~Rev.~Lett. {\bf 121}, 192501 (2018).

\bibitem{Wit18} W. Witt, V. Werner, N. Pietralla, M. Albers, A.D. Ayangeakaa, B. Busher,
                M.P. Carpenter, D. Cline, H.M. David, A. Hayes, {\it et al},
                Phys. Rev. C {\bf 98}, 041302(R) (2018).

\bibitem{Kar20} V. Karayonchev, J. Jolie, A. Blazhev, A. Devald, A. Esmaylzadeh, C. Fransen,
                G. H\"afner, L. Knafla, J. Litzinger, C. M\"uller-Gatermann, J.-M. regis,
                K. Schomacker, A. Vogt, and N. Warr, Phys. Rev. C {\bf 102}, 064314 (2020).

\bibitem{Pas23} G. Pasqualato, S. Ansari, J.S. Heines, V. Modamio, A. G\"oren, W. Korten,
                J. Ljunvall, A. Cl\'ement, J. Dudouet, A. Lemasson, {\it et al.},
                Eur. Phys. J. A {\bf 59}:276 (2023).

\bibitem{NDS96} D. Abriola, and A.A. Sonzogni, Nucl. Data Sheets {\bf 109}, 2501 (2008).

\bibitem{Ber19} G.F. Bertsch, Eur. Phys. J. A {\bf 55 }: 248 (2019).

\bibitem{Bor16} P.F. Bortignon, and R.A. Broglia, Eur. Phys. J. A {\bf 52 }: 280 (2016).

\bibitem{Wal11} C. Walz, H. Fujita, A. Krugmann, P. von Neumann-Cosle, N. Pietrala,
                V. Yu. Ponomarev, A. SheikhObeid, and J. Wambach,
                Phys. Rev. Lett {\bf 106}, 062501 (2011).

\bibitem{Rza13} T. Rz\c{a}ca-Urban, W.Urban,  A.G.Smith, I.Ahmad, F.Nowacki, and K. Sieja,
                Phys. Rev. C {\bf 88}, 034302 (2013).
\bibitem{Wis19} J. Wi\'sniewski, W. Urban, M. Czerwi\'nski, J. Kurpeta, A. P{\l}ochocki,
                M. Pomorski, T. Rz\c{a}ca-Urban, K. Sieja, L. Canete, T. Eronen,
                {\it et al.},  Phys.Rev. C {\bf 100}, 054331 (2019).

\bibitem{Urb25} W.Urban, A. Abramuk, J. Kurpeta,  T. Rz\c{a}ca-Urban,  A.G.Smith, and
                J.P. Greene, Phys.Rev. C {\bf 112}, 014316 (2025).

\bibitem{Kur72} A. Kuriyama, T. Maromuri, K. Matsuyanagi,
                Prog. Theor. Phys. {\bf 47}, 498 (1972);
                Prog. Theor. Phys. {\bf 51}, 779 (1974).

\bibitem{Kur75} A. Kuriyama, T. Maromuri, K. Matsuyanagi, R. Okamoto,
                Prog. Theor. Phys. {\bf 53}, 489 (1975).

\bibitem{Paa73} V. Paar, Nucl. Phys. A \textbf{211}, 29 (1973);
                            Phys. Lett. B \textbf{39}, 587 (1972).

\bibitem{Mat16} K. Matsuyanagi,  M. Matsuo, T. Nakatsukasa, K. Yoshida, N. Hinohara, and
                K. Sato, Phys. Scr. {\bf 91}, 063014 (2016).

\bibitem{Kis25} S. Kisyov, and S. Lalkovski, Symmetry {\bf17}, 1276 (2025).

\bibitem{Cak08} R.B. Cakirli, and R.F. Casten, Phys. Rev. C {\bf 78}, 041301(R) (2008).

\bibitem{Boy21} M. B\"oy\"ukata, C.E. Alonso, J.M. Arias, L. Fortunato, and A. Vitturi,
                Eur. Phys. J. A {\bf 57}: 2 (2021).

\bibitem{Rza07} T. Rz\c{a}ca-Urban, W.Urban, J.L. Durell, A.G.Smith, and I.Ahmad,
                Phys. Rev. C {\bf 76}, 027302 (2007).

\bibitem{Sie22} K. Sieja, Universe, {\bf 8}, 23 (2022).

\bibitem{Nya21} B.M Nyak\'o, J. Tim\'ar, M. Csatl\'os, Zs. Dombr\'adi, A Krasznohorkay,
                I. Kuti, D. Sohler, T.G. Tornyi, M. Czerwi\'nski, T. Rz\c{a}ca-Urban,
                {\it et al.}, Phys. Rev. C {\bf 103}, 034304 (2021).

\bibitem{Nya21b} B.M Nyak\'o, J. Tim\'ar, M. Csatl\'os, Zs. Dombr\'adi, A Krasznohorkay,
                I. Kuti, D. Sohler, T.G. Tornyi, M. Czerwi\'nski, T. Rz\c{a}ca-Urban,
                {\it et al.}, Phys. Rev. C {\bf 104}, 054305 (2021).

\bibitem{Toc25} G. Tocabens, D. Verney, S. P\'eru, S. Hilaire, E. Cantacuz\'ene, C. Delafosse,
                M. Cheikh Mhamed, I. Deloncle, W. Dong, C. Gaulard, {\it et al.},
                Phys. Rev. C {\bf 111}, 034306 (2025).

\bibitem{Chi70} E. Chiefetz, R.C. Jared, S.G. Thompson, J.B. Wilhelmy,
                Phys. Rev. Lett. {\bf 25}, 38 (1970).

\bibitem{She72} R.K.~Sheline, I. Ragnarsson, and .G. Nilsson,
                Phys.~Lett.~B {\bf 41}, 115 (1972).

\bibitem{Cas07} R.F. Casten, and E.A. McCutchan,
                J. Phys. G: Nucl. Part. Phys. {\bf 34}, R285 (2007).

\bibitem{Hey04} K. Heyde, J. Jolie, R. Fossion, S. De Baerdemacker, and V. Hellemans,
                Phys.Rev. C {\bf 69}, 054304 (2004).

\bibitem{Fed77} P.~Federman and S.~Pittel, Phys.~Lett.~B {\bf 69},  385 (1977);
                                           Phys.~Lett.~B {\bf 77}, 29 (1978);
                                           Phys.Rev. C {\bf 20}, 820 (1979).

\bibitem{Lhe94} G. Lhersonneau, B. Pfeiffer, K.-L. Kratz, T. Enqvist, P.P. Jauho,
                A. Jokinen, J. Kantele, M. Leino, J. M. Parmonen, H. Penttil\"a,
                and J. \"Ayst\"o, Phys.~Rev.~C {\bf 49}, 1379 (1994).

\bibitem{Ham95} J.H. Hamilton, A.V.Ramayya, S.J. Zhu, G.M. Ter-Akopian, Yu.Ts. Oganessian,
                J.D.Cole, J.O.Rasmussen, and M.A.Stoyer,
                Prog. Nucl. Part. Phys. {\bf 35}, 635 (1995).

\bibitem{WH1992} J.L. Wood, K. Heyde, W. Nazarewicz, M. Huyse, and P. van Duppen,
                Phys. Reports {\bf 215}, 101 (1992).

\bibitem{Gav22} N. Gavrielov, A.~Leviatan, and F. Iachello,
                Phys.~Rev.C  {\bf 105}, 014305 (2022).

\bibitem{Mas25} K.R. Mashtakov, P.E. Garrett, B. Olaizola, C. Andreoiu, G.C. Ball, P. Bender,
                V. Bildstein, A. Chester, D.S. Cross, H. Dawkins, {\it et al.},
                EPJ Web of Conferences {\bf 329}, 01012 (2025).

\bibitem{Hey88} K. Heyde, and R.A. Meyer, Phys. Rev. C {\bf 37}, 2170 (1988).

\bibitem{Mac90a} H. Mach, M. Moszy\'nski, R.L. Gill, G. Molna\'ar, F.K. Wohj, J.A. Winger,
                and J.C. Hill, Phys. Rev. C {\bf 41}, 350 (1990).

\bibitem{Hey90} K. Heyde, and R.A. Meyer, Phys. Rev. C {\bf 42}, 790 (1990).

\bibitem{Mac90b} H. Mach, M. Moszy\'nski, R.L. Gill, G. Molna\'ar, F.K. Wohj, J.A. Winger,
                and J.C. Hill, Phys. Rev. C {\bf 42}, 793 (1990).

\bibitem{Woo99} J.L. Wood, E.F. Zganjar, C. De Coster, and K. Heyde,
                Nucl. Phys. A {\bf 651}, 323 (1999).

\bibitem{Leo24} S. Leoni, B. Fornal, A. Bracco, Y. Tsunoda, and T. Otsuka,
                Prog. Part. Nucl. Phys. {\bf 139}, 104119 (2024).

\bibitem{HW2011} K. Heyde, J.L. Wood, Rev. Mod. Phys. {\bf 83}, 1467 (2011).

\bibitem{Bra90} F. Barranco, G.F. Bertsch, R.A. Broglia, and E. Vigezzi,
                Nucl. Phys {\bf A 512}, 253 (1990).

\bibitem{Bro94} R.A. Broglia, F. Barranco, G.F. Bertsch, and E. Vigezzi,
                Phys. Rev.~C {\bf 49}, 552 (1994).

\bibitem{Wu024} J. Wu, M. P. Carpenter, F.G. Kondev, R.V.F. Janssens, S. Zhu, E.A. McCutchan1,
                A.D. Ayangeakaa, J. Chen, J. Clark,   {\it et al.},
                Phys. Rev. C {\bf 109}, 0423174(2024).

\bibitem{Wer94} T.R. Werner, J. Dobaczewski, M.W. Guidry, W. Nazarewicz, and J.A. Sheikh,
                Nucl.~Phys. A {\bf 578}, 1 (1994).

\bibitem{Cru18} S. Cruz, P.C. Bender, R. Kr\"ucken, K. Wimmer, F. Ames, C. Andreoiu,
                R.A.E. Austin, C.S. Bancroft, R. Braid, T. Bruhn,  {\it et al.},
                Phys.~Lett.~B {\bf 786}, 94 (2018).

\bibitem{Urb26} W. Urban, T. Rz\c{a}ca-Urban, J. Wi\'sniewski, A.G. Smith, and J.P. Greene,
                arXiv:2606.06319v1 [nucl-ex].

\bibitem{Kib22} T. Kib\'edi, A.B. Garnsworthy, and J.L. Wood,
                Prog. Part. Nucl. Phys. {\bf 123}, 103930 (2022).

\bibitem{Mey85} R.A. Meyer, E. Monnand, J.A. Pinston, F. Schussler, B. Pfeiffer,
                I. Ragnarsson, H. Lawin, G. Lhersonneau, and K. Sistemich,
                Nucl.~Phys. {\bf A439}, 510 (1985).

\bibitem{Ang09} S. Anghel, G. Cata-Danil, and N.V. Zamfir,\\
                Rom. Journ. Phys. {\bf 54} 301 (2009).

\bibitem{Spa25} A. Sp{\v a}taru, {\it et al.}, Phys. Rev.~C {\bf 111}, 054307 (2025).

\bibitem{NUDAT} NuDat 3.0, {\it www.nndc.bnl.gov/nudat3/}

\bibitem{Gar05} J.E. Garcia-Ramos, K. Heyde, R. Fossian, V. Hellemans, and S. De Baerdemacker,
                Eur. Phys J. A {\bf 26}, 221 (2005).

\bibitem{Urb03} W. Urban, J.A. Pinston, T. Rz\c{a}ca-Urban, A. Z{\l}omaniec,
                G. Simpson, J.L. Durell, W.R. Phillips, A.G. Smith, B.J. Varley,
                I. Ahmad, and N. Schulz,
                Eur. Phys. J.  A {\bf 16}, 11 (2003).

\bibitem{Urb04} W. Urban, J.A. Pinston, J. Genevey, T. Rz\c{a}ca-Urban, A. Z{\l}omaniec,
                G. Simpson, J.L. Durell, W.R. Phillips, A.G. Smith, B.J. Varley, I. Ahmad,
                and N. Schulz, Eur. Phys. J. A {\bf 22}, 241 (2004).

\bibitem{Pet12} A. Petrovici,  Phys.~Rev.C  {\bf 85}, 034337 (2012).

\bibitem{Pet20} A. Petrovici,  and A.S. Mare, Phys.~Rev.C  {\bf 101}, 04307 (2020).

\bibitem{Pet25} A. Petrovici, and O. Andrei, Phys.~Rev.C  {\bf 111}, 014307 (2025).

\bibitem{Ska97} J. Skalski, S. Mizutori, and W. Nazarewicz, Nucl. Phys. A {\bf 617}, 282 (1997).

\bibitem{Gor07} S. Goriely, M. Samyn, and J.M. Pearson, Phys.~Rev.C  {\bf 75}, 064312 (2007).

\bibitem{Kos26} A. K\"oseo\v{g}lu, {\it et al.} 2026 J. Phys. G: Nucl. Part. Phys. in press
                https://doi.org/10/1088/1361-6471/ae8462

\bibitem{NDS_Sr96} D. Abriola(a), A.A. Sonzogni Nucl. Data Sheets {\bf 109}, 2501 (2008).

\bibitem{Mar20} E.V. Mardyban, E.A. Kolganova, T.M. Shneidman, R.V. Jolos, and N. Pietralla,
                Phys. Rev. C {\bf 102}, 034308 (2020).

\bibitem{Jak99} G. Jakob, N. Bencher-Koller, J. Holden, G. Kumbartzki, T.J. Mertzimekis,
                K.-H. Speidel, C.W. Beausang, and R. Kr\"ucken,
                Phys.~Lett.~B {\bf 468}, 1 (1999).

\bibitem{Elh08} E. Elhami, J.N. Orce, M. Scheck,  S. Mukhopadhyay, S.N. Choudry,
                M.T. McEllistrem, S.W. Yates, {\it et al.},
                Phys.~Rev.~C {\bf 78}, 064303 (2008).

\bibitem{Gar01} P.E. Garrett, J. Phys. G : Nucl. Par. Phys. {\bf 27}, R1 (2001).

\bibitem{Sha11} J.F. Sharpey-Schafer, {\it et al.}, Eur. Phys. J. A {\bf 47}, 5 (2011).

\bibitem{Sha16} J.F. Sharpey-Schafer, PoS (Bormio2016) 018.

\bibitem{MTV75} J. Meyer-Ter-Vehn, Nucl. Phys. A {\bf 249}, 111 (1975).

\bibitem{Gre19} E.T. Gregor, N.N. Arsenyev, M. Scheck, T.M. Shneidman, M. Thurauf, C. Bernards,
                A. Blanc, R. Chapman, F. Drouet, A.A. Dzhioev, {\it et al},
                J. Phys. G: Nucl. Part. Phys. {\bf 46}, 075101 (2019).

\bibitem{Mey88} R.A. Meyer, Hyp. Int. {\bf 43} 331, (1988) Phys.~Lett.~B {\bf 177}, 271 (1986).

\bibitem{Saz19} D.A. Sazonov, E.A. Kolganova, T.M. Shneidman, R.V. Jolos, N. Pietralla,
                and W. Witt, Phys. Rev. C {\bf 99}, 031304(R) (2019).

\bibitem{Stu04} A.E. Stuchbery, N. Benczer-Koller, G. Kumbratzki, and T.J. Mertzimekis,
                Phys. Rev. C {\bf 69}, 044302 (2004).

\bibitem{Isk24} \L.W. Iskra, R. M\v{a}rginean, S. Bottoni, N. M\v{a}rginean, {\it et al.},
                Phys. Rev. C {\bf 110}, 064312 (2024).

\bibitem{Rza09} T. Rz\c{a}ca-Urban, K. Sieja, W. Urban, F. Nowacki, J.L. Durell, A.G. Smith,
                and I. Ahmad, Phys. Rev. C {\bf 79}, 024319 (2009).

\end{thebibliography}
\end{document}